\documentclass[12pt,oneside]{article}

\usepackage{bm}
\usepackage{psfrag}
\usepackage[mathscr]{eucal}
\usepackage{times}
\usepackage{epsfig,graphicx,color,pstricks}
\usepackage{amsmath,amssymb,amsthm,amsfonts,eucal,latexsym,mathcom}
\usepackage{cite}
\usepackage[all]{xy}
\usepackage{subfigure}
\usepackage{graphics}

\newtheorem{theorem}{Theorem}[section]

\newtheorem{corollary}[theorem]{Corollary}
\newtheorem{remark}[theorem]{Remark}

\begin{document}

%

\title{Interference Channel with Generalized Feedback (a.k.a. with source cooperation)
\\Part  I: Achievable Region.
\footnote{The work of S. Yang and D. Tuninetti was partially funded by NSF CAREER award 0643954.}}
\author{Shuang (Echo) Yang, and Daniela~Tuninetti\\
  University of Illinois, Chicago,\\
  Electrical and Computer Engineering Department,\\
  Chicago, IL 60607, USA,\\
  \texttt{syang9@uic.edu, danielat@uic.edu }}
\maketitle

\begin{abstract}
An Interference Channel with Generalized Feedback (IFC-GF) is a model for a
wireless network where several source-destination pairs compete for
the same channel resources, and where the sources have the ability to
sense the current channel activity. The signal overheard from the
channel provides information about the activity of the other users,
and thus furnishes the basis for cooperation.  In this two-part paper
we study achievable strategies and outer bounds for a
general IFC-GF with two source-destination pairs.
We then evaluate the proposed regions for the Gaussian channel.

Part I: Achievable Region.
We propose that the generalized feedback is used to gain knowledge about
the message sent by the other user and then exploited in two ways:
(a) to {\em relay} the messages that can be decoded at both destinations--thus
realizing the gains of beam-forming of a distributed multi-antenna system--and
(b) to {\em hide} the messages that can not be decoded at the non-intended
destination--thus leveraging the interference ``pre-cancellation'' property of
dirty-paper-type coding.  We show that our achievable region generalizes several
known achievable regions for IFC-GF and that it reduces to known achievable regions
for some of the channels subsumed by the IFC-GF model.

Part II: Outer Bounds.
We strengthen the cut-set bound in two ways.
We first derive two new sum-rate bounds by using Sato's
``receiver cooperation with worst possible correlation'' idea:
we enhance the channel by letting also the destinations cooperate,
but--as in broadcast channels--we choose the
correlation among the destination outputs
that gives the tightest bound.  We show that several
bounds known in the literature--each derived with an ad-hoc
technique--are in fact examples of this ``receiver cooperation
with worst possible correlation'' idea.  We then derive a
sum-rate outer bound for a class of channels
with a special interference structure.
This class includes the Gaussian  channel.
When evaluating the proposed outer bound for the Gaussian channel,
we observe that our bound is the tightest among existing bounds
for certain range of parameters.
\end{abstract}

\section{Introduction}
\label{sec:intro inner}
The practical bottleneck of today's communication networks is interference.
The solution of commercial available networks is thus to avoid interference
through division of the resources, i.e., time, spectrum, space and waveforms,
among the competing users.  This approach is appealing in practice because
it results in a simple network architecture.  It might also appear a
good solution in theory since it is well known that in SISO (Single Input Single Output)
uplink and downlink wireless channels with perfect centralized channel
state information, it is sum-rate optimal to allocate all the available channel
resources to the user who experiences the instantaneous
highest channel gain~\cite{knopp_humblet:icc95}.
Perfect orthogonalization of the users is however not
possible in practice.  The practical approach to deal with residual interference
is to treat it as noise, thus completely neglecting its structure.
Whenever interference is treated as noise, the system becomes rate-limited even
when there is no power limitation at the transmitters.  This negative view
of interference was further reinforced by scaling law results of the early
00's that showed that the total rate of the wireless network with $K$ users
only scales as $\sqrt{K}$, thus yielding
a vanishing per-user rate as the network grows~\cite{gupta_kumar2:it001}.
It was also conjectured that, even with cooperation, the multiplexing gain of
an interference network would be one irrespectively of the number of users,
thus again the per-user rate would
vanish as the number of users in the network grows~\cite{HostMadsen-Nosratinia:isit2005}.

On the other hand, it has been known since the mid~70's that there exist
channels where interference does not reduce capacity~\cite{carleial:ifcstrong:it1975,
sato:IFCoutit1978}.
These channels are said
to have ``very strong interference'', that is, the power imbalance
between the useful signal and the interfering signal at a receiver
is so large that a given user can first decode the interfering signal
by treating its own signal as noise,
then strip the (now known) interference from its received signal, thus effectively
having  an interference-free channel.  This early example showed that interference
should not be treated as noise in general--in fact, as opposed to noise, it has a
structure that can be exploited--and that power imbalance must be leveraged upon
in the system design.  With these observations in mind, much progress
has been made recently in understanding the ultimate performance limits of
interference networks~\cite{etkin_tse_hua:withinonebit:subIt06,Jafar:2009:relays}.
In the authors' opinion, the major recent result has been to
show that in an interfering network with $K$ users the sum-rate capacity
scales as $\frac{K}{2}\log(1+{\rm SNR})$. This implies that each user can get
half the rate it would get if it were alone on the network, no matter how
how many users are present in the network.  The technique that achieves this
capacity scaling is referred to as {\em interference alignment}~\cite{Jafar:2009:relays}.
Under the interference alignment
paradigm, users do not avoid interference. Instead, they make sure that the
interference they collectively generate at a given receiver is neatly confined
in a specific signal subspace.  This leaves the complement of this subspace
interference free.  With interference alignment, a per-user
rate of $\frac{1}{2}\log(1+{\rm SNR})$ can be achieved
irrespectively of the number of users.


In this work we consider networks of {\em full-duplex} nodes, where several
source-destination pairs share the same channel.  The channel is assumed to be
static so that every node in the network has {\em perfect knowledge
of the channel sate}. Extensions to networks of half-duplex nodes and/or of
time-varying channels, such as fading channels, are planned as part of future work.
We focus on the case where all nodes can listen to the channel activity.
In particular, we are interested in the case where the sources can ``overhear''
what the other sources are sending, as in relay networks.
The key observation is that interference due to simultaneous communications
effectively spreads ``common information'' around the network.  This information
provides the basis for cooperation among otherwise uncoordinated nodes.

\subsection{Related Works}
The signal a node can ``overhear'' form the channel is a form
of feedback. To distinguish this feedback information
from the classical Shannon output feedback, Willems referred to it
as {\em Generalized Feedback} (GF)~\cite{willems:phd82}.
GF encompasses a wide range of possible situations, examples are:
{\em noisy output feedback} (with the non-feedback case
and the output feedback case at the two extremes); {\em conferencing
encoders} (where there are separate noise-free and interference-free links
between the sources, each of finite capacity); the case of
GF with independent noises, sometimes referred to as {\em user cooperation}.
The motivation to study GF comes from the work of Gaarder and Wolf~\cite{Gaarder_Wolf:it1975} who showed that output feedback can enlarge
the capacity  region of a MAC channel.
Here, we consider the general case of GF. Hence our results can
be specialized to all the above situations.  Our approach--especially
when in dealing with outer bounds--shows that all the above cases can be
dealt with in great generality. One of our contributions is to show a
unifying way of deriving some of the results available in the literature.

\paragraph{MAC-GF}
In~\cite{willems:phd82}, an achievable region for the MAC-GF
(Multiple Access Channels with Generalized Feedback) was derived.
The two main ingredients are {\em regular block-Markov superposition coding}
and {\em backward decoding}.  Block-Markov coding, also known as Decode-and-Forward
from the work by Cover and A.~El Gamal~\cite{Cover_AelGamal:relay:it1979}
in the context of relay channels, works as follows. Communication proceeds over
a frame of $N>1$ slots. The source splits the message to be transmitted in a
given slot into two parts. The first part is decoded by the destination,
but treated as noise by the relay. The second part is decoded also by the relay,
that then retransmits it to the destination in the next slot. Because the source knows
what the relay is going to send in the next slot, it can ``coordinate'' with the relay.
In each slot, the destination receives the superposition of the new information
(sent by the source) and the repetition of part of the old information (forwarded
by the relay).  The receiver waits until the whole frame has been
received.  Then appropriately combines the information sent in consecutive slots
to recover all the transmitted messages.

Willems's coding scheme for Gaussian channels was popularized by  Sendonaris et al.~\cite{Sendonaris_Erkip_Aazhang:coopdiv1:comm03}
under the name of {\em user cooperation diversity} in the context of cellular networks.
In~\cite{Sendonaris_Erkip_Aazhang:coopdiv1:comm03} it was showed that cooperation between
users achieves collectively higher data rates or, alternatively, allows to reach the same
data rates for less transmission power. Since the publication of~\cite{Sendonaris_Erkip_Aazhang:coopdiv1:comm03},
the interest in cooperative strategies has not ceased to increase (we do not
attempt here to review all the extensions of~\cite{Sendonaris_Erkip_Aazhang:coopdiv1:comm03} for
sake of space).

Although the MAC-GF has proved to be instrumental in understanding the potential
of user cooperation in networks, it is not as well suited for ad-hoc/peer-to-peer
networks, where the absence of coordination among users exacerbate the problem of
interference.  Host-Madsen~\cite{HostMadsen:it2006} first extended the Gaussian MAC-GF
model of~\cite{Sendonaris_Erkip_Aazhang:coopdiv1:comm03} to the case of Gaussian
IFC-GF. Before dwelling into the
literature of IFC-GF, we briefly revise the known results on IFC without feedback.

\paragraph{IFC without feedback}
The capacity region of a general IFC without feedback is still unknown.
The largest achievable region is due to Han and Kobayashi~\cite{Han_Kobayashi:it1981},
whose ``compact'' expression appeared~\cite{Fah_Garg_Motani:sub2006}.
In IFCs without feedback, communications is as follows.
Each transmitter splits its message in two parts: a {\em common message} and a
{\em private message}.  The two messages are superimposed and sent through the channel.
Each receiver decodes its intended common and private messages, as well as the common
message of the other user, by treating the other user's private message as noise.
The goal of this joint decoding is to reduce the interference level.  The idea of
information splitting was first proposed by Carleial~\cite{carleial:ifc:it1978}, to
whom many other early results on IFCs are due.

The Han-Kobayashi scheme is optimal in strong interference~\cite{carleial:ifcstrong:it1975, costa:awgnifc:it1985,costa_aelgamal:it1987,sato:IFCoutit1978,kramer:ifcout:it04}, and  it is shown to be sum-rate optimal in mixed interference~\cite{tuninettiweng:isit2008,canadanoisy}, in very-weak interference~\cite{chen-kramer:isit2008-out,canadanoisy,venunoisy2008}, for the
Z-IFC~\cite{sason:it2004} (where only one receiver experiences interference), and for
certain semi-deterministic channels~\cite{costa_aelgamal:it1982,telatar:tse:subIt07}.  Moreover, a simple rate-splitting choice
in the Han-Kobayashi scheme is optimal to within 1~bit for the Gaussian IFC~\cite{etkin_tse_hua:withinonebit:subIt06}.

{\em In this work, we propose an achievable region that
combines the idea of information splitting
with that of Block Markov Coding \& Backward Decoding.}

\paragraph{Early work on IFC-GF}
Host-Madsen~\cite{HostMadsen:it2006} first studied inner and outer bounds for the
sum-rate of IFC with both source and destination cooperation.  He showed
that in IFC networks with two SISO source-destination pairs, where the channel gains
stay constant and the user' powers increase, the multiplexing gain is one,
with both types of cooperation (instead of two, which would be the case if
cooperation were equivalent to a 2$\times$2 virtual MIMO channel).
In~\cite{HostMadsen:it2006}, several achievable
strategies are proposed  for the Gaussian channel only,
each relatively simple and tailored to a specific sets
of channel gains.

The work in~\cite{HostMadsen:it2006} was extended to a general DMC (Discrete Memoryless Channel) IFC-GF in~\cite{ifcgf-tuninetti-isit2007,CaoChen-IFCGF:DPC-ISIT2007,jiangoutputfeedbackciss2007}.  These works proposed to add one extra level of information splitting to the original Han-Kobayashi scheme.

\paragraph{IFC-GF: Cooperation on sending the common information}
In~\cite{ifcgf-tuninetti-isit2007}, we proposed to further split the common message in two parts: one part (referred to as {\em non-cooperative common information}) is as in the Han-Kobayashi scheme, while the other part (referred to as {\em cooperative common information}) is decoded at the other source too.  The sources use a block-Markov encoding scheme where in a given slot they ``retransmit'' the cooperative common information learnt in the previous slot.  This cooperation strategy aims to realize the beam-forming gain of a distributed MISO channel.

A scheme similar to ours in~\cite{ifcgf-tuninetti-isit2007} was independently proposed by Jiang et al. in~\cite{jiangoutputfeedbackciss2007} for the IFC with output feedback (i.e., not for a general GF setting). The difference lies in the way the cooperative  common information is dealt with, both at the sources and at the destination.
It is not clear which scheme achieves the largest achievable region (we will elaborate more on this point later on).

Recently, Prabhakaran and Viswanath~\cite{vinodSC2009} developed an outer bound for the sum-rate of a symmetric Gaussian IFC-GF with independent noises, inspired by the semi-deterministc model of~\cite{telatar:tse:subIt07}; they showed that their upperbound is achievable within a constant gap of 18~bits when the cooperation link gains are smaller than the direct link gains; their achievable strategy is a simple form of our region in~\cite{ifcgf-tuninetti-isit2007}.

{\em Cooperation on sending the common information is the essence of the first achievable region presented in Part~I this work.}

\paragraph{IFC-GF: Cooperation on sending the private information}
In~\cite{CaoChen-IFCGF:DPC-ISIT2007}, the authors proposed to further split the private message in two parts:  one part (which we shall refer to as {\em non-cooperative private information}) is as in the Han-Kobayashi scheme, while the other part (which we shall refer to as {\em cooperative private information}) is decoded at the other source too.  The sources use a block-Markov encoding scheme where in a given slot they ``hide'' the cooperative private information learnt in the previous slot to their intended receiver by using Gelfand-Pinsker coding~\cite{gelfand_pinsker_binning,Costa:dpc:it83}.  An approach similar to~\cite{CaoChen-IFCGF:DPC-ISIT2007}
(commonly referred to as ``dirty paper coding'' for Gaussian channels~\cite{Costa:dpc:it83}) was already used
in~\cite{HostMadsen:it2006} to characterize the high SNR sum-rate
capacity of the  Gaussian IFC-GF.  In~\cite{HostMadsen:it2006}, it was first noted that in the high SNR regime,
nulling the interference (i.e., an ancestor of interference alignment) is asymptotically optimal.

In~\cite{biaojournalofisit2007}, the ideas of~\cite{CaoChen-IFCGF:DPC-ISIT2007}
and of~\cite{ifcgf-tuninetti-isit2007} were merged into a scheme where
both the common and the private information are split into two parts.
In~\cite{ifcgf-tuninetti-ciss2008} we proposed a different (more structured)
coding strategy than in~\cite{biaojournalofisit2007}.
As opposed to~\cite{biaojournalofisit2007}, in~\cite{ifcgf-tuninetti-ciss2008}
we did not use independent Gelfand-Pinsker binning codes to pre-cancel the effect
of the cooperative private information.  Instead, we proposed to
superimpose several Gelfand-Pinsker binning codes
(inspired by the work in~\cite{maricett} for cognitive IFCs)
in such a way that the different binning stages are
performed sequentially and conditionally on the previous ones.
In addition, we also added a binning step similar to
Marton's achievable scheme for a general two-user broadcast
channel~\cite{marton:ach_reg}. The broadcast-type binning step
is possible in IFC-GFs because each encoder
knows part of the message sent by the other encoder, i.e.,
each transmitter is partially cognitive in the sense of~\cite{maricett}.
A simple form of~\cite{ifcgf-tuninetti-ciss2008}'s region was shown to be optimal within a constant gap of 18~bits for the symmetric Gaussian IFC-GF with independent noises when the cooperation link gains are larger than the direct link gains.

{\em The second achievable region presented in Part~I of this work is a further enhancement of~\cite{ifcgf-tuninetti-ciss2008}.}

\paragraph{IFC with degraded output feedback}
The GF setting covers a wide range of situations.
In particular, the case of degraded output feedback has been studied by a number
of groups.  Degraded output feedback refers to the case where the GF
signal received at a source is a noisier version of the signal received
at the intended destination.  When the variance of the extra noise on
the GF signal is zero, we have the so called output feedback.
Kramer in~\cite{kramer7,kramer8} developed inner and outer bounds for the Gaussian channel
with output feedback. Kramer and Gastpar~\cite{kramergastpar9}, and more recently
Tandon and Ulukus~\cite{tandonulukus2008}, derived an outer bound for the degraded output feedback
case based on the dependance balance idea of Hekstra and Willems~\cite{Hekstra_Willems:1989}.
Suh and Tse~\cite{suhtse} developed a novel outer bound for the Gaussian case with
output feedback and showed it to be achievable to within 1.7~bits.
In~\cite{suhtse} it was also shown that the achievable region of~\cite{jiangoutputfeedbackciss2007}
is optimal for the case of deterministic channels with output feedback.

{\em As mentioned earlier, our achievable region is applicable to all forms of GF,
thus also to the case of output feedback.}


\paragraph{Other channel models}
The IFC-GF reduces to some well known channel models.  Under certain conditions, it reduces to a Broadcast Channel (BC)~\cite{marton:ach_reg}, or to MAC-GF~\cite{willems:phd82}, or to a Relay Channel (RC)~\cite{Cover_AelGamal:relay:it1979}, or to a Cognitive IFC (C-IFC)~\cite{rini2010sifcdmc}.

{\em We shall also describe how our results encompass known results for these channels.}

\subsection{Summary of Contributions}
In this first part of the paper we present our achievable regions.
We first present a region where cooperation is on sending the common
information only.  The purpose is to highlight the key elements of our
encoding and  decoding scheme before proceeding to describe our more
general scheme, where the sources cooperate in sending both the common
and the private messages.  Our contributions are as follows.

Cooperation to send the common information only:
\begin{enumerate}
\item
We propose a structured way of superimposing the
different codebooks that greatly simplifies the error analysis.
Our codebook ``nesting'' is such that the ``cloud center'' codebook is the
one all terminals will be decoding, i.e., the cooperative common codebook,
to which we superimpose the non-cooperative common codebook
(to be decoded by the two destinations but not by the other source)
and finally we superimpose the non-cooperative private codebook
(to be decoded at the intended receiver only).  We shall see that
although the destinations have to decode five messages, only 5 out
of the possible $2^5-1=31$ error events matter.

\item
We then perform Fourier Motzking elimination to obtain a region
with only five types of rate bounds, as in the case of IFC without feedback.
This was not immediately obvious since, with GF, each message is split into three
parts, and not in two as for the case without feedback.

\item
We also show that when cooperation is on sending the common
information only, the achievable rate region cannot be enlarged
if the sources are required to decode more information than they
will eventually ``relay'' to the destinations.

\item
We show how our region does reduce to the achievable regions
of the channels subsumed by the IFC-GF model.
\end{enumerate}

For the more general form of cooperation
(to send both the common information and the private information):
\begin{enumerate}
\item
We extend our previous achievable region so as to include cooperation on
sending the private information too.  Our achievable region uses superposition
of several Gelfand-Pinsker binning codes.  We propose a structured way of binning
that reduces the number of rate constraints necessary to guarantee arbitrary
small probability of error at the encoders.

Moreover, we propose to ``jointly bin'' the different codebooks,
rather than perform several binning steps, one per codebook.
Our proposed joint binning is similar in spirit to multiple description
source coding.
\item
The error analysis at the decoders could be very messy and lengthy with so
many messages to decode.  Our proposed method for error analysis
leverages the way the codebooks are superimposed.  The type of error analysis
we use could be of interest in its own.  In particular, with our encoding
structure, we show that only 28 error events matter, out of the
$2^8-1=255$ possible.
\item
We show how our new region does reduce to the achievable regions
of the channels subsumed by the IFC-GF model.
Our region however does not give the largest possible achievable
rate for the relay channel because it does not include Compress-and-Forward
~\cite{Cover_AelGamal:relay:it1979}.~\footnote{We however notice here that the performance loss due to
Decode-and-Forward is less than within 1~bit for
the Gaussian relay channel.}
Extensions of the current achievable region so as to include Compress-and-Forward are left for future work.
\item
For the Gaussian channel, by means of a numerical example, we show that cooperation greatly increases the achievable rates with respect to the case without GF.
\end{enumerate}


The rest of the paper is organized as follows.
Section~\ref{sec:model} introduces the channel model and the notation.
Section~\ref{sec:ifc without feedback} revises known results for IFC without feedback.
Section~\ref{sec:inner} presents our achievable regions; in particular,
subsection~\ref{sec:superposition-only} describes the region where cooperation is on sending the common message only, and
subsection~\ref{sec:superposition+binning} describes the region where cooperation is also on sending the private message.
Section~\ref{sec:example suponly} evaluates the achievable region in subsection~\ref{sec:superposition-only} for the Gaussian channel (the evaluation of the region in subsection~\ref{sec:superposition+binning} is postponed to Part-II of this paper where we will also compare it with an outer bound). Section~\ref{sec:conc inner} concludes Part~I of this paper.
All the proofs are in the Appendix.

\section{Network Model and Definitions}
\label{sec:model}
Fig.~1 shows an IFC-GF with two source-destination pairs.
It consists of a channel with two input alphabets $(\Xc_1, \Xc_2)$,
four output alphabets $(\Yc_1, \Yc_2,\Yc_3, \Yc_4)$,
and a transition probability
$P_{Y_1\,Y_2\,Y_3\,Y_4|X_1\,X_2}$.
We assume that all the alphabets are finite sets
(the extension to continuous alphabets follows from
standard argument~\cite{book:cover_thomas:it})
and that the channel is memoryless.
Source~$u$, $u\in\{1,2\}$, has a message $W_{u}$ for destination~$u$.
The messages $W_{1}$ and $W_{2}$ are independent and uniformly distributed over the set
$\{1,\cdots,\eu^{n\,R_1}\}\times \{1,\cdots,\eu^{n\,R_2}\}$, where $n$ denotes
the codeword length and $R_u$ the transmission rate for user~$u$, $u\in\{1,2\}$.
At time $t$, $t\in\{1,\cdots,n\}$, source~$u$ maps its message $W_{u}$
and its past channel observations $Y_{u}^{t-1}$ into a channel input symbol
\[
X_{u,t} = f^{(n)}_{u,t}(W_{u},Y_{u}^{t-1}):
\quad
f^{(n)}_{u,t}:  \{1,\cdots,\eu^{n\,R_u}\} \times  \Yc_{u}^{t-1} \to \Xc_{u}.
\]
At time $n$, destination~$u$ outputs the estimate
of its intended message $W_u$ based on all its channel observations $Y_{u+2}^{n}$,
i.e.,
\[
\widehat{W}_u = g^{(n)}_u(Y_{u+2}^{n}):
\quad
g^{(n)}_{u}: \Yc_{u+2}^n \to \{1,\cdots,\eu^{n\,R_u}\}.
\]
The capacity region is the closure
of all rate pairs $(R_1,R_2)$ such that
\[
\max_{u\in\{1,2\}}\Pr[\widehat{W}_{u}\not = W_{u}] \to 0
\quad\text{as}\quad n\to\infty.
\]

\begin{figure}
\begin{center}
\includegraphics[width=10cm]{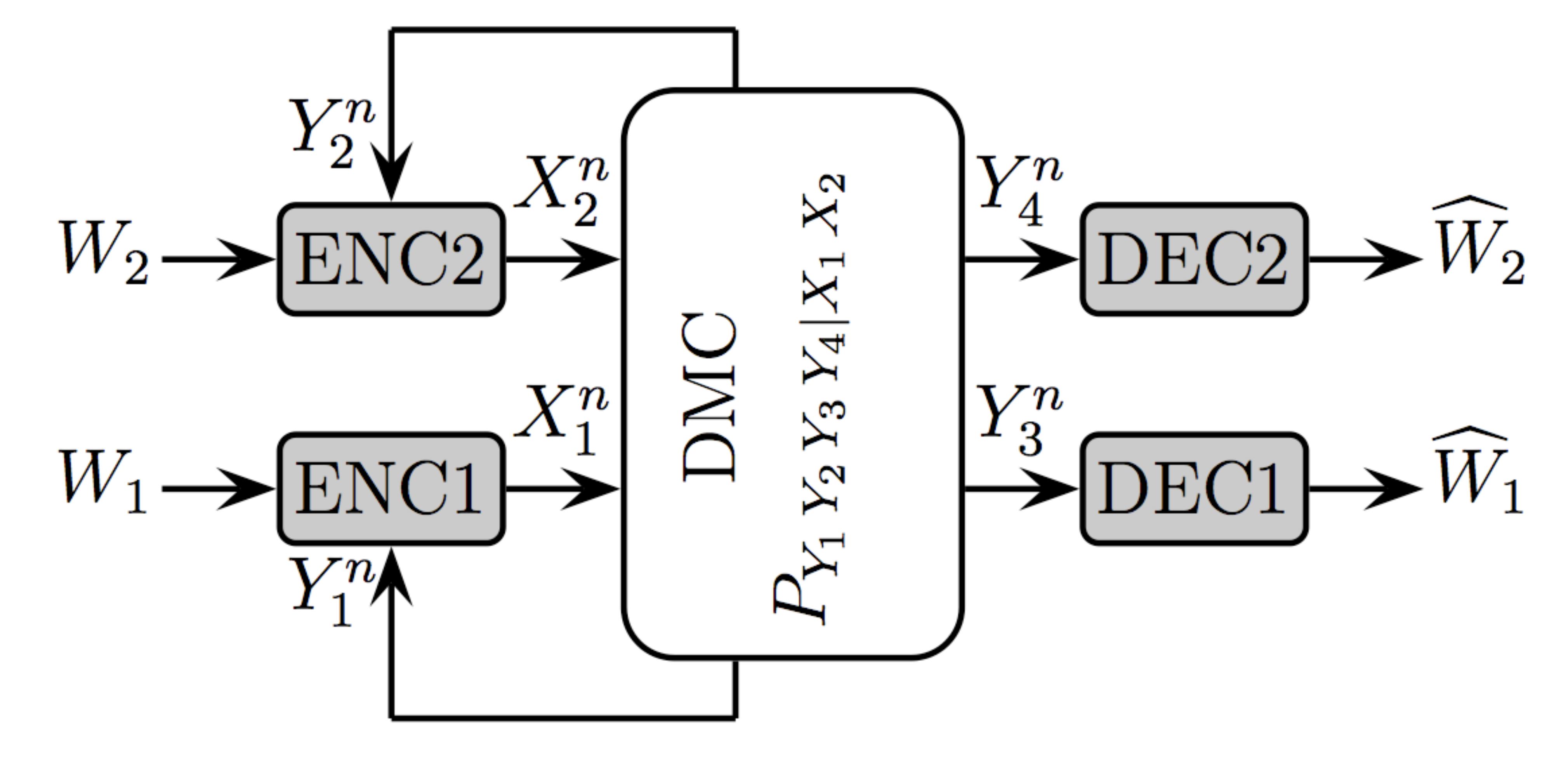}
\label{fig:ifcgf:model}
\caption{An Interference Channel with Generalized Feedback with two source-destination pairs.}
\end{center}
\end{figure}

\subsection{Notation}
The symbol $R_{xyz}$ is used to indicate the information
rate ``from source~$x$ to destination~$y$ with the help of~$z$'',
with $x\in\{1,2\}$,  $y\in\{0,1,2\}$ and $z\in\{c,n\}$.
In particular,
$y=0$ means that the message is decoded at both destinations (common message),
$y=1$ means that the message is decoded only at destination~1 (private message for user~1), and
$y=2$ means that the message is decoded only at destination~2 (private message for user~2);
$z=c$ means that the message is sent cooperatively by both sources, and
$z=n$ means that the message is sent non-cooperatively.

$T_\epsilon^{(n)}(P|\Sc)$ indicates the set
of length-$n$ sequences that are
strongly $\epsilon$-typical with respect to
the distribution $P$, conditioned on the sequences in $\Sc$~\cite{book:csiszar_korner:it}.

\section{IFC without feedback}
\label{sec:ifc without feedback}
In this section we briefly revise the best known inner and outer bound regions
for an IFC {\em without feedback}. This will
make the exposition of our coding
strategy and outer bound easier. The channel transition probability of an
IFC without feedback is
\[
P_{Y_1\,Y_2\,Y_3\,Y_4|X_1\,X_2}
=P_{Y_1\,Y_2}P_{Y_3\,Y_4|X_1\,X_2},
\]
that is, the GF signals $Y_1$ and $Y_2$ are independent of the inputs
(for example $\Yc_{1}=\Yc_{2}=\emptyset$).

\subsection{Inner Bound Region}
The largest known achievable region for an IFC without feedback
is due to Han and Kobayashi~\cite{Han_Kobayashi:it1981}
and is as follows~\cite{Fah_Garg_Motani:sub2006}.

\noindent
{\em Class of Input Distributions:}
Consider a distribution from the class
\begin{align}
\label{eq:inpdf no gf}
  &P_{Q U_1 T_1 X_1 U_2 T_2 X_2 Y_3 Y_4}=
\\&P_{Q}P_{U_1 T_1 X_1|Q}
        P_{U_2 T_2 X_2|Q}
        P_{Y_3 Y_4|X_1 X_2}.\nonumber
\end{align}
The channel transition probability $P_{Y_3 Y_4|X_1 X_2}$
is fixed, while the other factors in~\reff{eq:inpdf no gf} can be varied.

\noindent
{\em Rate Splitting:}
The message $W_u\in\{1,...,\eu^{n R_u}\}$, $u\in\{1,2\}$, is split into two parts $(W_{u0n}, W_{uun})$:
$W_{u0n}\in\{1,...,\eu^{n R_{u0n}}\}$ is the common information decoded at both receivers while
$W_{uun}\in\{1,...,\eu^{n R_{uun}}\}$ is the private information decoded only at the intended receiver,
with $R_u=R_{u0n}+R_{uun}$.
Without feedback, all messages are sent non-cooperativly.

\noindent
{\em Coodbook Generation:}
Consider a distribution in~\reff{eq:inpdf no gf}.
Pick uniformly at random a
length-$n$ sequences $Q^n$
from the typical set $T_\epsilon^{(n)}(P_Q)$.
For the codeword $Q^n=q^n$,
pick uniformly at random $\eu^{n R_{10n}}$
length-$n$ sequences $U_1^n(k)$,
$k\in\{1,\cdots,\eu^{n R_{10n}}\}$,
from the typical set $T_\epsilon^{(n)}(P_{U_1|Q}|q^n)$.
For each codeword $Q^n=q^n$
and each codeword $U_1^n(k)=u_1^n(k)$,
pick uniformly at random $\eu^{n R_{11n}}$
length-$n$ sequences $T_1^n(m,k)$,
$m\in\{1,\cdots,\eu^{n R_{11n}}\}$,
from the typical set $T_\epsilon^{(n)}(P_{T_1|Q U_1}|q^n,u_1^n(k))$.
For each $Q^n=q^n,\ U_1^n(k)=u_1^n(k)$ and $T_1^n(m,k)=t_1^n(m,k)$,
choose uniformly at random a sequence $X_1^n(m,k)$
from the typical set $T_\epsilon^{(n)}(P_{X_1|Q U_1 T_1}|q^n,u_1^n(k),t_1^n(m,k))$.

The generation of the codebooks at source~2 proceeds similarly.

\noindent
{\em Encoding:}
In order to send the messages $W_u=(W_{u0n}, W_{uun})$,
source~$u$, $u\in\{1,2\}$, transmits
$X_u^n(W_{u0n}, W_{uun})$.

\noindent
{\em Decoding:}
Destination~1 decodes from $Y_{3}^n$ the triplet $(W_{10n},W_{20n},W_{11n})$
by searching for a unique pair $(i_1,j_1)$, $j_1\in\{1,...,\eu^{n R_{10n}}\}$ and
$i_1\in\{1,...,\eu^{n R_{11n}}\}$, and some index $j_2$, $j_2\in\{1,...,\eu^{n R_{20n}}\}$, such that
\[
(U_1^n(j_1),T_1^n(i_1,j_1),U_2^n(j_2),Y_3^n)\in T_\epsilon^{(n)}(P^{(\rm dec1)}_{U_1 T_1 U_2 Y_3|Q}|Q^n),
\]
where
\begin{align*}
P^{(\rm dec1)}_{U_1 T_1 U_2 Y_3|Q}
&=
\frac{\sum_{X_1,T_2,X_2}
P_{Q}
P_{U_1 T_1 X_1|Q}
P_{U_2 T_2 X_2|Q}
P_{Y_3|X_1 X_2}}{P_{Q}}
\\&=
P_{U_1 T_1|Q}
P_{U_2|Q}
\Big(\sum_{X_1,X_2}
P_{X_1|Q U_1 T_1}
P_{X_2|Q U_2}
P_{Y_3|X_1 X_2}\Big).
\end{align*}
If no pair $(i_1,j_1)$ is found, or more than one pair is found, the receiver sets $(i_1,j_1)=(1,1)$;
in this case we say that an error at destination~1 has occurred.

Decoding at destination~2 proceeds similarly.

\noindent
{\em Error Analysis:}
The error analysis can be found in~\cite{Fah_Garg_Motani:sub2006}.
The probability of error at destination~1 can be driven to zero if the rates
$R_u=R_{u0n}+R_{uun}$, $u\in\{1,2\}$, are such that
\begin{subequations}
\begin{align}
                R_{11n}&\leq I(Y_3\wedge   T_1| U_1,U_2,Q)\label{anof:y3:t1}
\\R_{20n}        +R_{11n}&\leq I(Y_3\wedge   T_1,U_2| U_1,Q)\label{anof:y3:t1,u2}
\\        R_{10n}+R_{11n}&\leq I(Y_3\wedge   T_1,U_1| U_2,Q)\label{anof:y3:t1,u1}
\\R_{20n}+R_{10n}+R_{11n}&\leq I(Y_3\wedge   T_1,U_1, U_2|Q)\label{anof:y3:t1,u1,u2},
\end{align}
\label{eq:ifcnogf at dec1}
\end{subequations}
and similarly, the probability of error at destination~2 can be driven to zero if
\begin{subequations}
\begin{align}
                R_{22n}&\leq I(Y_4\wedge   T_2| U_1,U_2,Q)\label{bnof:y4:t2}
\\R_{10n}        +R_{22n}&\leq I(Y_4\wedge   T_2,U_1| U_2,Q)\label{bnof:y4:t2,u1}
\\        R_{20n}+R_{22n}&\leq I(Y_4\wedge   T_2,U_2| U_1,Q)\label{bnof:y4:t2,u2}
\\R_{20n}+R_{10n}+R_{22n}&\leq I(Y_4\wedge   T_2,U_1,U_2| Q)\label{bnof:y4:t2,u2,u1}.
\end{align}
\label{eq:ifcnogf at dec2}
\end{subequations}

\noindent
{\em Achievable Region:}
The region given by the intersection of~\reff{eq:ifcnogf at dec1} and~\reff{eq:ifcnogf at dec2}
can be compactly expressed after Fourier-Motzkin elimination as~\cite{Fah_Garg_Motani:sub2006}:
\begin{theorem}[\cite{Fah_Garg_Motani:sub2006}]
\label{theorem:in for general IFC without feedback}
For any distribution in~\reff{eq:inpdf no gf} the following region is achievable:
\begin{subequations}
\begin{align}
  R_1      &\leq {\rm \reff{anof:y3:t1,u1}}, \label{nogf a}
\\R_2      &\leq {\rm \reff{bnof:y4:t2,u2}}, \label{nogf c}
\\R_1 + R_2&\leq \min\{{\rm \reff{anof:y3:t1,u1,u2}+ \reff{bnof:y4:t2}},\
                       {\rm \reff{anof:y3:t1}      + \reff{bnof:y4:t2,u2,u1}},\
                       {\rm \reff{anof:y3:t1,u2}   + \reff{bnof:y4:t2,u1}}
                       \},   \label{nogf g}
\\2R_1 + R_2&\leq{\rm \reff{anof:y3:t1,u1,u2}}
               + {\rm \reff{anof:y3:t1}}
               + {\rm \reff{bnof:y4:t2,u1}},   \label{nogf h}
\\R_1 + 2R_2&\leq{\rm \reff{anof:y3:t1,u2}}
               + {\rm \reff{bnof:y4:t2}}
               + {\rm \reff{bnof:y4:t2,u2,u1}}\label{nogf j},
\end{align}
\label{eq:ifcnogf}
\end{subequations}
Without loss of generality,
one can set $T_1=X_1$ and $T_2=X_2$ in~\reff{eq:ifcnogf},
and, by Caratheodory's theorem, choose the auxiliary random variables $(Q,U_1,U_2)$
from alphabets with cardinality $|\Qc|\leq 7$, $|\Uc_1|\leq |\Xc_1|+4$
and $|\Uc_2|\leq |\Xc_2|+4$.
\end{theorem}

\begin{remark}
\label{rem:in for general IFC without feedback redundancy of two constrints}
It was remarked in~\cite{Fah_Garg_Motani:sub2006} that
after Fourier-Motzkin elimination, the following bounds also appear:
\begin{align*}
  R_1      &\leq {\rm \reff{anof:y3:t1}}    
               + {\rm \reff{bnof:y4:t2,u1}},
\\R_2      &\leq {\rm \reff{bnof:y4:t2}}    
               + {\rm \reff{anof:y3:t1,u2}},
\end{align*}
however they can be shown to be redundant. Intutively,
the reason is as follows: if
\[
  I(Y_3\wedge   T_1| U_1,U_2,Q)
 +\underbrace{I(Y_4\wedge   T_2,U_1| U_2,Q)}_{\text{at destination~2}}
 ={\rm \reff{anof:y3:t1}}
 +{\rm \reff{bnof:y4:t2,u1}}
< {\rm \reff{anof:y3:t1,u1}}
 =\underbrace{I(Y_3\wedge   T_1,U_1| U_2,Q)}_{\text{at destination~1}},
\]
then decoding at destination~2 constrains the rate of source~1 too much;
in this case, destination~2 should not be required to
decode the common information $U_1$.
Indeed, it can be shown~\cite{Fah_Garg_Motani:sub2006}
(see also Appendix~\ref{Proof of the redundancy of two single-rate constraints})
that the rate points for which
${\rm \reff{anof:y3:t1}}+ {\rm \reff{bnof:y4:t2,u1}}< R_1$,
but otherwise satisfy all other rate constraints in~\reff{eq:ifcnogf},
are contained in the sub-region of~\reff{eq:ifcnogf} obtained by choosing
$U_1=\emptyset$; thus the constraint
$R_1\leq {\rm \reff{anof:y3:t1}}+ {\rm \reff{bnof:y4:t2,u1}}$
can be removed without enlarging the achievable region.
A similar reasoning holds for the constraint
$R_2 \leq {\rm \reff{bnof:y4:t2}}+ {\rm \reff{anof:y3:t1,u2}}$.
\end{remark}

\subsection{Outer Bounds}
As shown in the second part of this paper,
the outer bound techniques of~\cite{kramer:it2004} for the Gaussian channel,
and the bound in~\cite{telatar:tse:subIt07} for the semi-determinisic channel,
give:

\begin{theorem}[\cite{kramer:it2004} and \cite{telatar:tse:subIt07}]
\label{theorem:out for general IFC without feedback}
\begin{subequations}
For any $P_{U X_1 X_2}=P_{U}P_{X_1|U}P_{X_2|U}$, the following region is an outer bound for a general IFC without feedback:
\begin{align}
R_1      &\leq  I(X_1;Y_3|X_2,U) \label{nof-relay-r1},\\
    R_2  &\leq  I(X_2;Y_4|X_1,U) \label{nof-relay-r2},\\
R_1+R_2  &\leq  I(X_1;Y_3| {Y'}_4,X_2,U) + I(X_1,X_2;Y_4|U)\label{nof-kramer-sr a},\\
R_1+R_2  &\leq  I(X_2;Y_4| {Y'}_3,X_1,U) + I(X_1,X_2;Y_3|U)\label{nof-kramer-sr b},
\end{align}
\label{eq:out ifcnof part1}
where $P_{Y_3,{Y'}_4|X_1,X_2}$ (and similarly for $P_{{Y'}_3,Y_4|X_1,X_2}$)
is such that $P_{{Y'}_4|X_1,X_2}=P_{Y_4|X_1,X_2}$, i.e., same marginal distribution,
but otherwise $P_{Y_3,{Y'}_4|X_1,X_2}$ has arbitrary joint distribution.

Furthermore, if the channel is semi-deterministic as defined in
in~\cite{telatar:tse:subIt07}, that is,
if there exist deterministic functions $f_3,f_4,v_{32},v_{41}$
and noise random variables $Z_3,Z_4$ such that
\begin{align*}
  Y_3 &= f_3(X_1,v_{32}(X_2,Z_3))\ \text{where $f_3$ is invertible given $X_1$},
\\Y_4 &= f_4(X_2,v_{41}(X_1,Z_4))\ \text{where $f_4$ is invertible given $X_2$},
\end{align*}
then
\begin{align}
  R_1+R_2  &\leq
     H(Y_{3}| \widetilde{V}_{41},U)
   + H(Y_{4}| \widetilde{V}_{32},U) \nonumber\\&
   - H(\widetilde{V}_{41}|X_{1},U)
   - H(\widetilde{V}_{32}|X_{2},U) \label{nof-tt-r1+r2}
\\ 2R_1+R_2 &\leq
     H(Y_{3}| \widetilde{V}_{41},X_2,U)
   + H(Y_{3}|U)
   + H(Y_{4}| \widetilde{V}_{32},U) \nonumber\\&
   - H(\widetilde{V}_{41}|X_{1},U)
   -2H(\widetilde{V}_{32}|X_{2},U)
      \label{nof-tt-2r1+r2},
\\ R_1+2R_2 &\leq
     H(Y_{4}| \widetilde{V}_{32},X_1,U)
   + H(Y_{4}|U)
   + H(Y_{3}| \widetilde{V}_{41},U) \nonumber\\&
   -2H(\widetilde{V}_{41}|X_{1},U)
   - H(\widetilde{V}_{32}|X_{2},U)
     \label{nof-tt-r1+2r2}.
\end{align}
\label{eq:out ifcnof}
where $\widetilde{V}_{41}$ and $\widetilde{V}_{32}$ are independent copies
of $V_{41}=v_{41}(X_1,Z_4)$ and $V_{32}=v_{32}(X_2,Z_3)$ conditioned on the input $(X_1,X_2)$.
\end{subequations}
\end{theorem}

\begin{remark}
\label{rem:out for general IFC without feedback}
Other outer bounds exist for specific channels. For example, for the Gaussian channel
in weak interference,
Kramer~\cite{kramer:it2004} developed a bound based on the idea of enhancing the IFC and make it
equivalent to a degraded broadcast channels.  Shang et at~\cite{chen-kramer:isit2008-out}, simultaneously
and independently of~\cite{venunoisy2008,canadanoisy}, showed that for the Gaussian channel
in very weak interference, it is sum-rate optimal to treat the interference as noise.
Outer bounds based on the idea of dependance balance appeared in~\cite{tandonulukus2008,kramergastpar9}.
Extensions of these bounding techniques to non-Gaussian channels and/or their
evaluation is not straightforward and it will not be attempted in this paper.
\end{remark}

\section{Inner Bounds}
\label{sec:inner}
In this section we derive our achievable region.
We divide the section into two parts. In the first part
(Section~\ref{sec:superposition-only}), we propose
a scheme where the sources cooperate by ``beam-forming''
part of the common messages. 
In this scheme, all messages are superimposed and thus we refer to it
as {\em superposition-only} achievable region. In the second part
(Section~\ref{sec:superposition+binning}),
we propose a scheme where the sources also cooperate on
sending part of the private messages by using binning, or dirty paper coding.
We refer to this schemes as {\em superposition \& binning} achievable region.
The {\em superposition \& binning} achievable region includes the
{\em superposition-only} achievable region as special case.  We present them both to
guide the reader into the two possible cooperation mechanisms.

\subsection{Superposition-only achievable region}
\label{sec:superposition-only}
\noindent
{\em Class of Input Distributions:}
Consider a distribution from the class
\begin{align}
\label{eq:inpdf}
  &P_{Q V_1 U_1 T_1 X_1 V_2 U_2 T_2 X_2 Y_1 Y_2 Y_3 Y_4}=
\\&P_{Q}P_{V_1 U_1 T_1 X_1|Q}
        P_{V_2 U_2 T_2 X_2|Q}
        P_{Y_1 Y_2 Y_3 Y_4|X_1 X_2},\nonumber
\end{align}
that is, conditioned on $Q$, the random variables $(V_1, U_1, T_1, X_1)$
generated at source~1 are independent of
the random variables $(V_2, U_2, T_2, X_2)$
generated at source~2.
The channel transition probability $P_{Y_1 Y_2 Y_3 Y_4|X_1 X_2}$
is fixed, while the other factors in the distribution in~\reff{eq:inpdf} can be varied.

\noindent
{\em Rate Splitting and Transmission Strategy:}
The message $W_u\in\{1,...,\eu^{n R_u}\}$, $u\in\{1,2\}$,
is divided into three parts $(W_{u0c}, W_{u0n}, W_{uun})$:
$W_{u0c}\in\{1,...,\eu^{n R_{u0c}}\}$ is the part of the common message
sent cooperatively by the sources,
$W_{u0n}\in\{1,...,\eu^{n R_{u0n}}\}$ is the part of the common message
sent by source~$u$ alone (i.e., non-cooperatively), and
$W_{uun}\in\{1,...,\eu^{n R_{uun}}\}$ is the private private message
sent non-cooperatively, with $R_u=R_{u0c}+R_{u0n}+R_{uun}$.

We propose that the cooperative common message sent by a source is decoded at the
other source thanks to the generalized feedback. Then,
the sources send both cooperative common messages
to the receivers as in a virtual MIMO channels, thus realizing the gain
of beamforming. This is possible by using regular block Markov superposition encoding~\cite{coverleung:macperfectfeed:IT98}
at the sources and backward decoding~\cite{willems:phd82} at the destinations. In particular,
transmission occurs over a frame of $N$ slots of $n$ channel uses each.
Source~1 in slot $b$, $b\in\{1,\cdots,N-1\}$, has an estimate $W^{'}_{20c,b-1}$
of the cooperative common message sent by source~2 in the previous slot.
Similarly, source~2 in slot $b$ has an estimate $W^{''}_{10c,b-1}$
of the cooperative common message sent by source~1 in the previous slot.
The random variable $Q$ in~\reff{eq:inpdf} conveys the two cooperative common messages
$(W_{10c},W_{20c})$ from the previous time slot to the destinations. In slot $b$,
$b\in\{1,\cdots,N-1\}$, the random variables
$V_u$, $U_u$, and $T_u$ in~\reff{eq:inpdf}, $u\in\{1,2\}$, convey
the new cooperative common message $W_{u0c,b}$,
the new non-cooperative common message $W_{u0n,b}$, and
the new non-cooperative private message $W_{uun,b}$, respectively.

By setting $V_1=V_2=\emptyset$, i.e., no cooperative common messages,
our proposed encoding schemes reduces to the Han and Kobayashi~\cite{Han_Kobayashi:it1981}
for the IFC without feedback in Th.~\ref{theorem:in for general IFC without feedback}.
In this case, $Q$ acts as a simple time-sharing random variable.

\noindent
{\em Codebook Generation:}
Pick uniformly at random $\eu^{n (R_{10c}+R_{20c})}$
length-$n$ sequences $Q^n([i,j])$,
$i\in\{1,\cdots,\eu^{n R_{10c}}\}$ and $j\in\{1,\cdots,\eu^{n R_{20c}}\}$,
from the typical set $T_\epsilon^{(n)}(P_Q)$.
For each $Q^n([i,j])=q^n([i,j])$,
pick uniformly at random $\eu^{n R_{10c}}$
length-$n$ sequences $V_1^n(k,[i,j])$,
$k\in\{1,\cdots,\eu^{n R_{10c}}\}$,
from the typical set $T_\epsilon^{(n)}(P_{V_1|Q}|q^n([i,j]))$.
For each $Q^n([i,j])=q^n([i,j])$
and $V_1^n(k,[i,j])=v_1^n(k,[i,j])$,
pick uniformly at random $\eu^{n R_{10n}}$
length-$n$ sequences $U_1^n(\ell,k,[i,j])$,
$\ell\in\{1,\cdots,\eu^{n R_{10n}}\}$,
from the typical set $T_\epsilon^{(n)}(P_{U_1|Q V_1}|q^n([i,j]),v_1^n(k,[i,j]))$.
For each $Q^n([i,j])=q^n([i,j])$, $V_1^n(k,[i,j])=v_1^n(k,[i,j])$,
and $U_1^n(\ell,k,[i,j])=u_1^n(\ell,k,[i,j])$,
pick uniformly at random $\eu^{n R_{11n}}$
length-$n$ sequences $T_1^n(m,\ell,k,[i,j])$,
$m\in\{1,\cdots,\eu^{n R_{11n}}\}$,
from the typical set $T_\epsilon^{(n)}(P_{T_1|Q V_1 U_1}|q^n([i,j]),v_1^n(k,[i,j]),u_1^n(\ell,k,[i,j]))$.
For each $Q^n([i,j])=q^n([i,j])$, $V_1^n(k,[i,j])=v_1^n(k,[i,j])$,
 $U_1^n(\ell,k,[i,j])=u_1^n(\ell,k,[i,j])$,
and $T_1^n(m,\ell,k,[i,j])=t_1^n(m,\ell,k,[i,j])$,
pick uniformly at random one sequence  $X_1^n(m,\ell,k,[i,j])$
from the typical set
\[
T_\epsilon^{(n)}(P_{X_1|Q V_1 U_1 T_1}|q^n([i,j]),v_1^n(k,[i,j]),u_1^n(\ell,k,[i,j]),t_1^n(m,\ell,k,[i,j])).
\]

The generation of the codebooks at source~2 is similar.

\noindent
{\em Encoding:}
In slot $b$, $b\in\{1,\cdots,N\}$, given the new message
$W_{u,b}=(W_{u0c,b}, W_{u0n,b}, W_{uun,b})$ for $u\in\{1,2\}$,
the transmitted codewords are
\begin{align*}
X_1^n(W_{11n,b},W_{10n,b},W_{10c,b},[W_{10c,b-1}, W^{'}_{20c,b-1}]),\\
X_2^n(W_{22n,b},W_{20n,b},W_{20c,b},[W^{''}_{10c,b-1},W_{20c,b-1}]),
\end{align*}
with the ``boundary'' conditions $W_{u0c,0} = W_{u0c,N} = 1$, $u=\{1,2\}$, i.e.,
on the first slot of the frame there is no cooperative common information from
a previous slot to relay, and on the last slot of the frame there is
no new cooperative common information to send because there will not be a future slot to relay it.
With this scheme, user $u\in\{1,2\}$ transmits at an actual rate
$R'_u = (1-1/N)R_{u0c}+R_{u0n}+R_{uun}<R_u$ (because no new cooperative common
information is sent on the last slot). The rate $R'_u$ can be made arbitrarily
close to $R_u$ by taking the frame length $N$ to be sufficiently large.

\begin{figure}
\begin{center}
\includegraphics[width=8cm]{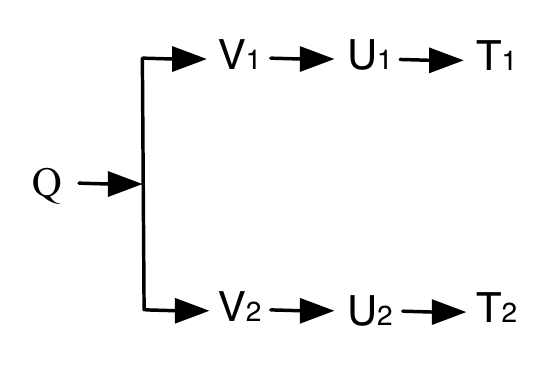}
\caption{A visual representation of the codebook generation for the
superposition-only achievable scheme. An arrow to a codebook/random variable indicates
that the codebook is superimposed to {\em all} the codebooks that precede it, and
codebooks linked by a vertical line are conditionally independent given
everything that precedes them.}
\label{fig:encoding isit 2007}
\end{center}
\end{figure}

Fig.~\ref{fig:encoding isit 2007} visualizes the proposed superposition
coding scheme: an arrow to a codebook/random variable indicates
that the codebook is superimposed to {\em all} the codebooks that precede it, and
codebooks linked by a vertical line are conditionally independent given
everything that precedes them.  For example, codebook $U_1$ is superimposed to
$Q$ and $V_1$, and it is conditionally independent of any codebook with index~2
when conditioned on $Q$.

\noindent
{\em Cooperation:}
In slot $b$, $b\in\{1,\cdots,N-1\}$, we can assume that the users' estimate of
the previous cooperative common messages is exact, that is,
$W^{''}_{10c,b-1}=W_{10c,b-1}$ and $W^{'}_{20c,b-1}=W_{20c,b-1}$,
since the total average probability of error at the receivers can be upper bounded
by the sum of the decoding error probabilities at each step, under the assumption
that no error propagation from the previous steps has occurred~\cite{willems:phd82}.
Next we describe how source~2 cooperates with source~1.
Source~1 proceeds similarly.

Source~2 at the end of slot $b$ decodes user~1's new cooperative common message
$W_{10c,b}$ carried by $V_1^n$ from its channel output
$Y_{2,b}^n$, knowing everything that was generated at source~2
at the beginning of the slot.
Formally, at the end of slot $b$, $b\in\{1,\cdots,N-1\}$,
source~2 has received $Y_{2,b}^n$ and looks for the unique
index $i\in\{1,...,\eu^{n R_{10c}}\}$ such that
\begin{align*}
(&
V_1^n(i,[\cdots]),
Y_{2,b}^n
)
\\&
\in T_\epsilon^{(n)}(P^{(\rm enc2)}_{V_1 Y_2|Q X_2}|\underline{X}_2^n),
\nonumber
\end{align*}
where the dots indicate known message indices,
where all that is known at source~2 is represented by
\[
\underline{X}_2 =(Q,V_2,U_2,T_2,X_2),
\]
and where
\begin{align*}
P^{(\rm enc2)}_{V_1 Y_2|Q X_2}
&=
\frac{
\sum_{U_1,T_1,X_1}
P_{Q}
P_{V_1 U_1 T_1 X_1|Q}
P_{V_2 U_2 T_2 X_2|Q}
P_{Y_2|X_1 X_2}}{P_{Q V_2 U_2 T_2 X_2}}
\\&=
P_{V_1|Q}
\Big(\sum_{X_1}
P_{X_1|Q V_1}
P_{Y_2|X_1 X_2}\Big).
\end{align*}
If none or more than one index $i$ is found, then source~2
sets $i=1$; in this case we say that an error has occurred at source~2.

\noindent
{\em Error Analysis:}
By standard arguments~\cite{ifcgf-tuninetti-isit2007} (see Appendix~\ref{app:proof of eq:enc1-pe}),
the probability of error at source~2
can be made as small as desired if
\begin{subequations}
\begin{align}
R_{10c} \leq I(V_1 \wedge   Y_2| Q,V_2,U_2,T_2,X_2).
\label{a:y2;v1}
\end{align}

\noindent
{\em Decoding:}
The destinations wait until the last slot of the frame (i.e. slot $N$)
has been received and then perform backward decoding. In slot $b\in\{N,\cdots,1\}$
destination~1 looks for the unique triplet
$(i_1,j_1,m_1)\in\{1,...,\eu^{n R_{10c}}\}\times \{1,...,\eu^{n R_{10n}}\} \times \{1,...,\eu^{n R_{11n}}\}$
and some pair $(i_2,j_2)\in\{1,...,\eu^{n R_{20c}}\}\times \{1,...,\eu^{n R_{20n}}\}$ such that
\begin{align*}
\Big(
Q^n(                    [i_1,i_2]),&
V_1^n(        W'_{10c,b},[i_1,i_2]),
U_1^n(    j_1,W'_{10c,b},[i_1,i_2]),
T_1^n(m_1,j_1,W'_{10c,b},[i_1,i_2]),\nonumber\\&
V_2^n(        W'_{20c,b},[i_1,i_2]),
U_2^n(    j_2,W'_{20c,b},[i_1,i_2]),\nonumber\\&
Y_{3,b}^n
\Big)\in T_\epsilon^{(n)}
(P^{(\rm dec1)}_{Q V_1 U_1 T_1 V_2 U_2 Y_3})
\end{align*}
where the pair $(W'_{10c,b},W'_{20c,b})$  was decoded in the previous step
($W_{u0c,B}=1$ by assumption, hence $W'_{u0c,B}=1$ too) and where
\begin{align*}
P^{(\rm dec1)}_{Q V_1 U_1 T_1 V_2 U_2 Y_3}
&= \sum_{X_1,T_2,X_2}
P_{Q}
P_{V_1 U_1 T_1 X_1|Q}
P_{V_2 U_2 T_2 X_2|Q}
P_{Y_3|X_1 X_2}
\\&=
P_{Q}
P_{V_1 U_1 T_1|Q}
P_{V_2 U_2|Q}
\Big(\sum_{X_1,X_2}
P_{X_1|Q V_1 U_1 T_1}
P_{X_2|Q V_2 U_2}
P_{Y_3|X_1 X_2}\Big).
\end{align*}
In words, destination~1 decodes the old cooperative
common messages in $Q^n$,
the current non-cooperative common messages in $(U_1^n,U_2^n)$, and
the current non-cooperative private message in $T_1^n$,
from its channel output $Y_{3,b}^n$.  The current
cooperative common messages in $(V_1^n,V_2^n)$ are known
from the decoding of slot $b+1$.
The current non-cooperative private message of user~2 in $T_2^n$
is treated  as noise.

If none or more than one triplet $(i_1,j_1,m_1)$ is found, then destination~1
sets $(i_1,j_1,m_1)=(1,1,1)$; in this case we say that an error has occurred
at destination~1.

\noindent
{\em Error Analysis:}
By standard arguments~\cite{ifcgf-tuninetti-isit2007}
(see Appendix~\ref{app:proof of eq:ifc-gf}),
the probability of error at destination~1
can be made as small as desired if
\begin{align}
  R_{11n}&\leq I(Y_3\wedge  T_1|Q,V_1,V_2,U_1,U_2)                 \label{a:y3;t1}
\\R_{11n}        +R_{20n}&\leq I(Y_3\wedge  T_1,U_2|Q,V_1,V_2,U_1) \label{a:y3;t1,u2}
\\R_{11n}+R_{10n}        &\leq I(Y_3\wedge  T_1,U_1|Q,V_1,V_2,U_2) \label{a:y3;t1,u1}
\\R_{11n}+R_{10n}+R_{20n}&\leq I(Y_3\wedge  T_1,U_1,U_2|Q,V_1,V_2) \label{a:y3;t1,u1,u2}
\\R_{11n}+R_{10n}+R_{20n}+(R_{20c}+R_{10c})
                         &\leq I(Y_3\wedge  T_1,U_1,U_2,Q,V_1,V_2) \label{a:y3;t1,u1,u2,q,v1,v2}.
\end{align}
\label{eq:gf at dec1}
Notice that
$\rm \reff{a:y3;t1}
\leq \min\{\reff{a:y3;t1,u2},\reff{a:y3;t1,u1}\}
\leq \max\{\reff{a:y3;t1,u2},\reff{a:y3;t1,u1}\}
\leq \reff{a:y3;t1,u1,u2}
\leq \reff{a:y3;t1,u1,u2,q,v1,v2}.$
\end{subequations}

By similar arguments, the probability of error at source~1
can be made as small as desired if
\begin{subequations}
\begin{align}
R_{20c} \leq I(V_2 \wedge   Y_1| Q,V_1,U_1,T_1,X_1),
\label{b:y1;v2}
\end{align}
and the probability of error at destination~2
can be made as small as desired if
\begin{align}
  R_{22n}                &\leq I(Y_4\wedge  T_2|Q,V_2,V_1,U_2,U_1) \label{b:y4;t2}
\\        R_{22n}+R_{10n}&\leq I(Y_4\wedge  T_2,U_1|Q,V_2,V_1,U_2) \label{b:y4;t2,u1}
\\R_{20n}+R_{22n}        &\leq I(Y_4\wedge  T_2,U_2|Q,V_2,V_1,U_1) \label{b:y4;t2,u2}
\\R_{20n}+R_{22n}+R_{10n}&\leq I(Y_4\wedge  T_2,U_2,U_1|Q,V_1,V_2) \label{b:y4;t2,u1,u2}
\\R_{22n}+R_{20n}+R_{10n}+(R_{20c}+R_{10c})
                         &\leq I(Y_4\wedge  T_2,U_1,U_2,Q,V_1,V_2).\label{b:y4;t2,u1,u2,q,v1,v2}
\end{align}
\label{eq:gf at dec2}
\end{subequations}

\noindent
{\em Achievable region:}
The intersection of the region in~\reff{eq:gf at dec1} with the
region in~\reff{eq:gf at dec2} can be compactly expressed
after Fourier-Motzkin elimination as follows:
\begin{theorem}
\label{theorem:in superposition for general IFC-GF}
For any distribution in~\reff{eq:inpdf}, the following region is achievable:
\begin{subequations}
\begin{align}
  &R_1 \leq     {\rm \reff{a:y3;t1,u1,u2,q,v1,v2}}                    \label{c1forgotten}
\\&R_1 \leq     {\rm \reff{a:y2;v1}+\reff{a:y3;t1,u1}}                \label{c1}
\\&R_2 \leq     {\rm \reff{b:y4;t2,u1,u2,q,v1,v2}}                    \label{c2forgotten}
\\&R_2 \leq     {\rm \reff{b:y1;v2}+\reff{b:y4;t2,u2}}                \label{c2}
\\&R_1	+R_2	\leq		{\rm \reff{a:y3;t1,u1,u2,q,v1,v2}+\reff{b:y4;t2}}                \label{c3}
\\&R_1	+R_2	\leq		{\rm \reff{a:y3;t1}              +\reff{b:y4;t2,u1,u2,q,v1,v2}}  \label{c4}
\\&R_1	+R_2	\leq		{\rm \reff{a:y2;v1}+\reff{b:y1;v2}
                         +\reff{a:y3;t1,u1,u2}+\reff{b:y4;t2}}        \label{c5}
\\&R_1	+R_2	\leq		{\rm \reff{a:y2;v1}+\reff{b:y1;v2}
                         +\reff{a:y3;t1}      +\reff{b:y4;t2,u1,u2}}  \label{c6}
\\&R_1	+R_2	\leq		{\rm \reff{a:y2;v1}+\reff{b:y1;v2}
                         +\reff{a:y3;t1,u2}   +\reff{b:y4;t2,u1}}     \label{c7}
\\&2R_1 +R_2 \leq		{\rm \reff{a:y2;v1}
                         +\reff{a:y3;t1}+\reff{a:y3;t1,u1,u2,q,v1,v2}+\reff{b:y4;t2,u1}}    \label{c8}
\\&2R_1	+R_2	\leq		{\rm 2\!\cdot\!\reff{a:y2;v1}+\reff{b:y1;v2}
                         +\reff{a:y3;t1}+\reff{a:y3;t1,u1,u2}+\reff{b:y4;t2,u1}}            \label{c9}
\\&R_1 +2R_2\leq		{\rm \reff{b:y1;v2}+
                          \reff{a:y3;t1,u2}+\reff{b:y4;t2}+\reff{b:y4;t2,u1,u2,q,v1,v2}}    \label{c10}
\\&R_1 +2R_2\leq		{\rm \reff{a:y2;v1}+2 \!\cdot\!\reff{b:y1;v2}
                         +\reff{a:y3;t1,u2}+\reff{b:y4;t2}+\reff{b:y4;t2,u1,u2}}            \label{c11}
\end{align}
\label{eq:ifcwithgf}
\end{subequations}
Without loss of generality, one can take $X_1=T_1$ and $X_2=T_2$ in~\reff{eq:ifcwithgf}.
\end{theorem}

We remark here that the proposed structured way of superimposing the
codebooks greatly simplifies the error analysis.
Our codebook ``nesting'', in fact, is such that the ``cloud center'' codebook is the
one all terminals will be decoding, i.e., the cooperative common coodebok,
to which we superimposed the non-cooperative common codebook
(to be decoded by the two destinations but not by the other source)
and finally we superimposed the non-cooperative private codebook
(to be decoded at the intended receiver only).  As a consequence,
although a destination has to decode five messages, only 5 out
of the possible $2^5-1=31$ error events matter (see~\reff{eq:gf at dec1} for destination~1
and~\reff{eq:gf at dec2} for destination~2).

\begin{remark}
\label{rem:in for general IFC-GF redundancy of two constrints}
The following rate constraints also appear after Fourier-Motzkin elimination:
\begin{align}
  &R_1 \leq{{\rm \reff{a:y2;v1}+\reff{a:y3;t1}+\reff{b:y4;t2,u1}}}\label{c1bis},
\\&R_2 \leq{{\rm \reff{b:y1;v2}+\reff{b:y4;t2}+\reff{a:y3;t1,u2}}}\label{c2bis}.
\end{align}
It can be shown (see Appendix~\ref{Proof of the redundancy of two single-rate constraints})
that~\reff{c1bis} and~\reff{c2bis} can be removed without enlarging the region in~\reff{eq:ifcwithgf}.
The intuitive argument is as for the channel without feedback in
Remark~\ref{rem:in for general IFC without feedback redundancy of two constrints}.
\end{remark}

The achievable region in~\reff{eq:ifcwithgf} subsumes achievable regions for other
multiuser channels. For example:
\begin{enumerate}

\item
By setting ${\rm \reff{a:y2;v1}=\reff{b:y1;v2}}=0$ in~\reff{eq:ifcwithgf},
we get the achievable region for an {\em\bf IFC without feedback} in~\reff{eq:ifcnogf}.
 In fact,~\reff{c3} is redundant because of~\reff{c5};
         ~\reff{c4} is redundant because of~\reff{c6};
         ~\reff{c8} is redundant because of~\reff{c9}; and
         ~\reff{c10} is redundant because of~\reff{c11}.
With $\rm \reff{a:y2;v1}=\reff{b:y1;v2}=0$, the region in~\reff{eq:ifcwithgf}
is unchanged if we set $V_1=V_2=Q$ (since the random variables
$Q$, $V_1$ and $V_2$ always appear together).

\item
By setting $Y_1=Y_3$ and $Y_2=Y_4$ we obtain an {\em\bf IFC with output feedback}
as studied in~\cite{kramer7,jiangoutputfeedbackciss2007,suhtse}.  In particular,
the region in~\cite[eq.(18)-(29)]{jiangoutputfeedbackciss2007} has the same codebook
structure of our Th.~\ref{theorem:in superposition for general IFC-GF}.
However, they differ in the encoding and decoding of the messages in $Q$.
In our achievable region, the sources repeat in $Q$ the whole past cooperative
common messages. In~\cite{jiangoutputfeedbackciss2007}, the sources repeat in $Q$
a quantized version of the past cooperative common message indices. In principle,
thus the encoding in~\cite{jiangoutputfeedbackciss2007} is more general.
However, in our achievable region, the destinations decode $Q$ jointly with all other
messages. In~\cite{jiangoutputfeedbackciss2007}, $Q$ is decoded first by treating
all the rest as noise, and then all the
other messages are jointly decoded. It is thus not clear a priori whether the
rate-saving due to sending a quantized version of $Q$ are wiped out by the
low rate imposed by decoding $Q$ first.
A formal comparison among the two regions is however difficult, since
the two could appear different but be actually the same when the union over
all possible input distributions is taken.

\item
By setting ${\rm \reff{a:y2;v1}}=\infty$, and ${\rm \reff{b:y1;v2}}=0$ in~\reff{eq:ifcwithgf},
we obtained a {\em\bf cognitive interference channel},
in which source~2 knows the message of source~1, while source is unaware of
the messages sent by source~2.  The cognitive channel has also been referred to as
{\em\bf IFC with degraded message set}~\cite{Wu_Vishwanath_Arapostathis:it2006}
and as {\em\bf IFC with unidirectional cooperation}~\cite{MaricYatesKramer:IFCdegmsgset:ita06}.

With ${\rm \reff{a:y2;v1}}=\infty$, and ${\rm \reff{b:y1;v2}}=0$
(and with $U_1=V_1=V_2=Q$ without loss of generality) the achievable
region  in~\reff{eq:ifcwithgf} reduces to:
\begin{subequations}
\begin{align}
  &R_1 \leq     {\rm \reff{a:y3;t1,u1,u2,q,v1,v2}}                    \label{c1forgotten {a:y2;v1}=infty {b:y1;v2}=0}
\\&R_2 \leq     {\rm \reff{b:y4;t2,u2}}                \label{c2 {a:y2;v1}=infty}
\\&R_1	+R_2	\leq		{\rm \reff{a:y3;t1,u1,u2,q,v1,v2}+\reff{b:y4;t2}}                \label{c3 {a:y2;v1}=infty {b:y1;v2}=0}
\\&R_1	+R_2	\leq		{\rm \reff{a:y3;t1}              +\reff{b:y4;t2,u1,u2,q,v1,v2}}  \label{c4 {a:y2;v1}=infty {b:y1;v2}=0}
\\&R_1 +2R_2\leq		{\rm \reff{a:y3;t1,u2}+\reff{b:y4;t2}+\reff{b:y4;t2,u1,u2,q,v1,v2}}.    \label{c10 {a:y2;v1}=infty {b:y1;v2}=0}
\end{align}
\label{eq:ifcwithgf {a:y2;v1}=infty {b:y1;v2}=0}
\end{subequations}
The region in~\reff{eq:ifcwithgf {a:y2;v1}=infty {b:y1;v2}=0} is only a subset of the best known achievable
region for a cognitive IFC by Rini et at~\cite{rini2010sifcdmc} because the region
in~\reff{eq:ifcwithgf} does not use binning.

\item
By setting ${\rm \reff{a:y2;v1}}={\rm \reff{b:y1;v2}}=\infty$
(and with $U_1=V_1=U_2=V_2=Q$ without loss of generality) in~\reff{eq:ifcwithgf},
we obtain the achievable region with superposition-only for a {\em\bf broadcast channel} (BC),
namely
\begin{subequations}
\begin{align}
  &R_1 \leq     {\rm \reff{a:y3;t1,u1,u2,q,v1,v2}}                    \label{c1forgotten {a:y2;v1}={b:y1;v2}=infty}
\\&R_2 \leq     {\rm \reff{b:y4;t2,u1,u2,q,v1,v2}}                    \label{c2forgotten {a:y2;v1}={b:y1;v2}=infty}
\\&R_1	+R_2	\leq		{\rm \reff{a:y3;t1,u1,u2,q,v1,v2}+\reff{b:y4;t2}}                \label{c3 {a:y2;v1}={b:y1;v2}=infty}
\\&R_1	+R_2	\leq		{\rm \reff{a:y3;t1}              +\reff{b:y4;t2,u1,u2,q,v1,v2}}.  \label{c4 {a:y2;v1}={b:y1;v2}=infty}
\end{align}
\label{eq:ifcwithgf {a:y2;v1}={b:y1;v2}=infty}
\end{subequations}
The region in~\reff{eq:ifcwithgf {a:y2;v1}={b:y1;v2}=infty} is only a subset of the largest known achievable
region for a general BC by Marton~\cite{marton:ach_reg} because the region
in~\reff{eq:ifcwithgf} does not use binning.
The region in~\reff{eq:ifcwithgf {a:y2;v1}={b:y1;v2}=infty} is however optimal for the case of
``more capable BC channels''~\cite{ElGamal:it1979} and for the case of ``BC with degraded message set''~\cite{Korner_Marton:it1977}.

\item
By setting $Y_3=Y_4=Y$ (and thus $T_1=T_2=\emptyset$ without loss of generality)
we obtain the achievable region for a {\em\bf multiple access channel with
GF}~\cite{willems:phd82}.
With $Y_3=Y_4=Y$ and $T_1=T_2=\emptyset$ , we have that
\reff{a:y3;t1} = \reff{b:y4;t2} = 0,
\reff{a:y3;t1,u1} = \reff{b:y4;t2,u1},
\reff{a:y3;t1,u2} = \reff{b:y4;t2,u2},
\reff{a:y3;t1,u1,u2} = \reff{b:y4;t2,u1,u2}, and
\reff{a:y3;t1,u1,u2,q,v1,v2} = \reff{b:y4;t2,u1,u2,q,v1,v2},
and thus the region in~\reff{eq:ifcwithgf} reduces to
\begin{subequations}
\begin{align}
  &R_1 \leq     {\rm \reff{a:y2;v1}+\reff{a:y3;t1,u1}}                \label{c1 mac}
\\&R_2 \leq     {\rm \reff{b:y1;v2}+\reff{a:y3;t1,u2}}                \label{c2 mac}
\\&R_1	+R_2	\leq		{\rm \reff{a:y3;t1,u1,u2,q,v1,v2}}                \label{c3 mac}
\\&R_1	+R_2	\leq		{\rm \reff{a:y2;v1}+\reff{b:y1;v2}+\reff{a:y3;t1,u1,u2}},        \label{c5 mac}
\end{align}
\label{eq:ifcwithgf mac}
\end{subequations}
as derived in~\cite{willems:phd82}.

\item
The region in~\reff{eq:ifcwithgf mac} is for a MAC-GF ``without common message''.
The case {\em\bf with common message}, that is, a message $W_0$ available at both sources
and to be decoded at both destinations,
can be easily incorporated by having the codebook/random~varaible $Q$
also carry the common message $W_0$.
If the rate of the common message $W_0$ is $R_0$, the region in~\reff{eq:ifcwithgf mac} must
be modified as follows: the rate constraint~\reff{c3 mac} becomes
\[
R_0+R_1+R_2	\leq		{\rm \reff{a:y3;t1,u1,u2,q,v1,v2}}.
\]
This trick (i.e., the random variable $Q$ carries the common message) can be used whenever a
common message has to be included in the IFC-GF setting.  In particular, with a common message,
our Th.~\ref{theorem:in superposition for general IFC-GF} must be modified as follows:
the left hand side of the inequalities in~\reff{a:y3;t1,u1,u2,q,v1,v2} and ~\reff{b:y4;t2,u1,u2,q,v1,v2}
must also include the rate of the common message.

\item
Further setting $R_2=0$ in~\reff{eq:ifcwithgf mac} gives
\[
R_1 \leq \min\{ {\rm \reff{a:y2;v1}+\reff{a:y3;t1,u1}}, {\rm \reff{a:y3;t1,u1,u2,q,v1,v2}}\},
\]
which is the achievable rate for a
{\em\bf full-duplex  relay channel} with partial decode and forward.
The above rate does not include the case of
Compress-and-Forward~\cite{Cover_AelGamal:relay:it1979}.

\item
Channels with {\em\bf conferencing encoders} are also a special case of GF.
A two-source conferencing model~\cite{willems:phd82} assumes that
there are two non-interferring, noise-free channels of finite capacity
between the communicating nodes, one for each direction of communication.
Let $C_{ij}$ be the capacity of the conferencing channel from node $j$ to node $i$.
Following~\cite{Kramer:multiuserITnotes} and with some abuse notation,
the conferencing model is captured as follows.
Let the inputs be $\Xm_1 = [F_1;X_1]$ and $\Xm_2 = [F_2;X_2]$,
where $F_1$ and $F_2$ have alphabet sizes $\log(C_{12})$ and $\log(C_{21})$, respectively.
Further set $Y_1 = F_2$ and $Y_2 = F_1$ and define the channel transition probability to be
\[
P_{Y_1,Y_2,Y_3,Y_4|[F_1;X_1],[F_2;X_2]} =
P_{Y_3,Y_4|X_1,X_2} \,1_{\{Y_1=F_2\}} \,1_{\{Y_2=F_1\}}.
\]
In this model, the choice $V_1=F_1$ and $V_2=F_2$ gives ${\rm \reff{a:y2;v1}}=C_{21}$ and
${\rm \reff{b:y1;v2}}=C_{12}$ is capacity achieving~\cite{willems:phd82}.

\end{enumerate}

\begin{remark}\label{re:onvinod}
In~\cite{vinodSC2009} it is show that the region in Th.~\ref{theorem:in superposition
for general IFC-GF} is sum-rate optimal to within 18~bits for Gaussian channels with
independent noises and symmetric cooperation links when the gain of the cooperation links
are smaller than the gain of the interfering links.

In the same work, it is show that when the gain of the cooperation links are larger than
the gain of the interfering links, then the transmitters should decode more information
from their received generalized feedback signal than they will actually use for cooperation.
In our setting this amounts to:  at the end of slot $b$, $b\in\{1,\cdots,N-1\}$,
source~1 looks for a unique index $i\in\{1,...,\eu^{n R_{20c}}\}$
and some index $j\in\{1,...,\eu^{n R_{20n}}\}$ such that the sequences
\begin{align*}
(&
V_2^n(i,[\cdots]),
U_2^n(j,i,[\cdots]),
Y_{1,b}^n
)
\\&
\in T_\epsilon^{(n)}(P^{(\rm enc1)}_{V_2 U_2 Y_1|Q,X_1}|\underline{X}_1^n),
\end{align*}
where the dots indicate known message indices,
where all that is known at source~2 is represented by
\[
\underline{X}_1 =(Q,V_1,U_1,T_1,X_1),
\]
and where
\begin{align*}
P^{(\rm enc1)}_{V_2 U_2 Y_1|Q X_1}
&=
\frac{\sum_{T_2,X_2}
P_{Q}
P_{V_1 U_1 T_1 X_1|Q}
P_{V_2 U_2 T_2 X_2|Q}
P_{Y_1|X_1 X_2}}
{P_{Q V_1 U_1 T_1 X_1}}
\\&=
P_{V_2 U_2|Q}
\Big(\sum_{X_2}P_{X_2|Q V_2 U_2} P_{Y_1|X_1 X_2}\Big).
\end{align*}
Decoding is successful with high probability if
\begin{align}
R_{20n}+R_{20c} &\leq I(V_2,U_2 \wedge   Y_1| Q,V_1,U_1,T_1,X_1).
\label{b1 vinodextension1}
\end{align}
Notice that the constraint in~\reff{b1 vinodextension1} allows
$R_{20c}\leq I(V_2,U_2 \wedge   Y_1| Q,V_1,U_1,T_1,X_1)$ while
the constraint in~\reff{b:y1;v2} only allowed for
$R_{20c}\leq I(V_2\wedge   Y_1| Q,V_1,U_1,T_1,X_1)$. However,
the constraint in~\reff{b1 vinodextension1} constrains $R_{20n}$ to satisfies
$R_{20n}\leq I(V_2,U_2 \wedge   Y_1| Q,V_1,U_1,T_1,X_1) -R_{20c}$
while the constraint in~\reff{b:y1;v2} does not constrain $R_{20n}$ at all.
It is not clear at priori which cooperation strategy is better, i.e.,
whether~\reff{b:y1;v2} or~\reff{b1 vinodextension1}.
We will show next--as intuition suggests--that the two are
equivalent, that is, with superposition-only, the sources should
relay to the destinations all the common information they have
acquired through the generalized feedback.

\begin{corollary}\label{cor:onvinod}
After FM elimination of the regions in~\reff{eq:gf at dec1} and~\reff{eq:gf at dec2}
with~\reff{b:y1;v2} replaced by~\reff{b1 vinodextension1}, we get
\begin{subequations}
\begin{align}
   R_1     &\leq \min\{ {\rm \reff{a:y3;t1,u1,u2,q,v1,v2}}, {\rm \reff{a:y2;v1}+\reff{a:y3;t1,u1}}\}
\\     R_2 &\leq \min\{{\rm \reff{b:y4;t2,u1,u2,q,v1,v2}}, {\rm \reff{b1 vinodextension1}+\reff{b:y4;t2}}\}
\\ R_1+R_2 &\leq \min\{{\rm \reff{a:y3;t1,u1,u2,q,v1,v2}+\reff{b:y4;t2}},
                       {\rm \reff{a:y3;t1}              +\reff{b:y4;t2,u1,u2,q,v1,v2}},
                       {\rm \reff{a:y2;v1}+\reff{b1 vinodextension1}
                         +\reff{a:y3;t1}   +\reff{b:y4;t2,u1}}\}
\\2R_1+R_2 &\leq {\rm \reff{a:y2;v1}+\reff{a:y3;t1}+\reff{a:y3;t1,u1,u2,q,v1,v2}+\reff{b:y4;t2,u1}}.
\end{align}
\label{eq:ifcwithgf source1 decodes v2 u2}
\end{subequations}
It can be easily verified that the region in~\reff{eq:ifcwithgf source1 decodes v2 u2}
is the same as region~\reff{eq:ifcwithgf}
computed for $P_{Q V_1 U_1 T_1 X_1 V'_2 U'_2 T_2 X_2}$ where $U'_2=\emptyset$ and $V'_2=(V_2,U_2)$ (i.e.,
in the case of no feedback this choice corresponds to sending only private information for user~2).
Hence, requiring source~1 to decode more information than actually used for cooperation
does not enlarge the achievable region in~\reff{eq:ifcwithgf}.
\end{corollary}

Along the same line of reasoning, 
one could also require that
at the end of slot $b$, $b\in\{1,\cdots,N-1\}$,
source~1 looks for a unique indices $i\in\{1,...,\eu^{n R_{20c}}\}$,
and some pair of indices $j\in\{1,...,\eu^{n R_{20n}}\}$
and $k\in\{1,...,\eu^{n R_{22n}}\}$,
such that the sequences
\begin{align*}
(&
V_2^n(i,[\cdots]),
U_2^n(j,i,[\cdots]),
T_2^n(k,j,i,[\cdots]),
Y_{1,b}^n
)
\\&
\in T_\epsilon^{(n)}(P^{(\rm enc1)}_{V_2 U_2 T_2 Y_1|Q X_1}|\underline{X}_1^n),
\nonumber
\end{align*}
where
\[
P^{(\rm enc1)}_{V_2 U_2 T_2 Y_1|Q X_1}
=
P_{V_2 U_2 T_2|Q}
\Big(\sum_{X_2}P_{X_2|Q V_2 U_2 T_2} P_{Y_1|X_1 X_2}\Big).
\]

Decoding is successful if
\begin{align}
R_2=R_{22n}+R_{20n}+R_{20c} &\leq I(T_2,U_2,V_2 \wedge   Y_1| Q,V_1,U_1,T_1,X_1).
\label{b1 vinodextension2}
\end{align}
The constraint in~\reff{b1 vinodextension2}
can be too restrictive for rate $R_2$ when the cooperation link is weak.
Indeed, consider the extreme case
where $Y_1$ is independent of everything else (i.e., unilateral cooperation), then
the constraint in~\reff{b1 vinodextension2} implies $R_2=0$, which can be easily beaten
by ignoring the generalized feedback.
\end{remark}

\subsection{Superposition \& binning achievable region}
\label{sec:superposition+binning}
Because the achievable region with {\em superposition-only} does not reduce to the largest known
achievable region when the IFC-GF reduces to a cognitive channel or a broadcast
channel, we now introduce binning in the
superposition only achievable scheme. In this new achievable scheme, the sources
also cooperate in sending part of the private messages.

Each message is divided into four parts: two
common messages and two private messages.  The sources
cooperate in sending one part of the common messages and
one part of the private messages.  Communication again
proceeds on a frame on $N$ slots.  In any given slot, the
sources decode the cooperative messages from
their GF signal; these message are then
relayed in the next slot. Since the common messages
are decoded at both destinations, the sources cooperate by
forwarding the cooperative common messages to the destinations
as in a virtual MIMO channel.
On the other hand, the private messages are decoded at the intended destination
only and treated as noise at the non-indened destination.  In this case,
a source treats the other source's cooperative private message as
``non-causally know interference'' in the next slot.  Cooperation
is then in the form of repetition/forwarding for one destination
and  pre-coding/binning for the other destination.
The details of our proposed superposition \& binning scheme are
as follows.

\noindent
{\em Class of Input Distributions:}
Consider a distribution from the class
\begin{align}
  &P_{Q V_1 U_1 T_1 S_1 Z_1 X_1 V_2 U_2 T_2 S_2 Z_2 X_2 Y_1 Y_2 Y_3 Y_4}
\nonumber\\&
=P_{Q S_1 S_2}
P_{V_1 U_1 T_1 Z_1 X_1|Q S_1 S_2}
P_{V_2 U_2 T_2 Z_2 X_2|Q S_1 S_2}
P_{Y_1 Y_2 Y_3 Y_4|X_1 X_2}.
\label{eq:inpdfnew}
\end{align}
that is, conditioned on $(Q,S_1,S_2)$, the random variables $(V_1, U_1, T_1, Z_1, X_1)$
generated at source~1 are independent of
the random variables $(V_2, U_2, T_2, Z_2, X_2)$
generated at source~2.
The channel transition probability $P_{Y_1 Y_2 Y_3 Y_4|X_1 X_2}$
is fixed, while the other factors in the distribution in~\reff{eq:inpdfnew} can be varied.

The random variables $S_u$ and $Z_u$, $u=\{1,2\}$,
which did not appear in~\reff{eq:inpdf},
convey the previous and the new,
respectively, cooperative private information
in a block-Markov encoding scheme.

\noindent
{\em Codebook Generation:}
From the input distribution in~\reff{eq:inpdfnew}, compute the
marginals $P_{V_1 U_1 T_1|Q}$ and $P_{V_2 U_2 T_2|Q}$
(i.e., drop the dependance on $(S_1,S_2)$), and construct
codebooks $(Q^n,V_1^n,U_1^n,T_1^n,V_2^n,U_2^n,T_2^n)$ as described
in Section~\ref{sec:superposition-only}.

From the input distribution in~\reff{eq:inpdfnew}, compute
the marginals $P_{S_1|Q}$ and $P_{S_2|Q}$.
Conditioned on each $Q^n([i,j])=q^n([i,j])$,
pick uniformly at random length-$n$ sequences $S_1^n(k_1,[i,j])$
from the typical set $T_\epsilon^{(n)}(P_{S_1|Q}|q^n([i,j]))$.
We shall specify later the range of the index $k$ in $S_1^n(k_1,[i,j])$.
Similarly, generate a codebook $S_2^n(k_2,[i,j])$
by picking uniformly at random length-$n$ sequences
from the typical set $T_\epsilon^{(n)}(P_{S_2|Q}|q^n([i,j]))$.
The range of the index $k_2$ will be specified later.

From the input distribution in~\reff{eq:inpdfnew}, compute the
the marginal $P_{Z_1|Q S_1 V_1}$.
For each triplet
$(Q^n([i,j]),S_1^n(k_1,[i,j]),V_1^n(\ell_1,[i,j]))=(q^n([i,j]),s_1^n(k_1,[i,j]),v_1^n(\ell_1,[i,j]))$,
generate a codebook $Z_1^n(m_1,\ell_1,k_1,[i,j])$
by picking uniformly at random length-$n$ sequences
from the typical set $T_\epsilon^{(n)}(P_{Z_1|Q S_1 V_1}|q^n([i,j]),s_1^n(k_1,[i,j]),v_1^n(\ell_1,[i,j]))$.
The range of the indices $m_1$ will be specified later.
Similarly, but with the role of the users swapped, construct a codebook
$Z_2^n(m_2,\ell_2,k_2,[i,j])$.

Finally, for each set of codewords
$(Q^n,S_1^n,S_2^n,V_1^n,U_1^n,T_1^n,Z_1^n)=(q^n,s_1^n,s_2^n,v_1^n,u_1^n,t_1^n,z_1^n)$
(we omit here the list of indices for sake of space)
pick uniformly at random one sequence  $X_1^n$
from the typical set $T_\epsilon^{(n)}(P_{X_1|Q  S_1 S_2 V_1 U_1 T_1 Z_1}|q^n,s_1^n,s_2^n,v_1^n,u_1^n,t_1^n,z_1^n)$.
Similarly, but with the role of the users swapped, construct a codebook $X_2^n$.

\begin{figure}
\begin{center}
\includegraphics[width=8cm]{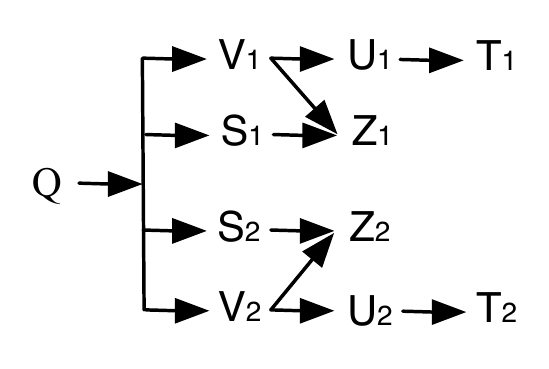}
\caption{A visual representation of the codebook generation for the
superposition \& binning achievable scheme. An arrow to a codebook/random variable indicates
that the codebook is superimposed to {\em all} the codebooks that precede it, and
codebooks linked by a vertical line are conditionally independent given
everything that precedes them.}
\label{fig:codebookgeneration cisslike 2008}
\end{center}
\end{figure}

\begin{figure}
\begin{center}
\includegraphics[width=8cm]{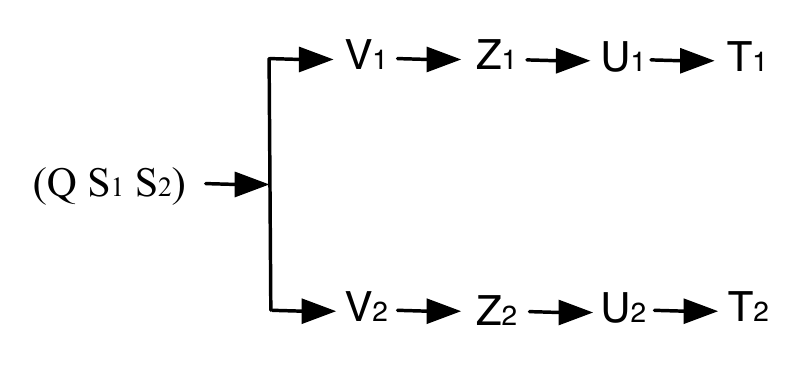}
\caption{A visual representation of the possible codebooks for the
superposition \& binning achievable scheme.}
\label{fig:codebookpossible cisslike 2008}
\end{center}
\end{figure}

Fig.~\ref{fig:codebookgeneration cisslike 2008} visualizes the proposed codebook generation
(the convention is the same used for Fig.~\ref{fig:encoding isit 2007}).
The class of input distributions in~\reff{eq:inpdfnew} however allows for codebooks
as depicted in Fig.~\ref{fig:codebookpossible cisslike 2008}. With binning we will force
a codebook generated as in Fig.~\ref{fig:codebookgeneration cisslike 2008} to look like
a codebook generated as in Fig.~\ref{fig:codebookpossible cisslike 2008}.

In order to complete the generalization of the scheme
in Section~\ref{sec:superposition-only}, we must consider each index $W_{xyz}$
(with the exception of the pair $[W_{10c,b-1}, W_{20c,b-1}]$ in $Q^n$)
as a pair of indices $[W_{xyz},B_{xyz}]$, where
$W_{xyz}\in\{1,\cdots,\eu^{n R_{xyz}}\}$
represents a message index, and
$B_{xyz}\in\{1,\cdots,\eu^{n R_{xyz}^{\prime}}\}$
represents a ``bin index''.  Notice that each bin
index $B_{xyz}$ has the same subscript of the corresponding
message index $W_{xyz}$ and has rate $R_{xyz}^{\prime}$
(i.e., notice the prime in the superscript).
The only exception is for $S_1$ and $S_2$ where the rate of the bin index is
indicated with a double prime as a superscript, that is,
$R_{11c}^{\prime\prime}$ and $R_{22c}^{\prime\prime}$, so as not to confuse them
with $R_{11c}^{\prime}$ in $Z_1$ and $R_{22c}^{\prime}$ in $Z_2$.

\noindent
{\em Encoding:}
We can assume correct decoding of the message indices
at the sources' side at the end of slot $b-1$
(no error propagation), i.e.,
\begin{align*}
W^{''}_{10c,b-1}=W_{10c,b-1} \ \text{(carried by $V_{1}^n$ and to be repeated in $Q^n$)},\\
W^{''}_{11c,b-1}=W_{11c,b-1} \ \text{(carried by $Z_{1}^n$ and to be repeated in $S_{1}^n$)},\\
W^{'}_{20c,b-1}=W_{20c,b-1}  \ \text{(carried by $V_{2}^n$ and to be repeated in $Q^n$)},\\
W^{'}_{21c,b-1}=W_{21c,b-1}  \ \text{(carried by $Z_{2}^n$ and to be repeated in $S_{1}^n$)},
\end{align*}
since the
total average probability of error can be upper bounded by the
sum of the decoding error probabilities at each step, under the
assumption that no error propagation from the previous steps
has occurred~\cite{buzo:backdec:IT89,Fah_Garg_Motani:sub2006}.

The encoding process at the beginning of slot $b$ consists of the following
binning and superposition steps, whose purpose is to allow the most
general possible class of input distributions:
\begin{itemize}
\item
{\em Binning codebooks $S_1^n$ and $S_2^n$ against each other.}
At the beginning of slot~$b$, given the past messages
$(W_{10c,b-1},W_{20c,b-1},W_{11c,b-1},W_{22c,b-1})$,
source~1 tries to find a pair $(b_{11c,b-1},b_{22c,b-1})$
such that
\begin{align*}
\Big(S_1^n ([W_{11c,b-1},b_{11c,b-1}],[\cdots])&,
\\   S_2^n ([W_{22c,b-1},b_{22c,b-1}],[\cdots])& \Big)
\\\in T_\epsilon^n(P_{S_1 S_2|Q}|Q^n),
\end{align*}
where the dots are in place of the know messages $[W_{10c,b-1}, W_{20c,b-1}]$
from the previous slot.  If more than one pair is found, then
source~1 chooses one at random.
If the search fails, source~1 sets $(b_{11c,b-1},b_{22c,b-1})=(1,1)$;
in this case we say that an error has occurred at source~1.

Error analysis.
The codewords $(Q^n,S_1^n,S_2^n)$ were sampled in an
i.i.d. fashion from the distribution $P_{Q S_1 S_2}=P_{Q}P_{S_1|Q}P_{S_2|Q}$.
The encoding process ``forces'' them to actually look as if they were sampled in an i.i.d.
fashion from the distribution $P_{Q}P_{ S_1 S_2|Q}$.  For that to be feasible
with high probability  we must have (see Appendix~\ref{app:proof of eq:like bc 0})
\begin{align}
R_{11c}^{\prime\prime} + R_{22c}^{\prime\prime} \geq I(S_1\wedge  S_2| Q).
\label{eq:like bc 0}
\end{align}
This encoding step is the same as the encoding
in Marton's achievable region for a general two-user broadcast
channel with common message~\cite{marton:ach_reg}.

This first encoding step is run in parallel at both sources,
so that the two sources have the same set of past
cooperative messages
$(W_{10c,b-1},W_{20c,b-1},W_{11c,b-1},W_{22c,b-1})$,
and of bin indices $(B_{11c,b-1},B_{22c,b-1})$.
This ``common knowledge'' furnishes the basis
for cooperation in slot $b$, where
\[
S_2^n([W_{22c,b-1},B_{22c,b-1}],[W_{10c,b-1}, W_{20c,b-1}])
\]
can be treated as ``non-causally known interference''
at source~1, and
\[
S_1^n([W_{11c,b-1},B_{11c,b-1}],[W_{10c,b-1}, W_{20c,b-1}])
\]
can be treated as ``non-causally known interference''
at source~2.

\item
{\em Joint conditional binning:}
Given the new messages  $(W_{10c,b},W_{10n,b},W_{11n,b})$,
source~1 tries to find  a set of bin indices
$(b_{10c,b},b_{10n,b},b_{11n,b})$ such that
\begin{align*}
\Big(                                         V_1^n ([W_{10c,b},b_{10c,b}],[\cdots])&,
\\                      U_1^n ([W_{10n,b},b_{10n,b}],[W_{10c,b},b_{10c,b}],[\cdots])&,
\\T_1^n ([W_{11n,b},b_{11n,b}],[W_{10n,b},b_{10n,b}],[W_{10c,b},b_{10c,b}],[\cdots])&,
\\\in T_\epsilon^n(P_{V_1 U_1 T_1|Q S_1 S_2}|Q^n,S^n_1,S^n_2)&,
\end{align*}
where the dots are in place of the know previous message $[W_{10c,b-1}, W_{20c,b-1}]$.
If more than one triplet is found, then source~1 chooses one at random.
If the search fails, source~1 sets $(b_{10c,b},b_{10n,b},b_{11n,b})=(1,1,1)$;
in this case we say that an error has occurred at source~1.

This encoding steps is a generalization of the
``sequential binning'' idea introduced in~\cite{maricett}.
Here, instead of doing several sequential binning steps,
we bin all the codewords at once.

Error analysis.
The triplet $(Q^n,S^n_1,S^n_2)$ found in the previous encoding step,
appears jointly typical according to $P_{Q S_1 S_2}$.  The set of codewords
$(Q^n,V_1^n,U_1^n,T_1^n)$ is jointly typical according to $P_{Q V_1 U_1 T_1}$
by construction.  However, codewords $(V_1^n,U_1^n,T_1^n)$ and $(S^n_1,S^n_2)$ were
generated independently conditioned on $Q^n$.  The purpose of this binning step
is to make the whole set $(Q^n,S^n_1,S^n_2,V_1^n,U_1^n,T_1^n)$ to look
jointly typical according to $P_{Q S_1 S_2 V_1 U_1 T_1}$.
By standard arguments, similar to those used in Multiple Description Coding~\cite{kramerMDC}
(see Appendix~\ref{app:proof of eq:like MDC}), the joint binning step is successful with
arbitrary high probability if
\begin{subequations}
\begin{align}
                                    R_{10c}^{\prime} &\geq I(        V_1\wedge S_1,S_2| Q)\label{eq:10c prime}
\\                 R_{10n}^{\prime}+R_{10c}^{\prime} &\geq I(    U_1,V_1\wedge S_1,S_2| Q) \label{eq:10n prime}
\\R_{11n}^{\prime}+R_{10n}^{\prime}+R_{10c}^{\prime} &\geq I(V_1,U_1,T_1\wedge S_1,S_2| Q). \label{eq:11n prime}
\end{align}
\label{eq:like MDC}

The rate constraint in~\reff{eq:10c prime}
can be understood as follows.  After the first
(successful) binning step, the codewords $(Q^n,S_1^n,S_2^n,V_1^n)$
look as if they sampled from the distribution
$P_{Q}P_{S_1 S_2|Q}P_{V_1|Q}$. The encoding step requires
them to look as if they were sampled from the distribution
$P_{Q}P_{S_1 S_2|Q}P_{V_1|Q S_1 S_2}$; hence we need to be able to
search among an exponential (in $n$)
number of codewords $V_1^n$, whose
exponent must be at least $H(V_1|Q)-H(V_1|Q S_1 S_2)=I(V_1\wedge   S_1,S_2| Q)$.
The rate constraints in~\reff{eq:10n prime} and in~\reff{eq:11n prime}
have a similar interpretation.

\item
{\em Final binning step:}
Given the new messages  $W_{11c,b}$,
source~1 tries to find a bin index
$b_{11c,b}$ such that
\begin{align*}
\Big(Q^n(\cdots),S^n_1(\cdots),S^n_2(\cdots),V^n_1(\cdots),U_1^n(\cdots),T_1^n(\cdots),&\\
Z_1^n([W_{11c,b},b_{11c,b}],\cdots)&\Big)
\\\in T_\epsilon^n(P_{Z_1| Q S_1 S_2 V_1 U_1 T_1}|Q^n,S_1^n,S_2^n,V_1^n,U_1^n,T_1^n)&,
\end{align*}
where the dots are in place of the know messages and bin indices.
If more than one index is found, then
source~1 chooses one at random.
If the search fails, source~1 sets $b_{11c,b}=1$;
in this case we say that an error has occurred at source~1.

Error analysis.
The set $(Q^n,S^n_1,S^n_2,V_1^n,U_1^n,T_1^n)$ found from the previous encoding steps,
appears jointly typical according to $P_{Q S_1 S_2 V_1 U_1 T_1}$.  By construction,
$(Q^n,S^n_1,V_1^n,Z_1^n)$ are jointly typical according to $P_{Q S_1 V_1 Z_1}$, that is,
conditioned on $(Q,S_1,V_1)$, $Z_1$ is independent of $(S_2,U_1,T_1)$.
The purpose of this binning step
is to make the whole set 
jointly typical according to $P_{Q S_1 S_2 V_1 U_1 T_1 Z_1}$.
By standard arguments
(see Appendix~\ref{app:proof of eq:like MDC}), this binning step is successful
with  arbitrary high probability if
\begin{align}
R_{11c}^{\prime} \geq I(Z_1\wedge  S_2,U_1,T_1| Q,S_1,V_1).
\label{eq:like MDC new}
\end{align}
\label{eq:like MDC enc1}
\end{subequations}

\item
Finally, source~1 sends a codeword $X_1^n$ that is
jointly typical with all the sequences found
in the previous binning steps.

Encoding at source~2 proceeds similarly.
\end{itemize}

\noindent
{\em Cooperation:}
If the encoding steps are successful at both sources,
all the transmitted and received signals
are jointly typical according to the
distribution in~\reff{eq:inpdfnew}.

At the end of slot~$b$, source~1 knows all the messages
generated at source~1 at the beginning of the slot, including
$(Q^n,S_2^n)$ (because we can assume successfully decoding of
all past cooperative messages).
Source~1 then searches for a unique pair of codeword indices $(i,j)$
and some bin index $(b_i,b_j)$ such that
\begin{align*}
&
\Big(V_2^n([j,b_j],\cdots),Z_2^n([i,b_i],\cdots),Y^n_{1,b}\Big)
\\&
\in T_\epsilon^n
(P^{(\rm enc1)}_{V_2 Z_2 Y_1| \underline{X}_1} |\underline{X}_1^n),
\end{align*}
where  all that is known at source~1 is represented by
\[
\underline{X}_1 =(Q,S_1,S_2,Z_1,V_1,U_1,T_1,X_1),
\]
where the dots indicate known old cooperative messages and related bin indices, and where
\begin{align*}
P^{(\rm enc1)}_{V_2 Z_2 Y_1| \underline{X}_1}
&=
\frac{\sum_{U_2,T_2,X_2}
P_{Q S_1 S_2}
P_{V_1 U_1 T_1 Z_1 X_1|Q S_1 S_2}
P_{V_2 U_2 T_2 Z_2 X_2|Q S_1 S_2}
P_{Y_1|X_1 X_2}}
{P_{Q S_1 S_2V_1 U_1 T_1 Z_1 X_1}}
\\&=
P_{V_2 Z_2|Q S_1 S_2}
\Big(\sum_{X_2}
P_{X_2|Q S_1 S_2 V_2 Z_2}
P_{Y_1|X_1 X_2}\Big)
\end{align*}
If the search fails, source~1 sets $(i,j)=(1,1)$;
in this case we say that an error has occurred at source~1.

\begin{figure}
\begin{center}
\includegraphics[width=8cm]{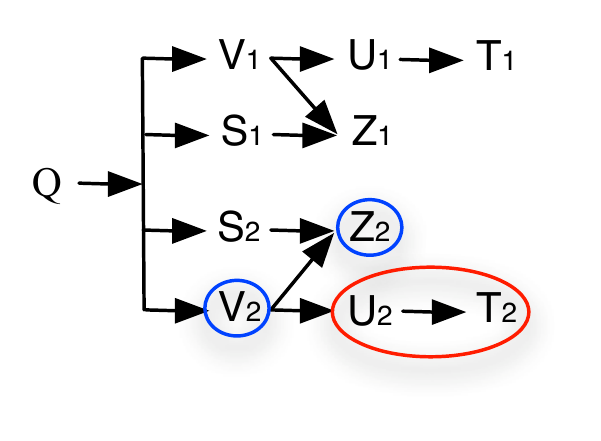}
\caption{Decoding at source~1:
the variables to be decoded are circled in blue,
the variables that are treated as noise are circled in red,
the remaining variables are known.}
\label{fig:source cooperation}
\end{center}
\end{figure}

Error analysis.
From the point of view of source~1, this decoding step
is equivalent to decoding the two
codewords $V_2^n([W_{20c,b},B_{20c,b}],\cdots)$ and
$Z_2^n([W_{22c,b},B_{22c,b}],\cdots)$ in a MAC-like
channel with output $Y^n_{1,b}$ and state $\underline{X}_1^n$ known
at the receiver only, similar in spirit to~\cite{GPMAC06}.
Decoding at source~1 is depicted in Fig.~\ref{fig:source cooperation}
where the variables circled in blue are to be decoded,
and the ones circled in red are treated as noise
(the rest is known).
By standard arguments (see Appendix~\ref{app:proof of eq:enc1-pe-gen all}),
decoding is correct  with high probability if
\begin{subequations}
\begin{align}
R_{Z_2}
   &\leq 
   I(Z_2\wedge  Y_1|\underline{X}_1,V_2)+I(Z_2\wedge  S_1|Q,S_2,V_2)
\label{eq:enc1-pe-gen-a}\\
R_{V_2}+
R_{Z_2} 
   &\leq 
    I(V_2,Z_2\wedge  Y_1|\underline{X}_1)+I(Z_2\wedge  S_1|Q,S_2,V_2)+I(V_2\wedge  S_1,S_2|Q),
\label{eq:enc1-pe-gen-c}
\end{align}
\label{eq:enc1-pe-gen all}
\end{subequations}
with $R_{Z_2}\defeq (R_{22c}+R_{22c}^{\prime})$
and  $R_{V_2}\defeq (R_{20c}+R_{20c}^{\prime})$.
Intuitively, the constraints in~\reff{eq:enc1-pe-gen all} are a consequence of the following observations.
Conditioned on $(Q,S_1,S_2)$, a wrong $V_2$ looks sampled from $P_{V_2|Q S_1 S_2}$
but it was actually sampled from $P_{V_2|Q}$; this accounts for the term $I(V_2\wedge  S_1,S_2|Q)$.
Similarly, conditioned on $(Q,S_1,S_2)$, a wrong $Z_2$ looks sampled from $P_{Z_2|Q S_1 S_2 V_2}$
but it was actually sampled from $P_{Z_2|Q,S_2,V_2}$; this accounts for the term $I(Z_2\wedge  S_1|Q,S_2,V_2)$.
The inequalities in~\reff{eq:enc1-pe-gen all} generalizes the
one in~\reff{a:y2;v1}, and reduce to~\reff{a:y2;v1}
when $S_1=Z_1=S_2=Z_2=Q$.

Decoding at source~2 proceeds similarly.

\begin{figure}
\begin{center}
\includegraphics[width=8cm]{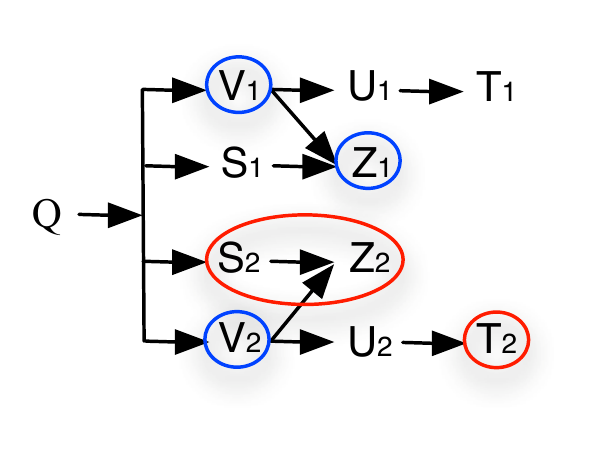}
\caption{Decoding at destination~1:
the variables whose message index is known (but the bin index is not) are circled in blue,
the variables that are treated as noise are circled in red,
the remaining variables are to be decoded.}
\label{fig:dest dec}
\end{center}
\end{figure}

\noindent
{\em Decoding:}
The receivers wait until the last slot has been received, and then
proceed to decode by using backward decoding.  We can assume that
when decoding the information sent in slot $b$, the decoding of
the information sent in the slots $b+1,\cdots,N$ was successful~\cite{willems:phd82,Fah_Garg_Motani:sub2006}.
When decoding slot $b$, destination~1 knows
the current cooperative messages $(W_{10c,b},W_{20c,b},W_{11c,b})$
carried by $(V_1,Z_1,V_2)$,
and tries to decode
the previous cooperative common messages $(W_{10c,b-1},W_{20c,b-1})$ in $Q$,
the previous cooperative private message $W_{11c,b-1}$ in $S_1$,
and the current non-cooperative messages $W_{10n,b}$ in $U_1$,
$W_{20n,b}$  in $U_2$, and
$W_{11n,b}$  in $T_1$.
Decoding at destination~1 is depicted in Fig.~\ref{fig:dest dec}
where the variables circled in blue are those
whose message index is known (but the bin index is not),
and the ones circled in red are treated as noise; the rest is to be decoded.

Formally,
destination~1 tries to find a unique set
of message indices
$({q_1},
  {q_2},
  {s_1},
  {u_1},
  {u_2},
  {t_1} 
)$
and some bin indices
$(
b_{v_1},
b_{v_2},
b_{z_1},
b_{s_1},
b_{u_1},
b_{u_2},
b_{t_1} 
)$
such that
\begin{align*}
\Big(
Q^n  (                                           [q_1,q_2])&,\nonumber\\
S_1^n(                   [{s_1},b_{s_1}]       , [q_1,q_2])&,\nonumber\\
V_1^n(                               [1,b_{v_1}],[q_1,q_2])&,\nonumber\\
Z_1^n(   [1,b_{z_1}],[{s_1},b_{s_1}],[1,b_{v_1}],[q_1,q_2])&,\nonumber\\
U_1^n(               [{u_1},b_{u_1}],[1,b_{v_1}],[q_1,q_2])&,\nonumber\\
T_1^n( [t_1,b_{t_1}],[{u_1},b_{u_1}],[1,b_{v_1}],[q_1,q_2])&,\nonumber\\
V_2^n(                               [1,b_{v_2}],[q_1,q_2])&,\nonumber\\
U_2^n(               [{u_2},b_{u_2}],[1,b_{v_2}],[q_1,q_2])&,\nonumber\\
Y_{3,b}^n
\Big)\in  T_\epsilon^{(n)}(P^{(\rm dec1)}_{Q S_1 V_1 U_1 T_1 Z_1 V_2 U_2 Y_3})& \nonumber
\end{align*}
where
\begin{align*}
&P^{(\rm dec1)}_{Q S_1 V_1 U_1 T_1 Z_1 V_2 U_2 Y_3}
=
\sum_{S_2,X_1,T_2,Z_2,X_2}
P_{Q S_1 S_2}
P_{V_1 U_1 T_1 Z_1 X_1|Q S_1 S_2}
P_{V_2 U_2 T_2 Z_2 X_2|Q S_1 S_2}
P_{Y_3|X_1 X_2}
\\&=
P_{Q S_1}
P_{V_1 U_1 T_1 Z_1|Q S_1}
P_{V_2 U_2|Q S_1}
\Big(\sum_{S_2,X_1,X_2}
\frac{P_{X_1 S_2|Q S_1 V_1 U_1 T_1 Z_1}P_{X_2 S_2|Q S_1 V_2 U_2}}{P_{S_2|Q S_1}}
P_{Y_3|X_1 X_2}\Big),
\end{align*}
and where, without loss of generality, we have assumed that the known messages
have index one.
Notice that the above expression for $P^{(\rm dec1)}_{Q S_1 V_1 U_1 T_1 Z_1 V_2 U_2 Y_3}$ does not imply
that $(V_1,U_1,T_1,Z_1)$ and $(V_2,U_2)$ are independent conditioned on $(Q S_1)$, i.e.,
\begin{align*}
P^{(\rm dec1)}_{Q S_1 V_1 U_1 T_1 Z_1 V_2 U_2}
&=\sum_{Y_3}
P^{(\rm dec1)}_{Q S_1 V_1 U_1 T_1 Z_1 V_2 U_2 Y_3}
\\&=
P_{Q S_1}
P_{V_1 U_1 T_1 Z_1|Q S_1}
P_{V_2 U_2|Q S_1}
\underbrace{\sum_{S_2}
\frac{P_{S_2|Q S_1 V_1 U_1 T_1 Z_1}P_{S_2|Q S_1 V_2 U_2}}{P_{S_2|Q S_1}}
}_{\text{not necessarily summing to 1 for all $(Q S_1 V_1 U_1 T_1 Z_1 V_2 U_2)$}}.
\end{align*}

The error analysis can be found in Appendix~\ref{app:proof of eq:dec1-pe-gen all}.
The probability of error at destination~1 can be made as small as desired
if the following rate constraints are satisfied:
\begin{subequations}
\begin{align}
R_{V_1}+R_{V_2}+R_{U_1}+R_{T_1}+R_{U_2}+R_{Z_1} &\leq E^{(1)}_{ 0}\\
      R_{U_1}+R_{T_1}+R_{U_2}+R_{Z_1} &\leq \min\{E^{(1)}_{ 1},E^{(1)}_{ 2},E^{(1)}_{ 4},E^{(1)}_{ 5}\} \\
      R_{U_1}+R_{T_1}        +R_{Z_1} &\leq \min\{E^{(1)}_{ 3},E^{(1)}_{ 6}\}\\
              R_{T_1}+R_{U_2}+R_{Z_1} &\leq \min\{E^{(1)}_{ 7},E^{(1)}_{ 8}\}\\
              R_{T_1}        +R_{Z_1} &\leq E^{(1)}_{ 9}\\
                      R_{U_2}+R_{Z_1} &\leq \min\{E^{(1)}_{10},E^{(1)}_{11}\}\\
                              R_{Z_1} &\leq E^{(1)}_{12}\\
      R_{U_1}+R_{T_1}+R_{U_2}         &\leq \min\{E^{(1)}_{13},E^{(1)}_{14},E^{(1)}_{16},E^{(1)}_{17},E^{(1)}_{22},E^{(1)}_{23}\}\\
      R_{U_1}+R_{T_1}                 &\leq \min\{E^{(1)}_{15},E^{(1)}_{18},E^{(1)}_{24}\}\\
              R_{T_1}+R_{U_2}         &\leq \min\{E^{(1)}_{19},E^{(1)}_{20},E^{(1)}_{25},E^{(1)}_{26}\}\\
              R_{T_1}                 &\leq \min\{E^{(1)}_{21},E^{(1)}_{27}\},
\end{align}
\label{eq:dec1-pe-gen all}
\end{subequations}
where the rates $R_{\star}$, $\star\in\{Q,V_1,U_1,T_1,S_1,Z_1,V_1,U_2,T_2,S_2,Z_2\}$, and
the quantities $E^{(1)}_{\ell}$, $\ell\in\{0,...,27\}$, are defined in Appendix~\ref{app:proof of eq:dec1-pe-gen all}.
Similarly, the rate constraints at destination~2 are as in~\reff{eq:dec1-pe-gen all} but with the role of the users swapped,

\noindent
{\em Achievable region:}
The achievable region with superposition \& binning as a function of $R_1$ and $R_2$ only,
can be obtained by applying the Fourier-Motzkin elimination procedure to the intersection
of~\reff{eq:like bc 0}, and
~\reff{eq:like MDC enc1},
~\reff{eq:enc1-pe-gen all},
~\reff{eq:dec1-pe-gen all},
and the regions corresponding to
~\reff{eq:like MDC enc1},
~\reff{eq:enc1-pe-gen all},
~\reff{eq:dec1-pe-gen all} but with the role of the users swapped.
However, the rates constraints are expressed as the minimum of several quantities
that we do not report it here for sake of space.
We note that when the binning rates are taken to satisfy the constraints in~\reff{eq:like MDC enc1}
and in~\reff{eq:enc1-pe-gen all} with equality for both users,
the achievable region has five types of rate bounds as for the non-feedback case, i.e.,
for $R_1$, for $R_2$, for $R_1+R_2$, for $2R_1+R_2$, and for $R_1+2R_2$.

As for the case of superposition-only, the error analysis for the case
of binning \& superposition is greatly simplified by the structured way
we performed superposition.   In particular, only 28 error events matter
out of the $2^8-1=255$ possible that result from the joint decoding of 8
messages.


\begin{remark}
In~\cite{ifcgf-tuninetti-ciss2008}, we constructed the codebooks as
depicted in Fig.~\ref{fig:codebookgeneration ciss2008}.
The difference with respect to the encoding proposed in this paper
is that $Z_1$ was superimposed to $S_2$ too (which carries the
private message from source~2--sent cooperatively by source~1 too--but
not decoded at destination~1), and $Z_2$ was superimposed to $S_1$ too.
Thus, with the encoding of~\cite{ifcgf-tuninetti-ciss2008}, $Z_1$ could not be decoded at destination~1,
since its decoding implies the decoding of all messages to which $Z_1$ is
superimposed, thus also $S_2$; but $S_2$ is a private message that must not
be decoded at destination~1. Similarly, $Z_2$ could be decoded at destination~2.

\begin{figure}
\begin{center}
\includegraphics[width=8cm]{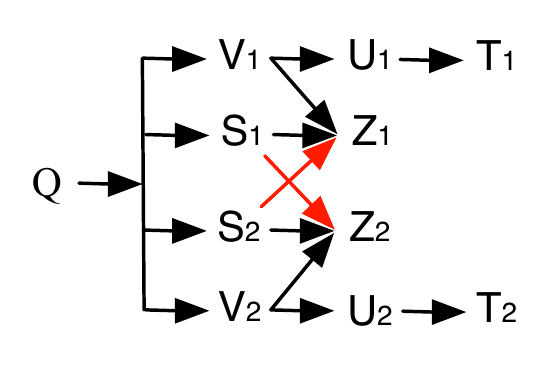}
\caption{A visualization of the codebook generation proposed in~\cite{ifcgf-tuninetti-ciss2008}.}
\label{fig:codebookgeneration ciss2008}
\end{center}
\end{figure}

With the encoding proposed in~\cite{ifcgf-tuninetti-ciss2008},
the rate constraint in~\reff{eq:like MDC new}
should be replaced by
\begin{align}
\label{eq:like MDC ciss2008 new}
R_{11c}^{\prime} \geq I(Z_1\wedge  U_1,T_1| Q,S_1,S_2,V_1),
\end{align}
and the rate constraints in~\reff{eq:enc1-pe-gen all}
should be replaced by
\begin{subequations}
\begin{align}
R_{Z_2}
   &\leq  I(Z_2\wedge  Y_1|\underline{X}_1,V_2)
\label{eq:enc1-pe-gen-a ciss}\\
R_{V_2}+
R_{Z_2}
   &\leq  I(V_2,Z_2\wedge  Y_1|\underline{X}_1),
        +I(V_2\wedge  S_1,S_2|Q)
\label{eq:enc1-pe-gen-c ciss}
\end{align}
\label{eq:enc1-pe-gen all ciss}
\end{subequations}
where the term $I(Z_2\wedge  S_1|Q,S_2,V_2)$ in~\reff{eq:enc1-pe-gen all}
does not appear in~\reff{eq:enc1-pe-gen all ciss} since $Z_2$ is superimposed to
$S_1$ by construction (thus it already has the desired joint distribution).
In other words, the encoding of~\cite{ifcgf-tuninetti-ciss2008}
is less stringent in terms of ``binning rates constraints''
but it is more stringent in terms on ``decoding rate constraints''.  However, these two effects
do compensate one another. Indeed, from~\reff{eq:enc1-pe-gen all ciss} and~\reff{eq:like MDC ciss2008 new}
we have:
\begin{align*}
R_{22c}
   &\leq I(Z_2\wedge  Y_1|\underline{X}_1,V_2)-R_{22c}^{\prime}\\
   &\leq I(Z_2\wedge  Y_1|\underline{X}_1,V_2) - I(Z_2\wedge  U_2,T_2| Q,S_1,S_2,V_2)\\
R_{20c}+
R_{22c}
   &\leq I(V_2,Z_2\wedge  Y_1|\underline{X}_1)
        +I(V_2\wedge  S_1,S_2|Q)-R_{20c}^{\prime}-R_{22c}^{\prime}\\
   &\leq I(V_2,Z_2\wedge  Y_1|\underline{X}_1) - I(Z_2\wedge  U_2,T_2| Q,S_1,S_2,V_2)
\end{align*}
which is the same we obtain from~\reff{eq:enc1-pe-gen all} and~\reff{eq:like MDC new}.

The two schemes are not the same.
With the encoding proposed in this paper, the codeword
in $Z_1$ and $Z_2$ can be decoded at the corresponding destination.
This improves on the performance since now the $Z$'s
do not longer act as noises at the intended destination.
\end{remark}

We showed in the previous section that the achievable region
with superposition-only in~\reff{eq:ifcwithgf} did not
reduced to the known achievable regions for some channels
subsumed by the IFC-GF model. With superposition \& binning we have:
\begin{enumerate}

\item
When the IFC-GF channel reduces to a {\em\bf broadcast channel},
our region with superposition \& binning with only $(Q,S_1,S_2)$
reduces to Marton's inner bound for a general broadcast channel~\cite{marton:ach_reg}.

\item
When the IFC-GF channel reduces to a {\em\bf cognitive channel},
our region with superposition \& binning achieves a subset of the
largest know achievable region for a cognitive channel~\cite{rini2010sifcdmc}.

The reason why our superposition \& binning region does not comprises the
region in~\cite{rini2010sifcdmc} is as follows.
Assume source~1 is the cognitive user and source~2 is the primary user.
Since the primary user is unaware of the message of the cognitive user,
we need to set $S_1=V_1=Z_1=\emptyset$.  Since there is not part of the
primary user's message that is not available 
at the cognitive user, we must set $Z_2=V_2=U_2=T_2=\emptyset$.  With these choices,
our encoding scheme only uses  $(Q,U_1,T_1,S_2)$ and is equivalent to the
scheme in~\cite{rini2010sifcdmc} with $U_{2pb}=\emptyset$ ($U_{2pb}$ in~\cite{rini2010sifcdmc}
carries part of the message of the primary user that is only sent by the
cognitive user; this feature is not present in our encoding scheme.)

\item
When the IFC-GF channel reduces to a {\em\bf relay channel},
our region with superposition \& binning does not seem to include
the compress-and-forward relaying strategy.

\end{enumerate}


\section{Example: the Gaussian IFC-GF}
\label{sec:example suponly}
In this section we provide an evaluation of our
{\em superposition-only} achievable region for the Gaussian channel.
We will provide an evaluation of our {\em superposition \& binning}
achievable region, as well as, a comparison with outer bounds, in
the second part of this paper.

We assume full duplex communication, perfect channel state information
at all terminals, and an average power constraint on the inputs.
A Gaussian channel in standard form has outputs:
\[
Y_{c} = h_{c1} X_{1} + h_{c2} X_{2} + N_{c},
\quad N_{c}\sim\Nc(0,1),
\,\,c\in\{1,...,4\}
\]
subject to $\E[|X_u|^2]\leq P_u$, $u\in\{1,2\}$. We assume that the
additive noises on the different channels are independent (the case
of correlated noises will be discussed in the second part of this paper).
Without loss of generality we assume that the {\em direct link}
channel gains $h_{31}$ and $h_{42}$ are real-valued since the destinations
can compensate for the phase of the intended signal.
Similarly, we assume that the {\em cooperation link} channel gains
$h_{21}$ and $h_{12}$ are also real-valued.
As opposed to the case without GF, the phase of the {\em
interfering link} channel gains $h_{32}$ and $h_{41}$ matter
because of transmitter cooperation.

In the following we are going to consider jointly Gaussian inputs
only when evaluating the achievable region in Th.~\ref{theorem:in superposition for general IFC-GF}.
Without loss of generality, let $Q\sim\Nc(0,1)$ and let
\begin{align*}
&   V_u = \alpha_u\,Q + X_{u0c}
\\& U_u = V_u + X_{u0n}
\\& T_u = X_u = U_u + X_{uun},
\end{align*}
for $u\in\{1,2\}$, where $(\alpha_1,\alpha_2)\in C^2$ and $X_{m}\sim\Nc(0,\sigma^2_m)$ for
$m\in\{{10c},{10n},{11n},{20c},{20n},{22n}\}$ are independent random variables whose
variances satisfy

\[
0\leq |\alpha_{u}|^2 + \sigma^2_{u0c} + \sigma^2_{u0n} + \sigma^2_{uun} \leq P_u, \quad u\in\{1,2\}.
\]
The right hand side of the rate bounds in region~\reff{eq:gf at dec1} are then
\begin{align*}
  &{\rm\reff{a:y2;v1}:\quad}\log\left(1+\frac{|h_{21}|^2 \sigma^2_{10c}}{1+|h_{21}|^2[\sigma^2_{10n}+\sigma^2_{11n}]}\right)
\\&{\rm\reff{a:y3;t1}:\quad}\log\left(1+\frac{|h_{31}|^2 \sigma^2_{11n}}{1+|h_{32}|^2 \sigma^2_{22n}}\right)
\\&{\rm\reff{a:y3;t1,u2}:\quad}\log\left(1+\frac{|h_{31}|^2 \sigma^2_{11n}+|h_{32}|^2 \sigma^2_{20n}}{1+|h_{32}|^2 \sigma^2_{22n}}\right)
\\&{\rm\reff{a:y3;t1,u1}:\quad}\log\left(1+\frac{|h_{31}|^2 [\sigma^2_{10n}+\sigma^2_{11n}]}{1+|h_{32}|^2 \sigma^2_{22n}}\right)
\\&{\rm\reff{a:y3;t1,u1,u2}:\quad}\log\left(1+\frac{|h_{31}|^2 [\sigma^2_{10n}+\sigma^2_{11n}]+|h_{32}|^2 \sigma^2_{20n}}{1+|h_{32}|^2 \sigma^2_{22n}}\right)
\\&{\rm\reff{a:y3;t1,u1,u2,q,v1,v2}:\quad}\log\left(1+\frac{|h_{31}|^2 [\sigma^2_{10c}+\sigma^2_{10n}+\sigma^2_{11n}]+|h_{32}|^2 [\sigma^2_{20c}+\sigma^2_{20n}]}{1+|h_{32}|^2 \sigma^2_{22n}}
  \right. \\&\quad\quad\quad\quad\quad +\left. \frac{|h_{31}\alpha_1+h_{32}\alpha_2|^2}{1+|h_{32}|^2 \sigma^2_{22n}}\right)
\end{align*}
and similarly for the other user.
Notice that, conditioning on $Q$ removes the
dependency between $X_1$ and $X_2$. The cooperation gain shows in eq.~\reff{a:y3;t1,u1,u2,q,v1,v2}
(where instead of $|h_{31}\alpha_1|^2+|h_{32}\alpha_2|^2$ one has
$|h_{31}\alpha_1+h_{32}\alpha_2|^2$) and in equation~\reff{a:y2;v1}.
The private message $X_{22n}$ ($X_{11n}$) acts as noise for destination~1 (destination~2).

\begin{figure}
        \begin{center}
        \includegraphics[width=8cm]{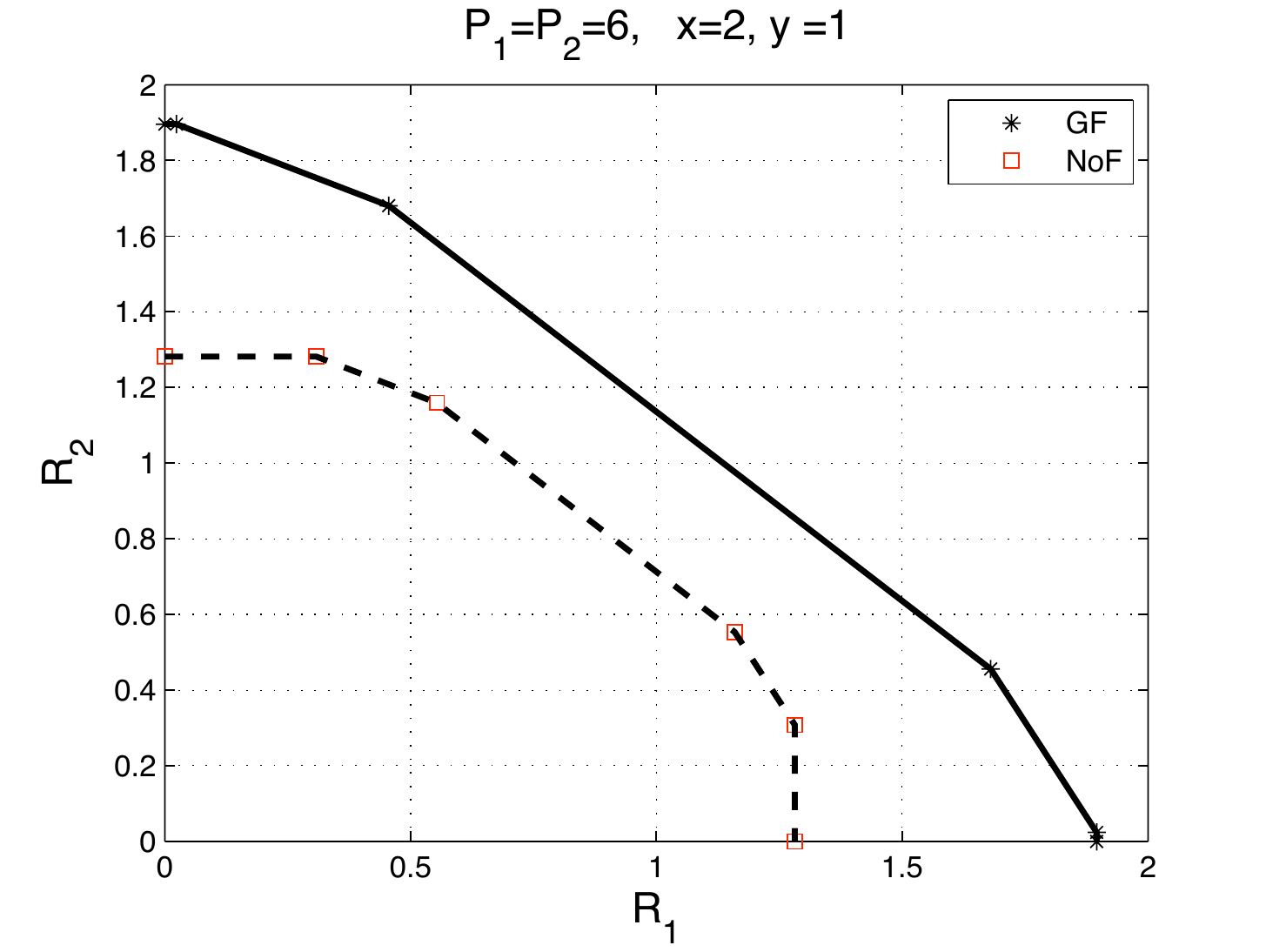}
        \caption{Performance comparison among standard IFC and IFC-GF.}
        \label{fig:suponly}
        \end{center}
    \end{figure}

For the purpose of numerical example, we consider a symmetric network where,
in a two dimensional Cartesian place with $x\geq0$ and $y\geq0$,
source~1 is in position $(-x/2,-y/2)$,
source~2 is in position $(-x/2,+y/2)$,
destination~1 is in position $(+x/2,-y/2)$, and
destination~2 is in position $(+x/2,+y/2)$.
Both users have the same power constraint $P$ and the channel gains
are inversely proportional to the distance between the terminals,  that is,
$h_{21}=h_{12}=h_{34}=h_{43}=1/y$,
$h_{31}=h_{13}=h_{24}=h_{42}=1/x$ and
$h_{41}=h_{14}=h_{23}=h_{32}=1/\sqrt{x^2+y^2}$.
Fig.~\ref{fig:suponly} shows the performance of our {\em superposition-only}
scheme and compares it with the case without generalized feedback,
for the case $P_1=P_2=6$ and $x=2$ and $y=1$. It can be seen that
great rate improvements are possible thanks to cooperation, when compared to the
case without GF. For example, the maximum rate for a given user improves from
1.3~bit/sec/Hz to 1.9~bit/sec/Hz (1.9=1.3$\times$(1+0.46), i.e., 46\% improvement),
while  the sum-rate improves from 1.7~bit/sec/Hz to 2.2~bit/sec/Hz
(2.2=1.7$\times$(1+0.29) , i.e., 29\% improvement).

\section{Conclusion}
\label{sec:conc inner}
In this paper we presented an achievable region for a general IFC-GF.
We built on the idea
of message splitting into common and private messages and proposed
a coding scheme where the sources cooperate to send part of the two
messages.  The cooperation in sending the common message aims to
realizing the gains of beam-forming, as in a distributed multi-antenna system,
while the cooperation in sending the private message aims to
leverage the interference ``pre-cancellation'' property of
dirty-paper-type coding.  We showed that our achievable region
generalizes several known achievable regions for IFC-GF and that
it reduces to known achievable regions
for some of the channels subsumed by the IFC-GF model.  Numerical
results show that cooperation can improve the achievable rate of all
the involved sources.  In the second part of this paper, we will
derive an outer bound for a general IFC-GF and use to assess the
performance of our proposed achievable strategy.

\bibliographystyle{plain}
\bibliography{career_biblio_v1}

\clearpage
\newpage

\appendix

\section{Proof of the redundancy of two single-rate constraints}
\label{Proof of the redundancy of two single-rate constraints}

For a fixed distribution
$P_{Q V_1 U_1 T_1 X_1 V_2 U_2 T_2 X_2}$ consider another distribution
$P_{Q' V'_1 U'_1 T'_1 X_1 V_2 U_2 T_2 X_2}$ with
\[
V'_1=U'_1=\emptyset,\
T'_1=(T_1,U_1),\
Q'=(Q,V_1),
\]
that is, source~1 does not send any common information.
With $P_{Q' V'_1 U'_1 T'_1 X_1 V_2 U_2 T_2 X_2}$,
the achievable region in~\reff{eq:ifcwithgf} reduces to:
\begin{subequations}
\begin{align}
  &R_1 \leq     {\rm \reff{a:y3;t1,u1}}                 \label{c1 u1=0}
\\&R_2 \leq     {\rm \reff{b:y4;t2,u1,u2,q,v1,v2}'}     \label{c2forgotten u1=0}
\\&R_2 \leq     {\rm \reff{b:y1;v2}+\reff{b:y4;t2,u2}'} \label{c2 u1=0}
\\&R_1 +R_2 \leq		{\rm \reff{a:y3;t1,u1,u2,q,v1,v2} +\reff{b:y4;t2}'}                \label{c3 u1=0}
\\&R_1 +R_2	\leq		{\rm \reff{b:y1;v2}
                         +\reff{a:y3;t1,u1,u2} +\reff{b:y4;t2}'}        \label{c5 u1=0}
\end{align}
\label{eq:ifcwithgf u1=0}
\end{subequations}
where a prime as a superscript indicates that the mutual information in the corresponding equation
must be computed for the distribution $P_{Q' V'_1 U'_1 T'_1 X_1 V_2 U_2 T_2 X_2}$ (rather than
for the distribution  $P_{Q V_1 U_1 T_1 X_1 V_2 U_2 T_2 X_2}$). Notice that all the mutual
informations in~\reff{eq:gf at dec1} are larger under
$P_{Q' V'_1 U'_1 T'_1 X_1 V_2 U_2 T_2 X_2}$ than under $P_{Q V_1 U_1 T_1 X_1 V_2 U_2 T_2 X_2}$
and satisfy
\begin{align*}
\rm
0 &\leq \reff{a:y3;t1}'   =\reff{a:y3;t1,u1}'    =\reff{a:y3;t1,u1}
\\&\leq \reff{a:y3;t1,u2}'=\reff{a:y3;t1,u1,u2}' =\reff{a:y3;t1,u1,u2}
\\&\leq \reff{a:y3;t1,u1,u2,q,v1,v2}'=\reff{a:y3;t1,u1,u2,q,v1,v2}
\\&\reff{a:y2;v1}'=0,
\end{align*}
while the mutual
informations in~\reff{eq:gf at dec2}  under
$P_{Q' V'_1 U'_1 T'_1 X_1 V_2 U_2 T_2 X_2}$ satisfy
\begin{align*}
\rm
0 &\leq  \reff{b:y4;t2}'              = \reff{b:y4;t2,u1}'   =I(Y_4\wedge  T_2|Q,V_1,V_2,U_2) 
\\&\leq  \reff{b:y4;t2,u2}'           = \reff{b:y4;t2,u1,u2}'=I(Y_4\wedge  T_2,U_2|Q,V_1,V_2) 
\\&\leq  \reff{b:y4;t2,u1,u2,q,v1,v2}'= I(Y_4\wedge  T_2,U_2,Q,V_1,V_2) 
\\&\reff{b:y1;v2}'=\reff{b:y1;v2}=I(V_2 \wedge   Y_1| Q,T_1,U_1,V_1,X_1).
\end{align*}

Consider the region in~\reff{eq:ifcwithgf} (under
$P_{Q V_1 U_1 T_1 X_1 V_2 U_2 T_2 X_2}$). If
\[
{\red \underbrace{\rm \reff{a:y2;v1}+\reff{a:y3;t1}+\reff{b:y4;t2,u1}}_{\rm in\ \reff{c1bis}} }
\geq
\min\{ \underbrace{\rm \reff{a:y3;t1,u1,u2,q,v1,v2}}_{\rm in\ \reff{c1forgotten}},
       \underbrace{\rm \reff{a:y2;v1}+\reff{a:y3;t1,u1}}_{\rm in\ \reff{c1}}   \}
\]
then the rate constraint in~\reff{c1bis} is redundant and can be omitted
from the region in~\reff{eq:ifcwithgf}. We will now show that
the rate constraint in~\reff{c1bis} can always be omitted
from the region in~\reff{eq:ifcwithgf} without enlarging the achievable region.
We will do so by showing that the rate points for which the rate
constraint in~\reff{c1bis} is violated, that is, when
\begin{align}
{\red \underbrace{\rm \reff{a:y2;v1}+\reff{a:y3;t1}+\reff{b:y4;t2,u1}}_{\rm in\ \reff{c1bis}} }
< R_1 \leq
\min\{ \underbrace{\rm \reff{a:y3;t1,u1,u2,q,v1,v2}}_{\rm in\ \reff{c1forgotten}},
       \underbrace{\rm \reff{a:y2;v1}+\reff{a:y3;t1,u1}}_{\rm in\ \reff{c1}}   \}
\label{c1bis violated}
\end{align}
holds together with all the rate constraint in~\reff{eq:ifcwithgf},
are contained in the region in~\reff{eq:ifcwithgf u1=0} (under
$P_{Q' V'_1 U'_1 T'_1 X_1 V_2 U_2 T_2 X_2}$). The region
in~\reff{eq:ifcwithgf u1=0} is a special case of the region in~\reff{eq:ifcwithgf}
for which the rate constraint in~\reff{c1bis} is redundant. This shows that
the region in~\reff{eq:ifcwithgf} is indeed achievable.

Assume now that~\reff{c1bis violated} holds together with all the rate constraint in~\reff{eq:ifcwithgf}
(under  $P_{Q V_1 U_1 T_1 X_1 V_2 U_2 T_2 X_2}$).  We will show that
that~\reff{c1bis violated} and~\reff{eq:ifcwithgf} imply~\reff{eq:ifcwithgf u1=0}.
We have:
\reff{c1 u1=0}~=~\reff{c1}.
Moreover
\begin{align*}
  &R_2\leq \underbrace{\rm \reff{a:y3;t1}+\reff{b:y4;t2,u1,u2,q,v1,v2}}_{\rm in\ \reff{c4}}
-\Big({\red{\rm \reff{a:y2;v1}+\reff{a:y3;t1}+\reff{b:y4;t2,u1}}}\Big)
\\&=I(Y_4;U_2,Q,V_1,V_2) - I(Y_2;V_1|\underline{X}_2)
\\&\leq I(Y_4;T_2,U_2,Q,V_1,V_2)-0
  = \underbrace{\rm \reff{b:y4;t2,u1,u2,q,v1,v2}'}_{\rm in\  \reff{c2forgotten u1=0}}
\end{align*}
and
\begin{align*}
  &R_2 \leq \underbrace{\rm \reff{a:y2;v1}+\reff{b:y1;v2}
               +\reff{a:y3;t1}+\reff{b:y4;t2,u1,u2}}_{\rm in\ \reff{c6}}
-\Big({\red{\rm \reff{a:y2;v1}
               +\reff{a:y3;t1}+\reff{b:y4;t2,u1}}}\Big)
\\&={\rm \reff{b:y1;v2}}+I(Y_4;U_2|Q,V_1,V_2)
\\&\leq {\rm \reff{b:y1;v2}}+I(Y_4;T_2,U_2|Q,V_1,V_2)
   =\underbrace{\rm  \reff{b:y1;v2}+\reff{b:y4;t2,u2}'}_{\rm in\ \reff{c2 u1=0}}
\end{align*}
and
\begin{align*}
  &R_1 +R_2 \leq \underbrace{\rm \reff{a:y2;v1}
                         +\reff{a:y3;t1}+\reff{a:y3;t1,u1,u2,q,v1,v2}+\reff{b:y4;t2,u1} }_{\rm in\ \reff{c8}}
-\Big({\red{\rm  \reff{a:y2;v1}+ \reff{a:y3;t1}               +\reff{b:y4;t2,u1}}}\Big)
\\&=   {\rm \reff{a:y3;t1,u1,u2,q,v1,v2}}
\\&\leq \underbrace{\rm\reff{a:y3;t1,u1,u2,q,v1,v2} +\reff{b:y4;t2,u1}' }_{\rm in\ \reff{c3 u1=0}}
\end{align*}
and
\begin{align*}
  &R_1	+R_2	\leq\underbrace{\rm 2\!\cdot\!\reff{a:y2;v1}+\reff{b:y1;v2}
                         +\reff{a:y3;t1}+\reff{a:y3;t1,u1,u2}+\reff{b:y4;t2,u1}}_{\rm in\ \reff{c9}}
-\Big({\red{\rm \reff{a:y2;v1}+\reff{a:y3;t1}               +\reff{b:y4;t2,u1}} }\Big)
\\&=     {\rm \reff{a:y2;v1}+\reff{b:y1;v2}+\reff{a:y3;t1,u1,u2}}
\\&\leq  \underbrace{\rm \reff{a:y2;v1}+\reff{b:y1;v2}+\reff{a:y3;t1,u1,u2}+\reff{b:y4;t2}'}_{\rm in\ \reff{c5 u1=0}}.
\end{align*}
This concludes the proof.

\section{Proof of~\reff{a:y2;v1} and of~\reff{b:y1;v2}}
\label{app:proof of eq:enc1-pe}
At the end of slot $b$, $b\in\{1,\cdots,N-1\}$,
transmitter~1 has received $Y_{1,b}^n$ and looks for the unique
index $i\in\{1,...,\eu^{n R_{20c}}\}$ such that the sequences
\begin{align}
(&
V_2^n(i,\cdots),
Y_{1,b}^n
)
\label{eq:dec enc 1}
\\&
\in T_\epsilon^{(n)}(P_{V_2|Q}P_{Y_1|Q,\underline{X}_1,V_2}|Q^n,\underline{X}_1^n),
\nonumber
\end{align}
where the dots indicate known message indices, and
where  all that is known at transmitter~1 is represented by
\[
\underline{X}_1^n =(Q^n,V_1^n,U_1^n,T_1^n,X_1^n).
\]
If none or more than one index $i\in\{1,...,\eu^{n R_{20c}}\}$ are found, then transmitter~1 sets $i=1$.

Assume message $i=1$ was sent.  An error occurs when
transmitter~1 declares $i\not=1$
in~\reff{eq:dec enc 1}, which occurs with probability
\begin{align*}
&\Pr[i\not=1| 1 \ \text{sent}]
\\&=
\Pr[\cup_{i>1}
(
V_2^n(i,\cdots),
Y_{1,b}^n)
\in T_\epsilon^{(n)}(P_{V_2|Q}P_{Y_1|Q,\underline{X}_1,V_2}|Q^n,\underline{X}_1^n)]
\\&\leq \sum_{i=2}^{\eu^{nR_{20c}}}|T_\epsilon^{(n)}(P_{V_2|Q}P_{Y_1|Q,\underline{X}_1,V_2}|Q^n,\underline{X}_1^n)|
P_{V_2^n|Q^n}P_{Y_1^n|Q^n,\underline{X}_1^n}
\\&\leq \eu^{n[ R_{20c} +  H(V_2|Q) +  H(Y_1|Q,\underline{X}_1,V_2) - H(V_2|Q) -  H(Y_1|Q,\underline{X}_1) + O(\epsilon)]}
\\&= \eu^{n[ R_{20c}  -  I(Y_1\wedge   V_2|Q,\underline{X}_1) + O(\epsilon)]}
\end{align*}
because $V_2$ and $Y_1$ are independent conditioned on $Q$ for any $i\not=1$.
The error probability goes to zero if~\reff{b:y1;v2} holds,
where $O(\epsilon)$ denotes a function of $\epsilon$ that goes to zero when $\epsilon \to 0$.

Similarly, we can prove~\reff{b1 vinodextension1} and~\reff{b1 vinodextension2} as follows.
We only give the proof for~\reff{b1 vinodextension1}.
Assume $i=1$ was sent. An error occur if the search for
\begin{align}
(&
V_2^n(i,\cdots),
U_2^n(j,i,\cdots),
Y_{1,b}^n
)
\label{eq:dec enc 1 v2}
\\&
\in T_\epsilon^{(n)}(P_{V_2,U_2|Q}P_{Y_1|Q,\underline{X}_1,V_2,U_2}|Q^n,\underline{X}_1^n),
\nonumber
\end{align}
results in an $i\not=1$. Thus an error occurs with probability
\begin{align*}
&\Pr[i\not=1| 1 \ \text{sent}]
\\&=
\Pr[\cup_{i>1,j\geq 1}
(
V_2^n(i,\cdots),
U_2^n(j,i,\cdots),
Y_{1,b}^n)
\in T_\epsilon^{(n)}(P_{V_2,U_2|Q}P_{Y_1|Q,\underline{X}_1,V_2,U_2}|Q^n,\underline{X}_1^n)]
\\&\leq \sum_{i=2}^{\eu^{nR_{20c}}}
        \sum_{j=1}^{\eu^{nR_{20n}}}|T_\epsilon^{(n)}
        (P_{V_2,U_2|Q}P_{Y_1|Q,\underline{X}_1,V_2,U_2}|Q^n,\underline{X}_1^n)|
P_{V_2^n,U_2^n|Q^n}P_{Y_1^n|Q^n,\underline{X}_1^n}
\\&\leq \eu^{n[ R_{20c}+ R_{20n}  +  H(V_2,U_2|Q) +  H(Y_1|Q,\underline{X}_1,V_2,U_2) - H(V_2,U_2|Q) -  H(Y_1|Q,\underline{X}_1) + O(\epsilon)]}
\\&= \eu^{n[ R_{20c} + R_{20n} -  I(Y_1\wedge   V_2,U_2|Q,\underline{X}_1) + O(\epsilon)]}
\end{align*}
because $(V_2,U_2)$ and $Y_1$ are independent conditioned on $Q$ for any $i\not=1$.
Hence, the probability of error vanishes as the block-length
increases if~\reff{b1 vinodextension1} holds.

\section{Proof of~\reff{eq:gf at dec1} and~\reff{eq:gf at dec2}}
\label{app:proof of eq:ifc-gf}

For the error analysis,
we consider only the probability of error in each block as the
total average probability of error can be upper bounded by the
sum of the decoding error probabilities at each step, under the
assumption that no error propagation from the previous steps
has occurred~\cite{buzo:backdec:IT89,Fah_Garg_Motani:sub2006}.

Receiver~1 looks for the unique triplet $(q_1,u_1,t_1)$
and some pair $(q_2,u_2)$ such that the sequences
\begin{align}
\Big(
Q^n(                              [q_1,q_2]),&
V_1^n(                  W_{10c,b},[q_1,q_2]),
U_1^n(              u_1,W_{10c,b},[q_1,q_2]),
T_1^n(          t_1,u_1,W_{10c,b},[q_1,q_2]),\nonumber\\&
V_2^n(                  W_{20c,b},[q_1,q_2]),
U_2^n(              u_2,W_{20c,b},[q_1,q_2]),
Y_{3,b}^n
\Big)\nonumber\\
\in &T_\epsilon^{(n)}
(P_{Q}P_{V_1,U_1,T_1|Q}P_{V_2,U_2|Q}P_{Y_3|Q,V_1,U_1,T_1,V_2,U_2}),
\label{eq:rx1 event E i j k l m}
\end{align}
where $(W_{10c,b},W_{20c,b})$ were decoded exactly in the previous step.
Let $E_{q_1 q_2 u_1 u_2 t_1}$ be the event that the sequences
in~\reff{eq:rx1 event E i j k l m} are strongly jointly typical.
Assume that $(q_1 q_2 u_1 u_2 t_1)=(11111)$ was sent.
The total probability of error at receiver 1 can be bounded as
\begin{align*}
&P_{e,1}^{(n)}
    = \Pr[E_{11111}^c \cup_{(q_1 u_1 t_1)\not=(111), \forall(q_2 u_2)} E_{q_1 q_2 u_1 u_2 t_1}]
\leq  \Pr[E_{11111}^c]
\\&
+\sum_{q_2>1,u_2>1} \Pr[\cup_{(q_1 u_1 t_1)\not=(111)} E_{q_1 q_2 u_1 u_2 t_1}]
+                   \Pr[\cup_{(q_1 u_1 t_1)\not=(111)} E_{q_1   1 u_1   1 t_1}]
\\&
+\sum_{u_2>1}       \Pr[\cup_{(q_1 u_1 t_1)\not=(111)} E_{q_1   1 u_1 u_2 t_1}]
+\sum_{q_2>1}       \Pr[\cup_{(q_1 u_1 t_1)\not=(111)} E_{q_1 q_2 u_1   1 t_1}]
\end{align*}
where all probabilities are conditioned on the event that
$(q_1 q_2 u_1 u_2 t_1)=(11111)$ was sent.

The probability of the event $E_{11111}^c$ vanishes because the
transmitted codewords are jointly typical with the received sequence with high
probability.

Although it seems we need to consider $7\times 4=28$ different error events,
all events with $(q_1 q_2)\not=(1,1)$, i.e., $Q^n$ wrong, are such that
the estimated codewords are independent of the actual transmitted ones,
and hence of the output. Hence, the probabilities that either $q_1\not=1$
or $q_2\not=1$ are dominated by
\begin{align}
\label{eq:prob E i j k l m}
\sum_{i>1, j>1,  k\geq 1,\ell\geq  1,  m\geq 1} \Pr[E_{i j k \ell m}]
\end{align}
for which we have
\begin{align*}
  &\Pr[E_{i j k \ell m}]
\\&\leq |T_\epsilon^n(P_{Q}P_{V_1,U_1,T_1|Q}P_{V_2,U_2|Q}P_{Y_3|V_1,U_1,T_1,V_2,U_2,Q})|
\\&P_{Q^n}\,P_{V_1^n,U_1^n,T_1^n|Q^n}\,P_{V_2^n,U_2^n|Q^n}\,P_{Y_3^n}
\\&\leq \eu^{-n(
+H(Q)+H(V_1,U_1,T_1,X_1|Q)+H(V_2,U_2|Q)+H(Y_3|Q,V_1,U_1,T_1,V_2,U_2)
)} \\& \eu^{-n(
-H(Q)-H(V_1,U_1,T_1,X_1|Q)-H(V_2,U_2|Q)-H(Y_3)
+O(\epsilon))}
\\&\leq \eu^{-n(I(Y_3\wedge  \,Q,V_1,U_1,T_1,V_2,U_2)+O(\epsilon))}.
\end{align*}
Hence, the probability in~\reff{eq:prob E i j k l m} vanishes as $n\to\infty$ if~\reff{a:y3;t1,u1,u2,q,v1,v2} holds.

Besides the events with $(q_1,q_2)\not=(1,1)$,
the other probability affecting $P_{e,1}^{(n)}$ are as follows.
From now on $(q_1,q_2)=(1,1)$, i.e., $Q^n$ correct.
Recall that when $Q^n$ is correct, also $V_1^n$ and $V_2^n$ are correct.

The events with $(u_1>1, t_1=1)$
and $(u_1>1, t_1>1)$, i.e., $U_1^n$ wrong,
have the same probability since, once $U_1^n$ is wrong,
the received signal is independent of the chosen
$(U_1^n(u_1,\cdots),T_1^n(t_1,u_1,\cdots))$ conditioned on
the transmitted/correct $(Q^n,V_1^n)$. This is so because,
even if $t_1=1$, the chosen $T_1^n$ is not the transmitted
one because superimposed to a wrong $U_1^n$.
Hence, if $u_2=1$, i.e., $U_2^n$ correct, we have
\begin{align*}
  &\sum_{k>1, m\geq 1} \Pr[ E_{1 1 k 1 m}]
\\&\leq \sum_{k>1, m>1}|T_\epsilon^n(P_{Q}P_{V_1,U_1,T_1|Q}P_{V_2,U_2|Q}P_{Y_3|V_1,U_1,T_1,V_2,U_2,Q})|
\\&P_{Q^n}\,P_{V_1^n,U_1^n,T_1^n|Q^n}\,P_{V_2^n,U_2^n|Q^n}\,P_{Y_3^n|Q^n,V_1^n,V_2^n,U_2^n}
\\&\leq \sum_{k>1, m>1}\eu^{n(
+H(Q)+H(V_1,U_1,T_1,X_1|Q)+H(V_2,U_2|Q)+H(Y_3|Q,V_1,U_1,T_1,V_2,U_2)
)} \\& \eu^{n(
-H(Q)-H(V_1,U_1,T_1,X_1|Q)-H(V_2,U_2|Q)-H(Y_3|Q,V_1,V_2,U_2)
+O(\epsilon))}
\\&=\eu^{n(R_{10n}+R_{11n}-I(Y_3\wedge  \,U_1,T_1|\,Q,V_1,V_2,U_2)+O(\epsilon))}
\end{align*}
which gives~\reff{a:y3;t1,u1};
while if $u_2>1$, i.e., $U_2^n$ wrong, we have
\begin{align*}
  &\sum_{\ell>1, k>1, m\geq1} \hspace*{-.7cm}\Pr[ E_{1 1 k \ell  m}]
\\&\leq \sum_{\ell>1, k>1, m>1}|T_\epsilon^n(P_{Q}P_{V_1,U_1,T_1|Q}P_{V_2,U_2|Q}P_{Y_3|V_1,U_1,T_1,V_2,U_2,Q})|
\\&P_{Q^n}\,P_{V_1^n,U_1^n,T_1^n|Q^n}\,P_{V_2^n,U_2^n|Q^n}\,P_{Y_3^n|Q^n,V_1^n,V_2^n,U_1^n,U_2^n}
\\&\leq\sum_{\ell>1, k>1, m>1} \eu^{n(
+H(Q)+H(V_1,U_1,T_1,X_1|Q)+H(V_2,U_2|Q)+H(Y_3|Q,V_1,U_1,T_1,V_2,U_2)
)} \\& \eu^{n(
-H(Q)-H(V_1,U_1,T_1,X_1|Q)-H(V_2,U_2|Q)-H(Y_3|Q,V_1,V_2)
+O(\epsilon))}
\\&=\eu^{n(R_{20n}+R_{10n}+R_{11n}-I(Y_3\wedge  \,T_1,U_1,U_2|\,Q,V_1,V_2)+O(\epsilon))}
\end{align*}
which gives~\reff{a:y3;t1,u1,u2}.

From now on $(q_1,q_2,u_1)=(1,1,1)$, i.e., $Q^n$ and $U_1^n$ correct.
If $u_2=1$, i.e., $U_2^n$ correct, and $t_1>1$, i.e., $T_1^n$ wrong, we have
\begin{align*}
  &\sum_{m>1}      \Pr[ E_{1 1 1 1 m}]
\\&\leq \sum_{m>1}|T_\epsilon^n(P_{Q}P_{V_1,U_1,T_1|Q}P_{V_2,U_2|Q}P_{Y_3|V_1,U_1,T_1,V_2,U_2,Q})|
\\&P_{Q^n}\,P_{V_1^n,U_1^n,T_1^n|Q^n}\,P_{V_2^n,U_2^n|Q^n}\,P_{Y_3^n|Q^n,V_1^n,V_2^n,U_1^n,U_2^n}
\\&\leq \sum_{m>1}\eu^{-n(
+H(Q)+H(V_1,U_1,T_1,X_1|Q)+H(V_2,U_2|Q)+H(Y_3|Q,V_1,U_1,T_1,V_2,U_2)
)} \\& \eu^{-n(
-H(Q)-H(V_1,U_1,T_1,X_1|Q)-H(V_2,U_2|Q)-H(Y_3|Q,V_1,V_2,U_1,U_2)
+O(\epsilon))}
\\&=\eu^{n(R_{11n}-I(Y_3\wedge  \,T_1|\,Q,V_1,U_1,V_2,U_2)+O(\epsilon))}
\end{align*}
which gives~\reff{a:y3;t1,u2},
while if $u_2>1$, i.e., $U_2^n$ wrong, and $t_1>1$, i.e., $T_1^n$ wrong, we have
\begin{align*}
  &\sum_{\ell>1, m>1}      \Pr[ E_{1 1 1 \ell  m}]
\\&\leq \sum_{\ell>1, m>1}|T_\epsilon^n(P_{Q}P_{V_1,U_1,T_1|Q}P_{V_2,U_2|Q}P_{Y_3|V_1,U_1,T_1,V_2,U_2,Q})|
\\&P_{Q^n}\,P_{V_1^n,U_1^n,T_1^n|Q^n}\,P_{V_2^n,U_2^n|Q^n}\,P_{Y_3^n|Q^n,V_1^n,V_2^n,U_1^n,U_2^n}
\\&\leq \sum_{\ell>1, m>1}\eu^{n(
+H(Q)+H(V_1,U_1,T_1,X_1|Q)+H(V_2,U_2|Q)+H(Y_3|Q,V_1,U_1,T_1,V_2,U_2)
)} \\& \eu^{n(
-H(Q)-H(V_1,U_1,T_1,X_1|Q)-H(V_2,U_2|Q)-H(Y_3|Q,V_1,V_2,U_1)
+O(\epsilon))}
\\&=\eu^{n(R_{20n}+R_{11n}-I(Y_3\wedge  \,T_1,U_2|\,Q,V_1,U_1,V_2)+O(\epsilon))}
\end{align*}
which gives~\reff{a:y3;t1}.


\begin{center}
\begin{table}
\caption{Error events at destination~1.}
\begin{tabular}{|l|llll|c|l|}
\hline
 &$1$&$2$&$3$&$4$& & \\
 &$[q_1,q_2]$&$u_1$&$t_1$&$u_2$& $N$&$\Cc$  \\
\hline
\hline
$\Ec^{(1)}_{0}$& 1 & * & * & * & $2^3$& $\emptyset$\\
\hline
\hline
$\Ec^{(1)}_{1}$& 0 & 1 & * & 1 & $2$& $Q V_1 V_2 $\\
\hline
$\Ec^{(1)}_{2}$& 0 & 1 & * & 0 & $2$& $Q V_1 V_2 U_2$\\
\hline
\hline
$\Ec^{(1)}_{3}$& 0 & 0 & 1 & 1 & $1$& $Q V_1 V_2 U_1$\\
\hline
$\Ec^{(1)}_{4}$& 0 & 0 & 1 & 0 & $1$& $Q V_1 V_2 U_1 U_2$\\
\hline
\hline
OK--not an error & 0 & 0 & 0 & * & $2$& all correct\\
\hline
\hline
\end{tabular}
\label{tab:dec1 suponly}
\end{table}
\end{center}
The developed analysis can be summarized as in Table~\ref{tab:dec1 suponly}.
In Table~\ref{tab:dec1 suponly} the symbols ``1'', ``0'' and ``$\star$''
have the following meaning.
A ``1'' indicates that the message index is in error.
A ``0'' indicates that the message index is correct.
A ``$\star$'' indicates that it does not matter whether the message index is in error;
this is so because of superposition coding;
in this case in fact, the codeword selected by the decoder--even though with the correct message
index--is superimposed to a wrong codeword and it is thus independent of the received signal.
In case of a ``$\star$'', the factorization of the joint probability needed for
the evaluation of the probability of error is as for the case
where the message is wrong; this implies that the error event that gives the
most stringent rate bound is that for which the message is wrong
(i.e., as far as error bounds are concerned, a ``$\star$'' is equivalent to a ``1'').
The second to last column in Table~\ref{tab:dec1 suponly} counts how many error events are
included in the corresponding row (i.e., each ``$\star$'' corresponds to two possible cases),
and the last column lists the elements of the sect $\Cc$, where $\Cc$ is the set of
correctly decoded codewords.

\section{Proof of~\reff{eq:like bc 0}}
\label{app:proof of eq:like bc 0}
The probability that encoder~1 fails to find a good
pair of indices $B_{11c,b-1}=i$, $B_{22c,b-1}=j$ is
\begin{align*}
  \Pr[\cup_{i\geq1,j\geq1}
                    \Big(
  & S_1^n ([W_{11c,b-1},i],[W_{10c,b-1}, W_{20c,b-1}]),
\\& S_2^n ([W_{22c,b-1},j],[W_{10c,b-1}, W_{20c,b-1}]) \Big)
    \not\in T_\epsilon^n(P_{Q S_1 S_2}|Q^n)]
\\&=\Big(1-p\Big)^{\eu^{n(R_{11c}^\prime+R_{22c}^\prime)}}
\leq \exp(-\eu^{n(R_{11c}^\prime+R_{22c}^\prime)}\,p)
\\&\leq \exp(-\eu^{n(R_{11c}^\prime+R_{22c}^\prime -I(S_1\wedge  S_2|Q)-O(\epsilon))})
\end{align*}
which goes to zero as $n\to\infty$ if~\reff{eq:like bc 0} holds, where
\begin{align*}
p = \Pr[
                    \Big(
  & S_1^n ([W_{11c,b-1},i],[W_{10c,b-1}, W_{20c,b-1}]),
\\& S_2^n ([W_{22c,b-1},j],[W_{10c,b-1}, W_{20c,b-1}]) \Big)
    \in T_\epsilon^n(P_{Q S_1 S_2}|Q^n)]
\\&\leq |T_\epsilon^n(P_{Q S_1 S_2}|Q^n)|P_{S_1^n|Q^n}P_{S_2^n|Q^n}
\\&\leq \eu^{n(H(S_1 S_2|Q) -H(S_1|Q)-H(S_2|Q)-O(\epsilon))}
\\&= \eu^{n(-I(S_1\wedge  S_2|Q)-O(\epsilon))},
\end{align*}
where $O(\epsilon)\to0$ as $\epsilon\to0.$

\section{Proof of~\reff{eq:like MDC}}
\label{app:proof of eq:like MDC}

Encoding fails if for all set of indices $(B_{10c,b},B_{10n,b},B_{11n,b})$
can be found no triplet $(V_1^n,U_1^n,T_1^n)$ can be found to be jointly
typical with $(Q^n,S_1^n,S_2^n)$.  The probability of this event can be
bounded as
\begin{align*}
\Pr[
\bigcap_{B_{10c}=1}^{\eu^{nR_{10c}^\prime}}
\bigcap_{B_{10n}=1}^{\eu^{nR_{10n}^\prime}}
\bigcap_{B_{11n}=1}^{\eu^{nR_{11n}^\prime}}&
\\                                        V_1^n ([W_{10c,b},B_{10c}],\cdots)&,
\\                    U_1^n ([W_{10n,b},B_{10n}],[W_{10c,b},B_{10c}],\cdots)&,
\\T_1^n ([W_{11n,b},B_{11n}],[W_{10n,b},B_{10n}],[W_{10c,b},B_{10c}],\cdots)&)
\\\not\in T_\epsilon^n(P_{Q,S_1,S_2,V_1,U_1,T_1}|Q^n, S_1^n, S_2^n)]&
\\= \Pr[K=0] \leq \frac{{\rm Var}[K]}{\mathbb{E}^2[K]}&,
\end{align*}
where
\begin{align*}
K=
\sum_{B_{10c}=1}^{\eu^{nR_{10c}^\prime}}
\sum_{B_{10n}=1}^{\eu^{nR_{10n}^\prime}}
\sum_{B_{11n}=1}^{\eu^{nR_{11n}^\prime}}
\, K_{B_{10c},B_{10n},B_{11n}}\\
\end{align*}
for
\begin{align*}
K_{B_{10c},B_{10n},B_{11n}}
= 1\{
\\                                        V_1^n ([W_{10c,b},B_{10c}],\cdots)&,
\\                    U_1^n ([W_{10n,b},B_{10n}],[W_{10c,b},B_{10c}],\cdots)&,
\\T_1^n ([W_{11n,b},B_{11n}],[W_{10n,b},B_{10n}],[W_{10c,b},B_{10c}],\cdots)&)
\\\in T_\epsilon^n(P_{Q,S_1,S_2,V_1,U_1,T_1}|Q^n, S_1^n, S_2^n)&
\}
\end{align*}
and $1_{\{A\}}$ is the indicator function that equals~one whenever the condition in $A$ is true.

The mean of the random variable $K$ is easily lower bounded as
\begin{align*}
\E[K]=
\sum_{B_{10c}=1}^{\eu^{nR_{10c}^\prime}}
\sum_{B_{10n}=1}^{\eu^{nR_{10n}^\prime}}
\sum_{B_{11n}=1}^{\eu^{nR_{11n}^\prime}}
\Pr[
\\                                        V_1^n ([W_{10c,b},B_{10c}],\cdots)&,
\\                    U_1^n ([W_{10n,b},B_{10n}],[W_{10c,b},B_{10c}],\cdots)&,
\\T_1^n ([W_{11n,b},B_{11n}],[W_{10n,b},B_{10n}],[W_{10c,b},B_{10c}],\cdots)&)
\\\in T_\epsilon^n(P_{Q,S_1,S_2,V_1,U_1,T_1}|Q^n, S_1^n, S_2^n)&
]
\\=
\sum_{B_{10c}=1}^{\eu^{nR_{10c}^\prime}}
\sum_{B_{10n}=1}^{\eu^{nR_{10n}^\prime}}
\sum_{B_{11n}=1}^{\eu^{nR_{11n}^\prime}}
|T_\epsilon^n(P_{Q,S_1,S_2,V_1,U_1,T_1}|Q^n, S_1^n, S_2^n)| P_{V_1^n,U_1^n,T_1^n|Q^n}
\\\geq \exp \{n[R_{10c}^\prime+R_{10n}^\prime+R_{11n}^\prime +
H(V_1,U_1,T_1|Q,S_1,S_2)  - H(V_1,U_1,T_1|Q) - O(\epsilon)] \} &
\\= \exp \{n[R_{10c}^\prime+R_{10n}^\prime+R_{11n}^\prime-I(S_1,S_2\wedge  V_1,U_1,T_1|Q)-O(\epsilon)] \},
\end{align*}
and upper bounded as
\begin{align*}
\E[K]\leq \exp \{n[R_{10c}^\prime+R_{10n}^\prime+R_{11n}^\prime-I(S_1,S_2\wedge  V_1,U_1,T_1|Q)+O(\epsilon)] \}.
\end{align*}

The variance of $K$ can be computed as
\begin{align*}
 {\rm Var}[K]=
\sum_{B_{10c}=1}^{e^{nR_{10c}^\prime}}
\sum_{B_{10n}=1}^{e^{nR_{10n}^\prime}}
\sum_{B_{11n}=1}^{e^{nR_{11n}^\prime}}
\sum_{B_{10c}^\prime=1}^{e^{nR_{10c}^\prime}}
\sum_{B_{10n}^\prime=1}^{e^{nR_{10n}^\prime}}
\sum_{B_{11n}^\prime=1}^{e^{nR_{11n}^\prime}}&
\\\Big(
 \Pr[K_{B_{10c},B_{10n},B_{11n}}=1,K_{B_{10c}^\prime,B_{10n}^\prime,B_{11n}^\prime}=1]&
\\
-\Pr[K_{B_{10c},B_{10n},B_{11n}}=1]\Pr[K_{B_{10c}^\prime,B_{10n}^\prime,B_{11n}^\prime}=1]&
\Big).
\end{align*}
When $B_{10c}\not=B_{10c}^\prime$, the random variables
$K_{B_{10c},B_{10n},B_{11n}}$ and
$K_{B_{10c}^\prime,B_{10n}^\prime,B_{11n}^\prime}$ are independent
by construction, hence they do not contribute to the summation.
When $B_{10c}=B_{10c}^\prime$, we upper-bound ${\rm Var}[K]$ by neglecting
the non-negative term $\Pr[K_{B_{10c},B_{10n},B_{11n}}=1]\Pr[K_{B_{10c}^\prime,B_{10n}^\prime,B_{11n}^\prime}=1]$.
Hence we have
\begin{align*}
{\rm Var}[K]
\leq
\sum_{B_{10c}=B_{10c}^\prime}
\sum_{B_{10n}=B_{10n}^\prime}
\sum_{B_{11n}=B_{11n}^\prime}
\Pr[K_{B_{10c},B_{10n},B_{11n}}=1]
&\\+
\sum_{B_{10c}=B_{10c}^\prime}
\sum_{B_{10n}=B_{10n}^\prime}
\sum_{(B_{11n},\ B_{11n}^\prime \not= B_{11n})}
\Pr[K_{B_{10c},B_{10n},B_{11n}}=1]
\underbrace{\Pr[K_{B_{10c},B_{10n},B_{11n}^\prime}=1|K_{B_{10c},B_{10n},B_{11n}}=1]}_{\leq \eu^{-nA}}
&\\+
\sum_{B_{10c}=B_{10c}^\prime}
\sum_{(B_{10n},\ B_{10n}^\prime\not=B_{10n})}
\sum_{(B_{11n},\ B_{11n}^\prime)}
\Pr[K_{B_{10c},B_{10n},B_{11n}}=1]
\underbrace{\Pr[K_{B_{10c},B_{10n}^\prime,B_{11n}^\prime}=1|K_{B_{10c},B_{10n},B_{11n}}=1]}_{\leq \eu^{-nB}}
&\\
\leq \eu^{n[R_{10c}^\prime+R_{10n}^\prime+R_{11n}^\prime-I(S_1S_2\wedge  V_1U_1T_1|Q) -O(\epsilon)]}
(1+\eu^{n[R_{11n}^\prime-A]}+\eu^{n[R_{10n}^\prime+R_{11n}^\prime-B]})
\end{align*}
We now evaluate $A$ and $B$. We have, neglecting the terms that go to zero as $\epsilon\to0$,
\begin{align*}
\eu^{-nB}
&=\Pr[
(U_1^n,
T_1^n)
\in T_\epsilon^n(P_{Q,S_1,S_2,V_1,U_1,T_1}|Q^n,S_1^n,S_2^n,V_1^n)] \\
&\leq \eu^{nH(U_1,T_1|Q,S_1,S_2,V_1)-nH(U_1,T_1|Q,V_1)} \\
&\leq \eu^{-n I(U_1,T_1\wedge  S_1,S_2|Q,V_1)},
\end{align*}
and
\begin{align*}
\eu^{-nA}
&=\Pr[
T_1^n
\in T_\epsilon^n(P_{Q,S_1,S_2,V_1,U_1,T_1}|Q^n,S_1^n,S_2^n,V_1^n,U_1^n)] \\
&\leq \eu^{nH(T_1|Q,S_1,S_2,V_1,U_1)-nH(T_1|Q,V_1,U_1)} \\
&\leq \eu^{-n I(T_1\wedge  S_1,S_2|Q,V_1,U_1)}.
\end{align*}
After having evaluated $A$ and $B$, we have that
\begin{align*}
\frac{{\rm Var}[K]}{\mathbb{E}^2[K]}
\leq \frac{1+\eu^{n[R_{11n}^\prime-A]}+\eu^{n[R_{10n}^\prime+R_{11n}^\prime-B]}}
{\eu^{n[R_{10c}^\prime+R_{10n}^\prime+R_{11n}^\prime-I(S_1S_2\wedge  V_1U_1T_1|Q)]}}
\end{align*}
goes to zero as $n\to \infty$ if~\reff{eq:like MDC} holds.


\section{Proof of~\reff{eq:enc1-pe-gen all}}
\label{app:proof of eq:enc1-pe-gen all}

Let
\[
E_{i j b_i b_j} =
(V_2^n([j, b_j],\cdots),Z_2^n([i, b_i]\cdots),Y^n_{1,b})
\in T_\epsilon^n
(P_{V_2 Z_2|Q S_1 S_2}
 P_{Y_1|Q S_1 S_2 V_2 Z_2 \underline{X}_1}|\underline{X}_1^n),
\]
where  all that is known at transmitter~1 is represented by
\[
\underline{X}_1^n =(Q^n,S_1^n,S_2^n,Z_1^n,V_1^n,U_1^n,T_1^n,X_1^n).
\]
Assume that $(i,j,b_i,b_j)=(1111)$ was sent.
The probability that the estimate of $(j,i)=(W_{20c,b},W_{22c,b})$
is wrong is bounded by
\begin{align*}
&P_{e,enc1}^{(n)}
    = \Pr[E_{1111}^c \cup_{(i,j)\not=(11), \forall (b_i,b_j)} E_{i j k}]
\\&
\leq  \Pr[E_{1111}^c]
+\sum_{i>1,j>1} \Pr[\cup_{\forall (b_i,b_j)} E_{i j b_i b_j}]
\\&
+\sum_{i>1}      \Pr[\cup_{\forall b_i} E_{i 1 b_i 1}]
+\sum_{i>1,b_j>1}\Pr[\cup_{\forall b_i} E_{i 1 b_i b_j}]
\\&
+\sum_{j>1}      \Pr[\cup_{\forall b_j} E_{1 j 1 b_j}]
+\sum_{j>1,b_i>1}\Pr[\cup_{\forall b_j} E_{1 j b_i b_j}]
\end{align*}
where all probabilities are conditioned on $(i,j,b_i,b_j)=(1111)$ being sent.

The probability of $E_{1111}^c$ is vanishing as $n\to\infty$ because the transmitted codewords are
jointly typical with the received signal with high probability.

For the other three terms we have:
if $i>1,j>1$, i.e., when $V_2^n$ is wrong then whether
$Z_2^n$ is correct or wrong does not change the distribution to use in the
error events and the most string error bound is for the case where both are wrong;
thus we have
\begin{align*}
  &\sum_{i>1,j>1} \Pr[\cup_{\forall (b_i,b_j)} E_{i j b_i b_j}]
\\&\leq \sum_{i>1,j>1,b_i\geq1,b_j\geq1}
|T_\epsilon^n(P_{V_2 Z_2|Q S_1 S_2} P_{Y_1|Q S_1 S_2 V_2 Z_2 \underline{X}_1}|\underline{X}_1^n)|
P_{V_2|Q} P_{Z_2|Q S_2 V_2} P_{Y_1|Q S_1 S_2  \underline{X}_1}
\\&\leq\eu^{n[R_{22c}+R_{20c}+R_{22c}^{\prime}+R_{20c}^{\prime}]}
\\&\eu^{n[
 H(V_2|Q S_1 S_2)
+H(Z_2|Q S_1 S_2 V_2 )
+H(Y_1|Q S_1 S_2 V_2 Z_2 \underline{X}_1)]}
\\&\eu^{-n[
 H(V_2|Q)
+H(Z_2|Q     S_2 V_2)
+H(Y_1|Q S_1 S_2 \underline{X}_1)]}
\\&\leq\eu^{n[R_{22c}+R_{20c}+R_{22c}^{\prime}+R_{20c}^{\prime}
-I(V_2\wedge S_1 S_2|Q )
-I(Z_2\wedge S_1 |Q S_2 V_2)
-I(Y_1\wedge V_2 Z_2 |Q S_1 S_2 \underline{X}_1)]}
\end{align*}
since by construction $V_2$ and $Z_2$ are independent conditioned on $Q$ and $Z_2$
is superimposed to $(S_2,V_2)$. This probability
can be driven to zero if~\reff{eq:enc1-pe-gen-c} holds.

Finally, if $i>1,j=1$, i.e., $V_2^n$ correct and $Z_2^n$ wrong, we have
\begin{align*}
  &\sum_{i>1} \Pr[\cup_{\forall b_i} E_{i 1 b_i 1}]
\\&\leq \sum_{i>1,b_i\geq1}
|T_\epsilon^n(P_{V_2 Z_2|Q S_1 S_2} P_{Y_1|Q S_1 S_2 V_2 Z_2 \underline{X}_1}|\underline{X}_1^n)|
P_{V_2|Q S_1 S_2} P_{Z_2|Q S_2 V_2} P_{Y_1|Q S_1 S_2 V_2 \underline{X}_1}
\\&\leq\eu^{n[R_{22c}+R_{22c}^{\prime}]}
\\&\eu^{n[
 H(V_2|Q S_1 S_2)
+H(Z_2|Q S_1 S_2 V_2 )
+H(Y_1|Q S_1 S_2 V_2 \underline{X}_1)]}
\\&\eu^{-n[
 H(V_2|Q S_1 S_2)
+H(Z_2|Q     S_2 V_2)
+H(Y_1|Q S_1 S_2 V_2 \underline{X}_1)]}
\\&\leq\eu^{n[R_{22c}+R_{22c}^{\prime}
-I(Z_2\wedge S_1 |Q S_2 V_2)
-I(Y_1\wedge Z_2 |Q S_1 S_2 V_2 \underline{X}_1)]}
\end{align*}
since now the $V_2$ has the marginal imposed by the binning step during the encoding process.
This probability can be driven to zero if~\reff{eq:enc1-pe-gen-a} holds.

\section{Proof of~\reff{eq:dec1-pe-gen all}}
\label{app:proof of eq:dec1-pe-gen all}

In slot $b$, $b=N, N-1,...,1$,
destination~1 tries to find a unique set
of message indices
$({q_1},
  {q_2},
  {s_1},
  {u_1},
  {u_2},
  {t_1} 
)$
and some bin indices
$(
b_{v_1},
b_{v_2},
b_{z_1},
b_{s_1},
b_{u_1},
b_{u_2},
b_{t_1} 
)$
such that
\begin{align*}
\Big(
Q^n  (                                           [q_1,q_2])&,\nonumber\\
S_1^n(                   [{s_1},b_{s_1}]       , [q_1,q_2])&,\nonumber\\
V_1^n(                               [1,b_{v_1}],[q_1,q_2])&,\nonumber\\
Z_1^n(   [1,b_{z_1}],[{s_1},b_{s_1}],[1,b_{v_1}],[q_1,q_2])&,\nonumber\\
U_1^n(               [{u_1},b_{u_1}],[1,b_{v_1}],[q_1,q_2])&,\nonumber\\
T_1^n( [t_1,b_{t_1}],[{u_1},b_{u_1}],[1,b_{v_1}],[q_1,q_2])&,\nonumber\\
V_2^n(                               [1,b_{v_2}],[q_1,q_2])&,\nonumber\\
U_2^n(               [{u_2},b_{u_2}],[1,b_{v_2}],[q_1,q_2])&,\nonumber\\
Y_{3,b}^n
\Big)\in  T_\epsilon^{(n)}(P^{(\rm dec1)}_{Q S_1 V_1 U_1 T_1 Z_1 V_2 U_2 Y_3})& \nonumber
\end{align*}
where
\begin{align}
  &P^{(\rm dec1)}_{Q S_1 V_1 U_1 T_1 Z_1 V_2 U_2 Y_3}  \label{eq:pdf_for_typicality_at_dec1}
\\&=
P_{Q S_1}
P_{V_1 U_1 T_1 Z_1|Q S_1}
P_{V_2 U_2|Q S_1}
\Big(\sum_{S_2,X_1,X_2}
P_{X_1 S_2|Q S_1 V_1 U_1 T_1 Z_1}
\frac{P_{X_2 S_2|Q S_1 V_2 U_2}}{P_{S_2|Q S_1}}
P_{Y_3|X_1 X_2}\Big).\nonumber
\end{align}
Notice that, given $(Q,S_1)$ the input variables for source~1 are
not independent of the input variables for source~2.

\begin{table}
\caption{Error events at destination~1.}
{\tiny
\begin{tabular}{|l|llllllll|c|l|}
\hline
 &$1$&$2$&$3$&$4$&$5$&$6$&$7$&$8$& & \\
 &$[q_1,q_2]$&$[1,b_{v_1}] $&$[u_1,b_{u_1}] $&$[t_1,b_{t_1}] $&$[s_1,b_{s_1}] $&$[  1,b_{z_1}] $&$[  1,b_{v_2}] $&$[u_2,b_{u_2}]$& $N$
 &$\Cc$  \\ 
\hline
\hline
$\Ec^{(1)}_{ 0}$& 1 & * & * & * & * & * & * & * & $2^7$& $\emptyset$\\
\hline
\hline
$\Ec^{(1)}_{ 1}$& 0 & 1 & * & * & 1 & * & 1 & * & $2^4$& $Q$\\
$\Ec^{(1)}_{ 2}$& 0 & 1 & * & * & 1 & * & 0 & 1 & $2^3$& $Q,V_2$\\
$\Ec^{(1)}_{ 3}$& 0 & 1 & * & * & 1 & * & 0 & 0 & $2^3$& $Q,V_2,U_2$\\
\hline
$\Ec^{(1)}_{ 4}$& 0 & 0 & 1 & * & 1 & * & 1 & * & $2^3$& $Q,V_1$\\
$\Ec^{(1)}_{ 5}$& 0 & 0 & 1 & * & 1 & * & 0 & 1 & $2^2$& $Q,V_2,V_1$\\
$\Ec^{(1)}_{ 6}$& 0 & 0 & 1 & * & 1 & * & 0 & 0 & $2^2$& $Q,V_2,U_2,V_1$\\
\hline
$\Ec^{(1)}_{ 7}$& 0 & 0 & 0 & 1 & 1 & * & 1 & * & $2^2$& $Q,V_1,U_1$\\
$\Ec^{(1)}_{ 8}$& 0 & 0 & 0 & 1 & 1 & * & 0 & 1 & $2^1$& $Q,V_2,V_1,U_1$\\
$\Ec^{(1)}_{ 9}$& 0 & 0 & 0 & 1 & 1 & * & 0 & 0 & $2^1$& $Q,V_2,U_2,V_1,U_1$\\
\hline
$\Ec^{(1)}_{10}$& 0 & 0 & 0 & 0 & 1 & * & 1 & * & $2^2$& $Q,V_1,U_1,T_1$\\
$\Ec^{(1)}_{11}$& 0 & 0 & 0 & 0 & 1 & * & 0 & 1 & $2^1$& $Q,V_2,V_1,U_1,T_1$\\
$\Ec^{(1)}_{12}$& 0 & 0 & 0 & 0 & 1 & * & 0 & 0 & $2^1$& $Q,V_2,U_2,V_1,U_1,T_1$\\
\hline
\hline
$\Ec^{(1)}_{13}$& 0 & 1 & * & * & 0 & * & 1 & * & $2^4$& $Q,S_1$\\
$\Ec^{(1)}_{14}$& 0 & 1 & * & * & 0 & * & 0 & 1 & $2^3$& $Q,S_1,V_2$\\
$\Ec^{(1)}_{15}$& 0 & 1 & * & * & 0 & * & 0 & 0 & $2^3$& $Q,S_1,V_2,U_2$\\
\hline
$\Ec^{(1)}_{16}$& 0 & 0 & 1 & * & 0 & 1 & 1 & * & $2^2$& $Q,S_1,V_1$\\
$\Ec^{(1)}_{17}$& 0 & 0 & 1 & * & 0 & 1 & 0 & 1 & $2^1$& $Q,S_1,V_2,V_1$\\
$\Ec^{(1)}_{18}$& 0 & 0 & 1 & * & 0 & 1 & 0 & 0 & $2^1$& $Q,S_1,V_2,U_2,V_1$\\
\hline
$\Ec^{(1)}_{19}$& 0 & 0 & 0 & 1 & 0 & 1 & 1 & * & $2^1$& $Q,S_1,V_1,U_1$\\
$\Ec^{(1)}_{20}$& 0 & 0 & 0 & 1 & 0 & 1 & 0 & 1 & $2^0$& $Q,S_1,V_2,V_1,U_1$\\
$\Ec^{(1)}_{21}$& 0 & 0 & 0 & 1 & 0 & 1 & 0 & 0 & $2^0$& $Q,S_1,V_2,U_2,V_1,U_1$\\
\hline
\hline
$\Ec^{(1)}_{22}$& 0 & 0 & 1 & * & 0 & 0 & 1 & * & $2^2$& $Q,S_1,Z_1,V_1$\\
$\Ec^{(1)}_{23}$& 0 & 0 & 1 & * & 0 & 0 & 0 & 1 & $2^1$& $Q,S_1,Z_1,V_2,V_1$\\
$\Ec^{(1)}_{24}$& 0 & 0 & 1 & * & 0 & 0 & 0 & 0 & $2^1$& $Q,S_1,Z_1,V_2,U_2,V_1$\\
\hline
$\Ec^{(1)}_{25}$& 0 & 0 & 0 & 1 & 0 & 0 & 1 & * & $2^1$& $Q,S_1,Z_1,V_1,U_1$\\
$\Ec^{(1)}_{26}$& 0 & 0 & 0 & 1 & 0 & 0 & 0 & 1 & $2^0$& $Q,S_1,Z_1,V_2,V_1,U_1$\\
$\Ec^{(1)}_{27}$& 0 & 0 & 0 & 1 & 0 & 0 & 0 & 0 & $2^0$& $Q,S_1,Z_1,V_2,U_2,V_1,U_1$\\
\hline
\hline
OK      & 0 & 0 & 0 & 0 & 0 & * & * & * & $2^3$& all correct \\
\hline
\end{tabular}
}
\label{table:Error events at decoder1}
\end{table}

The possible error events are listed in Table~\ref{table:Error events at decoder1}.
In Table~\ref{table:Error events at decoder1} the symbols ``1'', ``0'' and ``$\star$''
have the following meaning.
A ``1'' indicates that either the message index  or the bin index are in error.
A ``0'' indicates that both   the message index and the bin index are correct.
A ``$\star$'' indicates that it does not matter whether the message index is in error;
this is so because of superposition coding;
in this case in fact, the codeword selected by the decoder--even though with the correct message
index--is superimposed to a wrong codeword and it is thus independent of the received signal.
In case of a ``$\star$'', the factorization of the joint probability needed for
the evaluation of the probability of error is as for the case
where the message is wrong; this implies that the error event that gives the
most stringent rate bound is that for which the message is wrong
(i.e., as far as error bounds are concerned, a ``$\star$'' is equivalent to a ``1'').
The second to last column in Table~\ref{table:Error events at decoder1} counts how many error events are
included in the corresponding row (i.e., each ``$\star$'' corresponds to two possible cases).

\begin{figure}
\begin{center}
\includegraphics[width=8cm]{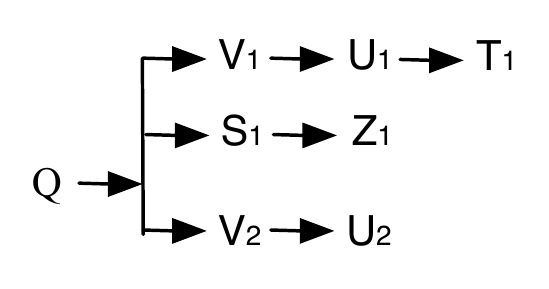}
\caption{From destination~1's perspective, conditioned on $Q$,
decoding is as for a 3-(virtual)user multiple access channel,
where each user sends a superposition of codebooks.  Here,
(virtual)user~1 sends $(V_1, U_1, T_1)$,
(virtual)user~2 sends $(S_1, Z_1)$,
and (virtual)user~3 sends $(V_2, U_2)$.}
\label{fig:decoding perspective s1 wrong}
\end{center}
\end{figure}

There are several groups of error events in Table~\ref{table:Error events at decoder1}:
For event $\Ec^{(1)}_{ 0}$: $Q$ is wrong, and hence all the decoded codewords are independent of the received signal.
For events from $\Ec^{(1)}_{ 1}$ to $\Ec^{(1)}_{12}$: $S_1$ is wrong, and thus also $Z_1$ is wrong (because superimposed to $S_1$).
For events from $\Ec^{(1)}_{13}$ to $\Ec^{(1)}_{21}$: $S_1$ is correct but $Z_1$ is wrong.
For events from $\Ec^{(1)}_{22}$ to $\Ec^{(1)}_{27}$: both $S_1$ and $Z_1$ are correct.
Notice that, because of the way codebooks are superimposed, out of the possible
$2^8-1=255$ error events, only 28 events matter.
A way to understand the error events listed
in Table~\ref{table:Error events at decoder1} is a follows.
From destination~1's perspective, given $Q$, there are three
``super-codebooks'' to decode.
Conditioned on $Q$, each ``super-codebook'' is the superposition of one or more codebooks;
each ``super-codebook'' is represented as a separate line in Fig.~\ref{fig:decoding perspective s1 wrong}.

\begin{figure}
\begin{center}
\includegraphics[width=8cm]{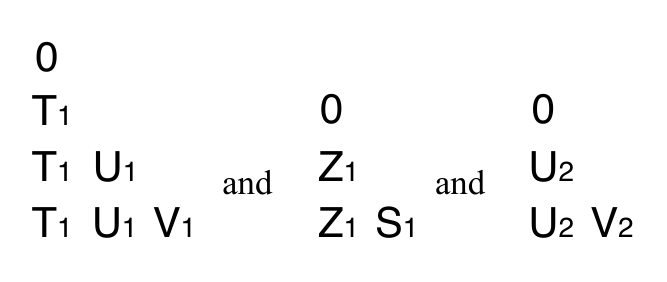}
\caption{Destination~1 must consider all possible combinations
of events that consist of jointly decoding
    a set of messages from the first  column 
and a set of messages from the second column 
and a set of messages from the third  column 
(even though not all combinations are actual errors for destination~1).}
\label{fig:all error combinations}
\end{center}
\end{figure}

Decoding proceeds as for a multiple access channel.
In particular, we need to consider all possible combinations
of events that consist of jointly decoding
    a set of messages from the first  column of Fig.~\ref{fig:all error combinations}
and a set of messages from the second column of Fig.~\ref{fig:all error combinations}
and a set of messages from the third  column of Fig.~\ref{fig:all error combinations}
(even though not all combinations are actual errors for destination~1).
In considering such ``joint-decoding events'', the messages that do not appear in
the ``set of jointly decoded messages'' must be considered as correctly decoded and
stripped form the received signal, as in a standard multiple access channel.

The last column in Table~\ref{table:Error events at decoder1} is used as follows.
Let $\mathbf{X}$ be the set of transmitted codewords (we do not write here the superscript $n$
that indicates the block-length in order to have a lighter notation), and
$\mathbf{X}'$ be the set of decoded codewords.
Let $\Cc$ be the subset of the correctly decoded message indices
such that $\mathbf{X}(\Cc)=\mathbf{X}'(\Cc)$ (recall that with superposition coding,
the decoder might select a codeword $\mathbf{X}'$ that is different from the transmitted
codeword $\mathbf{X}$ but with same message index; this
happens when an error is committed on one of the ``base layer'' codewords).
The last column of Table~\ref{table:Error events at decoder1}
lists the elements of $\Cc$, i.e., $\Cc$ contains the codewords that
have a ``0'' in the corresponding row.
The sets $\Cc$ are important for the factorization of the joint probability needed for
the evaluation of the probability of error.  The error analysis proceeds as follows.
The joint distribution of the decoded codewords and the received signal is
\begin{align*}
  &\sum_{\mathbf{X}(\Cc^c)} P_{\mathbf{X}(\Cc) \mathbf{X}(\Cc^c) \mathbf{X}'(\Cc^c) Y}
\\&=
\sum_{\mathbf{X}(\Cc^c)} P_{\mathbf{X}(\Cc)} P_{\mathbf{X}(\Cc^c)|\mathbf{X}(\Cc)}
P^{(g)}_{\mathbf{X}'(\Cc^c)| \mathbf{X}(\Cc)} P_{Y|\mathbf{X}(\Cc) \mathbf{X}(\Cc^c)}
\\&=
P^{(g)}_{\mathbf{X}'(\Cc^c)| \mathbf{X}(\Cc)} P_{\mathbf{X}(\Cc) Y},
\end{align*}
where $P$ is a distribution from the set of possible input distributions in~\reff{eq:inpdfnew}
and $P^{(g)}$ is computed from $P$ as described in the codebook generation paragraph in
Section~\ref{sec:superposition+binning}, that is,
\[
P^{(g)}_{Q V_1 U_1 T_1 S_1 Z_1 V_2 U_2}
=
P_{Q}
P_{S_1|Q}
P_{V_1 U_1 T_1|Q}
P_{Z_1|Q S_1 V_1}
P_{V_2 U_2|Q},
\]
where all the factors of $P^{(g)}$ are obtained from the corresponding marginalization of $P$.
In the following, we shall drop the prime superscript to distinguished between the wrongly decoded
codewords and the transmitted codewords. We will add a superscript ``(g)'' to the symbol for entropy
to indicate that the entropy must be evaluated by using the distribution $P^{(g)}$;
the symbol for entropy without any superscript indicates that the entropy must be evaluated by using
the distribution $P$.

Destination~1 searches for codewords that are joint typicality with the received signal
according to $P_{\mathbf{X}(\Cc) \mathbf{X}(\Cc^c) Y}$; however, assuming that the messages
in $\Cc$ are correctly decoded and those in $\Cc^c$ are wrongly decoded (this is the case that
gives the most stringent error bound), the
actual joint distribution is $P^{(g)}_{\mathbf{X}(\Cc^c)| \mathbf{X}(\Cc)} P_{\mathbf{X}(\Cc) Y}$.
The probability of the error for the messages in $\Cc^c$
(neglecting the terms that will eventually be taken to go to zero) is:
\begin{align*}
&\Pr[{\rm error}\ \Cc^c]
=\sum_{\mathbf{x}\in T^{(n)}_{\epsilon}(P_{\mathbf{X}(\Cc) \mathbf{X}(\Cc^c) Y}|\mathbf{X}(\Cc))}
   P^{(g)}_{\mathbf{X}(\Cc^c)| \mathbf{X}(\Cc)} P_{Y|\mathbf{X}(\Cc) }
\\&\leq \exp\Big(n \Big[R(\Cc^c)
         +H(\mathbf{X} (\Cc^c)| \mathbf{X}(\Cc))+H(Y|\mathbf{X}(\Cc) \mathbf{X}(\Cc^c))
\\&\quad -H^{(g)}(\mathbf{X}(\Cc^c)| \mathbf{X}(\Cc))-H(Y|\mathbf{X}(\Cc)) \Big]\Big)
\\&= \exp\Big(n \Big[R(\Cc^c)
         -I(Y\wedge \mathbf{X}(\Cc^c) |\mathbf{X}(\Cc))
         +H(\mathbf{X} (\Cc^c), \mathbf{X}(\Cc))-H^{(g)}(\mathbf{X}(\Cc^c), \mathbf{X}(\Cc))
\\&\quad -H(\mathbf{X}(\Cc)) +H^{(g)}(\mathbf{X}(\Cc)) \Big]\Big)
\end{align*}
where $R(\Cc^c)$ is the sum of the rates corresponding to the wrongly decoded messages
that are indexed by $\Cc^c$.
In the following, for any two distributions $P$ and $Q$, the notation $\E[\log(P/Q)]$ stands for the
Kullback-Leibler divergence $D(P||Q)$. 
Let
\begin{align*}
\Delta^{(1)}
  &\defeq H^{(g)}(\mathbf{X}(\Cc^c), \mathbf{X}(\Cc)) - H(\mathbf{X}(\Cc^c), \mathbf{X}(\Cc))
 =
 \E\left[\log \frac{      P_{\mathbf{X}(\Cc^c\cup\Cc)}}
                   {P^{(g)}_{\mathbf{X}(\Cc^c\cup\Cc)}} \right]
\\&=
\E\left[\log \frac{P_{Q V_1 U_1 T_1 S_1 Z_1 V_2 U_2}}{P_{Q} P_{V_1 U_1 T_1|Q} P_{S_1|Q} P_{Z_1|Q S_1 V_1} P_{V_2 U_2|Q}}\right]
\\&=
\E\left[\log \frac{P_{S_1|Q V_1 U_1 T_1} P_{Z_1|Q V_1 U_1 T_1 S_1} P_{V_2 U_2|Q V_1 U_1 T_1 S_1 Z_1}}{P_{S_1|Q} P_{Z_1|Q S_1 V_1} P_{V_2 U_2|Q}}\right]
\\&= I(S_1\wedge V_1 U_1 T_1|Q)
    +I(Z_1\wedge U_1 T_1|Q S_1 V_1)
    +I(V_2 U_2\wedge V_1 U_1 T_1 S_1 Z_1|Q)
\end{align*}   
Finally, $\Pr[{\rm error}\ \Cc^c]\to 0 $ as $n\to\infty$ if
\begin{align*}
R(\Cc^c)
  &\leq I(Y\wedge \mathbf{X}(\Cc^c) |\mathbf{X}(\Cc)) + \Delta^{(1)}
-\underbrace{
 \E\left[\log \frac{      P_{\mathbf{X}(\Cc)}}
                   {P^{(g)}_{\mathbf{X}(\Cc)}} \right]}_{\defeq\Delta^{(1)}_{\Cc}}.
\end{align*}

We now evaluate $\Delta^{(1)}_{\Cc}$ for all possible error events in  Table~\ref{table:Error events at decoder1}.
For $\Ec^{(1)}_{ 0}$: $Q$ is wrong and hence--because of superposition encoding--the most stringent
error event is when the messages carried by $Q$ and all the messages superimposed
to $Q$ are wrong. In this case $\Cc=\emptyset$ and thus $\Delta^{(1)}_{\emptyset}=0$.
It can be also easily verified that
\begin{align*}
\underbrace{\Delta^{(1)}_{\emptyset}}_{\text{for}\ \Ec^{(1)}_{ 0}}
  &=\underbrace{\Delta^{(1)}_{\{Q\}}}_{\text{for}\ \Ec^{(1)}_{ 1}}
   =\underbrace{\Delta^{(1)}_{\{Q,V_2\}}}_{\text{for}\ \Ec^{(1)}_{ 2}}
   =\underbrace{\Delta^{(1)}_{\{Q,V_2,U_2\}}}_{\text{for}\ \Ec^{(1)}_{ 3}}
   =\underbrace{\Delta^{(1)}_{\{Q,V_1\}}}_{\text{for}\ \Ec^{(1)}_{ 4}}
\\&=\underbrace{\Delta^{(1)}_{\{Q,V_1,U_1\}}}_{\text{for}\ \Ec^{(1)}_{ 7}}
   =\underbrace{\Delta^{(1)}_{\{Q,V_1,U_1,T_1\}}}_{\text{for}\ \Ec^{(1)}_{10}}
   =\underbrace{\Delta^{(1)}_{\{Q,S_1\}}}_{\text{for}\ \Ec^{(1)}_{13}}
   =0,
\end{align*}
because in these cases $P_{\mathbf{X}(\Cc)}=P^{(g)}_{\mathbf{X}(\Cc)}$.
Then we have:
\begin{align*}
\Ec^{(1)}_{ 5} &:{\Delta^{(1)}_{Q,V_1,V_2}}
     =  \E\left[\log \frac{P_{Q V_1 V_2}}
                         {P_{Q}P_{V_1|Q}P_{V_2|Q}} \right]
  \\&= I(V_1\wedge V_2|Q),
\\
\Ec^{(1)}_{ 6} &:{\Delta^{(1)}_{Q,V_1,V_2,U_2}}_{ }
     =  \E\left[\log \frac{P_{Q V_1 V_2 U_2}}
                         {P_{Q}P_{V_1|Q}P_{V_2U_2|Q}} \right]
  \\&= I(V_1\wedge V_2,U_2|Q),
\\
\Ec^{(1)}_{ 8} &:{\Delta^{(1)}_{Q,V_1,U_1,V_2}}
     =  \E\left[\log \frac{P_{Q V_1 U_1 V_2}}
                         {P_{Q}P_{V_1 U_1|Q}P_{V_2|Q}} \right]
  \\&= I(V_1,U_1\wedge V_2|Q),
\\
\Ec^{(1)}_{ 9} &:{\Delta^{(1)}_{Q,V_1,U_1,V_2,U_2}}
     =  \E\left[\log \frac{P_{Q V_1 U_1 V_2 U_2}}
                         {P_{Q}P_{V_1 U_1|Q}P_{V_2 U_2|Q}} \right]
  \\&= I(V_1,U_1\wedge V_2,U_2|Q),
\\
\Ec^{(1)}_{11} &:{\Delta^{(1)}_{Q,V_1,U_1,T_1,V_2}}
     =  \E\left[\log \frac{P_{Q V_1 U_1 T_1 V_2}}
                         {P_{Q}P_{V_1 U_1 T_1|Q}P_{V_2|Q}} \right]
  \\&= I(V_1,U_1,T_1\wedge V_2|Q),
\\
\Ec^{(1)}_{12} &:{\Delta^{(1)}_{Q,V_1,U_1,T_1,V_2,U_2}}
     =  \E\left[\log \frac{P_{Q V_1 U_1 T_1 V_2 U_2}}
                         {P_{Q}P_{V_1 U_1 T_1|Q}P_{V_2 U_2|Q}} \right]
  \\&= I(V_1,U_1,T_1\wedge V_2,U_2|Q),
\\
\Ec^{(1)}_{14} &:{\Delta^{(1)}_{Q,S_1,V_2}}
     =  \E\left[\log \frac{P_{Q S_1 V_2}}
                         {P_{Q}P_{S_1|Q}P_{V_2|Q}} \right]
  \\&= I(S_1\wedge V_2|Q),
\\
\Ec^{(1)}_{15} &:{\Delta^{(1)}_{Q,S_1,V_2,U_2}}
     =  \E\left[\log \frac{P_{Q S_1 V_2 U_2}}
                         {P_{Q}P_{S_1|Q}P_{V_2 U_2|Q}} \right]
  \\&= I(S_1\wedge V_2,U_2|Q),
\end{align*}
and
\begin{align*}
\Ec^{(1)}_{16} &:{\Delta^{(1)}_{Q,S_1,V_1}}
    =  \E\left[\log \frac{P_{Q S_1 V_1}}
                         {P_{Q}P_{S_1|Q}P_{V_1|Q}} \right]
 \\&= I(S_1\wedge V_1|Q),
\\
\Ec^{(1)}_{17} &:{\Delta^{(1)}_{Q,S_1,V_1,V_2}}
     =  \E\left[\log \frac{P_{Q S_1 V_1 V_2}}
                         {P_{Q}P_{S_1|Q}P_{V_1|Q}P_{V_2|Q}} \right]
     =  \E\left[\log \frac{P_{V_1|Q S_1}P_{V_2|Q S_1 V_1}}
                         {P_{V_1|Q}P_{V_2|Q}} \right]
  \\&= I(S_1\wedge V_1|Q)+I(S_1,V_1\wedge V_2|Q),
\\
\Ec^{(1)}_{18} &:{\Delta^{(1)}_{Q,S_1,V_1,V_2,U_2}}
   =  \E\left[\log \frac{P_{Q S_1 V_1 V_2 U_2}}
                         {P_{Q}P_{S_1|Q}P_{V_1|Q}P_{V_2 U_2|Q}} \right]
  \\&= I(S_1\wedge V_1|Q)+I(S_1,V_1\wedge V_2,U_2|Q),
\end{align*}
and
\begin{align*}
\Ec^{(1)}_{19} &:{\Delta^{(1)}_{Q,S_1,V_1,U_1}}
   =  \E\left[\log \frac{P_{Q S_1 V_1 U_1}}
                         {P_{Q}P_{S_1|Q}P_{V_1 U_1|Q}} \right]
 \\&= I(S_1\wedge V_1,U_1|Q),
\\
\Ec^{(1)}_{20} &:{\Delta^{(1)}_{Q,S_1,V_1,U_1,V_2}}
    =  \E\left[\log \frac{P_{Q S_1 V_1 U_1 V_2}}
                         {P_{Q}P_{S_1|Q}P_{V_1 U_1|Q}P_{V_2|Q}} \right]
 \\&= I(S_1\wedge V_1,U_1|Q)+I(S_1,V_1,U_1\wedge V_2|Q),
\\
\Ec^{(1)}_{21} &:{\Delta^{(1)}_{Q,S_1,V_1,U_1,V_2,U_2}}
    =  \E\left[\log \frac{P_{Q S_1 V_1 U_1 V_2 U_2}}
                         {P_{Q}P_{S_1|Q}P_{V_1 U_1|Q}P_{V_2 U_2|Q}} \right]
 \\&= I(S_1\wedge V_1,U_1|Q)+I(S_1,V_1,U_1\wedge V_2,U_2|Q),
\end{align*}
and
\begin{align*}
\Ec^{(1)}_{22} &:{\Delta^{(1)}_{Q,S_1,Z_1,V_1}}
    =  \E\left[\log \frac{P_{Q S_1 Z_1 V_1}}
                         {P_{Q}P_{S_1|Q}P_{V_1|Q}P_{Z_1|Q S_1 V_1}} \right]
    =  \E\left[\log \frac{P_{V_1|Q S_1}}
                         {P_{V_1|Q}} \right]
 \\&= I(S_1\wedge V_1|Q),
\\
\Ec^{(1)}_{23} &:{\Delta^{(1)}_{Q,S_1,Z_1,V_2,V_1}}
    =  \E\left[\log \frac{P_{Q S_1 Z_1 V_1 V_2}}
                         {P_{Q}P_{S_1|Q}P_{V_1|Q}P_{Z_1|Q S_1 V_1}P_{V_2|Q}} \right]
    =  \E\left[\log \frac{P_{V_1|Q S_1}P_{V_2|Q S_1 Z_1 V_1}}
                         {P_{V_1|Q}P_{V_2|Q}} \right]
\\&= I(S_1\wedge V_1|Q)+I(V_2\wedge S_1, Z_1, V_1|Q),
\\
\Ec^{(1)}_{24} &:{\Delta^{(1)}_{Q,S_1,Z_1,V_2,U_2,V_1}}
     =  \E\left[\log \frac{P_{Q S_1 Z_1 V_1 V_2 U_2}}
                         {P_{Q}P_{S_1|Q}P_{V_1|Q}P_{Z_1|Q S_1 V_1}P_{V_2 U_2|Q}} \right]
\\&= I(S_1\wedge V_1|Q)+I(V_2, U_2\wedge S_1, Z_1, V_1|Q),
\end{align*}
and
\begin{align*}
\Ec^{(1)}_{25} &:{\Delta^{(1)}_{Q,S_1,Z_1,V_1,U_1}}
   =  \E\left[\log \frac{P_{Q S_1 Z_1 V_1 U_1}}
                         {P_{Q}P_{S_1|Q}P_{V_1 U_1|Q}P_{Z_1|Q S_1 V_1}} \right]
   =  \E\left[\log \frac{P_{V_1|Q S_1} P_{U_1|Q S_1 V_1 Z_1}}
                         {P_{V_1|Q}     P_{U_1|Q V_1}        } \right]
\\&= I(S_1\wedge V_1|Q)+I(U_1\wedge S_1, Z_1|Q, V_1),
\\
\Ec^{(1)}_{26} &:{\Delta^{(1)}_{Q,S_1,Z_1,V_2,V_1,U_1}}
   =  \E\left[\log \frac{P_{Q S_1 Z_1 V_1 U_1 V_2}}
                         {P_{Q}P_{S_1|Q}P_{V_1 U_1|Q}P_{Z_1|Q S_1 V_1}P_{V_2|Q}} \right]
\\&= I(S_1\wedge V_1|Q)+I(U_1\wedge S_1, Z_1|Q, V_1)+I(V_2\wedge S_1, V_1, Z_1, U_1|Q),
\\
\Ec^{(1)}_{27} &:{\Delta^{(1)}_{Q,S_1,Z_1,V_2,U_2,V_1,U_1}}
  =  \E\left[\log \frac{P_{Q S_1 Z_1 V_1 U_1 V_2 U_2}}
                         {P_{Q}P_{S_1|Q}P_{V_1 U_1|Q}P_{Z_1|Q S_1 V_1}P_{V_2 U_2|Q}} \right]
\\&= I(S_1\wedge V_1|Q)+I(U_1\wedge S_1, Z_1|Q, V_1)+I(V_2, U_2\wedge S_1, V_1, Z_1, U_1|Q).
\end{align*}

Let
\begin{align*}
                                    &\quad R_{Q}   = R_{10c}+R_{20c} \\
R_{V_1}^{\prime} = R_{10c}^{\prime},&\quad R_{V_1} = R_{10c}+R_{10c}^{\prime}\\
R_{V_2}^{\prime} = R_{20c}^{\prime},&\quad R_{V_2} = R_{20c}+R_{20c}^{\prime}\\
                                    &\quad R_{U_1} = R_{10n}+R_{10n}^{\prime}\\
                                    &\quad R_{T_1} = R_{11n}+R_{11n}^{\prime}\\
R_{S_1}^{\prime} = R''_{11c},       &\quad R_{S_1} = R_{11c}+R''_{11c}       \\
R_{Z_1}^{\prime} = R_{11c}^{\prime},&\quad R_{Z_1} = R_{11c}+R_{11c}^{\prime}\\
                                    &\quad R_{U_2} = R_{20n}+R_{20n}^{\prime}\\
                                    &\quad R_{T_2} = R_{22n}+R_{22n}^{\prime}\\
R_{S_2}^{\prime} = R''_{22c},       &\quad R_{S_2} = R_{22c}+R''_{22c}       \\
R_{Z_2}^{\prime} = R_{22c}^{\prime},&\quad R_{Z_2} = R_{22c}+R_{22c}^{\prime},
\end{align*}
and hence
\begin{align*}
R_{V_1}+R_{V_2}&= R_{Q}+(R_{V_1}^{\prime}+R_{V_2}^{\prime})\\
R_{Z_1}&=R_{S_1}-R_{S_1}^{\prime}+R_{Z_1}^{\prime}\\
R_{Z_2}&=R_{S_2}-R_{S_2}^{\prime}+R_{Z_2}^{\prime}.
\end{align*}

Recall that we defined
\[
\Delta^{(1)} =
     I(S_1\wedge V_1, U_1, T_1|Q)
    +I(Z_1\wedge U_1, T_1|Q, S_1, V_1)
    +I(V_2, U_2\wedge V_1, U_1, T_1, Z_1|Q, S_1).
\]
In addition to the rate constraints from cooperation among the sources (see~\reff{eq:enc1-pe-gen all}), i.e.,
\begin{align*}
           R_{Z_1} &\leq C^{(1)}_{1}=I(Z_1\wedge  Y_2|\underline{X}_2,V_1)+I(Z_1\wedge  S_2|Q,S_1,V_1)
\\ R_{V_1}+R_{Z_1} &\leq C^{(1)}_{2}=I(V_1,Z_1\wedge  Y_2|\underline{X}_2)+I(Z_1\wedge  S_2|Q,S_1,V_1)+I(V_1\wedge  S_1,S_2|Q)
\\         R_{Z_2} &\leq C^{(2)}_{1}=I(Z_2\wedge  Y_1|\underline{X}_1,V_2)+I(Z_2\wedge  S_1|Q,S_2,V_2)
\\ R_{V_2}+R_{Z_2} &\leq C^{(2)}_{2}=I(V_2,Z_2\wedge  Y_1|\underline{X}_1)+I(Z_2\wedge  S_1|Q,S_2,V_2)+I(V_2\wedge  S_1,S_2|Q)
\end{align*}
we have the following rate constraints arising from decoding at destination~1:
\begin{align*}
\Ec^{(1)}_{ 0}&:R_{U_1}+R_{T_1}+
             \underbrace{[R_{S_1}+R'_{Z_1}-R'_{S_1}]}_{R_{Z_1}}
           && +R_{U_2}
              +\underbrace{[R_{Q}+R'_{V_1}+R'_{V_2}]}_{R_{V_1}+R_{V_2}}    = \nonumber\\
    &=R_{U_1}+R_{T_1}+R_{Z_1}+R_{U_2}+R_{V_1}+R_{V_2} &&\leq E^{(1)}_{ 0}=I(Y_3\wedge Q, V_1, U_1, T_1, S_1, Z_1, V_2, U_2)    -(R'_{S_1})        \\
    &&&+\Delta^{(1)} 
\end{align*}
%
\begin{align*}
\Ec^{(1)}_{ 1} &:       R_{U_1}+R_{T_1}+R_{Z_1}+R_{U_2} &&\leq E^{(1)}_{ 1}=I(Y_3\wedge    V_1, U_1, T_1, S_1, Z_1, V_2, U_2 | Q)            -(R'_{V_1}+R'_{S_1}+R'_{V_2})\\
    &&&+\Delta^{(1)}\\ 
\Ec^{(1)}_{ 2} &:       R_{U_1}+R_{T_1}+R_{Z_1}+R_{U_2} &&\leq E^{(1)}_{ 2}=I(Y_3\wedge    V_1, U_1, T_1, S_1, Z_1,      U_2 | Q, V_2)       -(R'_{V_1}+R'_{S_1})\\
    &&&+\Delta^{(1)}\\ 
\Ec^{(1)}_{ 3} &:       R_{U_1}+R_{T_1}+R_{Z_1}         &&\leq E^{(1)}_{ 3}=I(Y_3\wedge    V_1, U_1, T_1, S_1, Z_1           | Q, V_2, U_2)  -(R'_{V_1}+R'_{S_1})\\
    &&&+\Delta^{(1)} 
\end{align*}
\begin{align*}
\Ec^{(1)}_{ 4} &:       R_{U_1}+R_{T_1}+R_{Z_1}+R_{U_2} &&\leq E^{(1)}_{ 4}=I(Y_3\wedge         U_1, T_1, S_1, Z_1, V_2, U_2 | Q, V_1)           -(R'_{S_1}+R'_{V_2})\\
    &&&+\Delta^{(1)}\\ 
\Ec^{(1)}_{ 5} &:       R_{U_1}+R_{T_1}+R_{Z_1}+R_{U_2} &&\leq E^{(1)}_{ 5}=I(Y_3\wedge         U_1, T_1, S_1, Z_1,      U_2 | Q, V_1, V_2)      -(R'_{S_1})\\
    &&&+\Delta^{(1)}-I(V_1\wedge V_2|Q)\\ 
\Ec^{(1)}_{ 6} &:       R_{U_1}+R_{T_1}+R_{Z_1}         &&\leq E^{(1)}_{ 6}=I(Y_3\wedge         U_1, T_1, S_1, Z_1           | Q, V_1, V_2, U_2) -(R'_{S_1})\\
    &&&+\Delta^{(1)}-I(V_1\wedge V_2, U_2|Q)\\ 
\end{align*}
\begin{align*}
\Ec^{(1)}_{ 7} &:               R_{T_1}+R_{Z_1}+R_{U_2} &&\leq E^{(1)}_{ 7}=I(Y_3\wedge              T_1, S_1, Z_1, V_2, U_2 | Q, V_1, U_1 )          -(R'_{S_1}+R'_{V_2})\\
    &&&+\Delta^{(1)}\\ 
\Ec^{(1)}_{ 8} &:               R_{T_1}+R_{Z_1}+R_{U_2} &&\leq E^{(1)}_{ 8}=I(Y_3\wedge              T_1, S_1, Z_1,      U_2 | Q, V_1, U_1, V_2)      -(R'_{S_1})\\
    &&&+\Delta^{(1)}-I(V_1, U_1\wedge V_2|Q)\\ 
\Ec^{(1)}_{ 9} &:               R_{T_1}+R_{Z_1}         &&\leq E^{(1)}_{ 9}=I(Y_3\wedge              T_1, S_1, Z_1           | Q, V_1, U_1, V_2, U_2) -(R'_{S_1})\\
    &&&+\Delta^{(1)}-I(V_1, U_1\wedge V_2, U_2|Q)
\end{align*}
\begin{align*}
\Ec^{(1)}_{10} &:                       R_{Z_1}+R_{U_2} &&\leq E^{(1)}_{10}=I(Y_3\wedge                   S_1, Z_1, V_2, U_2 | Q, V_1, U_1, T_1)           -(R'_{S_1}+R'_{V_2}) \\
    &&&+\Delta^{(1)}\\ 
\Ec^{(1)}_{11} &:                       R_{Z_1}+R_{U_2} &&\leq E^{(1)}_{11}=I(Y_3\wedge                   S_1, Z_1,      U_2 | Q, V_1, U_1, T_1, V_2)      -(R'_{S_1})\\
    &&&+\Delta^{(1)}-I(V_1, U_1, T_1\wedge V_2|Q)\\
\Ec^{(1)}_{12} &:                       R_{Z_1}         &&\leq E^{(1)}_{12}=I(Y_3\wedge                   S_1, Z_1           | Q, V_1, U_1, T_1, V_2, U_2) -(R'_{S_1})\\
    &&&+\Delta^{(1)}-I(V_1, U_1, T_1\wedge V_2, U_2|Q)
\end{align*}
%
\begin{align*}
\Ec^{(1)}_{13} &:       R_{U_1}+R_{T_1}        +R_{U_2} &&\leq E^{(1)}_{13}=I(Y_3\wedge    V_1, U_1, T_1,      Z_1, V_2, U_2 | Q, S_1)           -(R'_{V_1}+R'_{Z_1}+R'_{V_2})\\
    &&&+\Delta^{(1)},\\
\Ec^{(1)}_{14} &:       R_{U_1}+R_{T_1}        +R_{U_2} &&\leq E^{(1)}_{14}=I(Y_3\wedge    V_1, U_1, T_1,      Z_1,      U_2 | Q, S_1, V_2)      -(R'_{V_1}+R'_{Z_1})\\
    &&&+\Delta^{(1)}-I(S_1\wedge V_2|Q),\\
\Ec^{(1)}_{15} &:       R_{U_1}+R_{T_1}                 &&\leq E^{(1)}_{15}=I(Y_3\wedge    V_1, U_1, T_1,      Z_1           | Q, S_1, V_2, U_2) -(R'_{V_1}+R'_{Z_1})\\
    &&&+\Delta^{(1)}-I(S_1\wedge V_2, U_2|Q),
\end{align*}
\begin{align*}
\Ec^{(1)}_{16} &:       R_{U_1}+R_{T_1}        +R_{U_2} &&\leq E^{(1)}_{16}=I(Y_3\wedge         U_1, T_1,      Z_1, V_2, U_2 | Q, S_1, V_1)           -(R'_{Z_1}+R'_{V_2})\\
    &&&+\Delta^{(1)}-I(S_1\wedge V_1|Q),\\
\Ec^{(1)}_{17} &:       R_{U_1}+R_{T_1}        +R_{U_2} &&\leq E^{(1)}_{17}=I(Y_3\wedge         U_1, T_1,      Z_1,      U_2 | Q, S_1, V_1, V_2)      -(R'_{Z_1})\\
    &&&+\Delta^{(1)}-I(S_1\wedge V_1|Q)-I(S_1, V_1\wedge V_2|Q),\\
\Ec^{(1)}_{18} &:       R_{U_1}+R_{T_1}                 &&\leq E^{(1)}_{18}=I(Y_3\wedge         U_1, T_1,      Z_1           | Q, S_1, V_1, V_2, U_2) -(R'_{Z_1})\\
    &&&+\Delta^{(1)}-I(S_1\wedge V_1|Q)-I(S_1, V_1\wedge V_2, U_2|Q),
\end{align*}
\begin{align*}
\Ec^{(1)}_{19} &:               R_{T_1}        +R_{U_2} &&\leq E^{(1)}_{19}=I(Y_3\wedge              T_1,      Z_1, V_2, U_2 | Q, S_1, V_1, U_1 )          -(R'_{Z_1}+R'_{V_2})\\
    &&&+\Delta^{(1)}-I(S_1\wedge V_1, U_1|Q),\\
\Ec^{(1)}_{20} &:               R_{T_1}        +R_{U_2} &&\leq E^{(1)}_{20}=I(Y_3\wedge              T_1,      Z_1,      U_2 | Q, S_1, V_1, U_1, V_2)      -(R'_{Z_1})\\
    &&&+\Delta^{(1)}-I(S_1\wedge V_1, U_1|Q)-I(S_1, V_1, U_1\wedge V_2|Q),\\
\Ec^{(1)}_{21} &:               R_{T_1}                 &&\leq E^{(1)}_{21}=I(Y_3\wedge              T_1,      Z_1           | Q, S_1, V_1, U_1, V_2, U_2) -(R'_{Z_1})\\
    &&&+\Delta^{(1)}-I(S_1\wedge V_1, U_1|Q)-I(S_1, V_1, U_1\wedge V_2, U_2|Q),
\end{align*}
\begin{align*}
\Ec^{(1)}_{22} &:       R_{U_1}+R_{T_1}        +R_{U_2} &&\leq E^{(1)}_{22}=I(Y_3\wedge         U_1, T_1,           V_2, U_2 | Q, S_1, Z_1, V_1)      -(R'_{V_2})\\
    &&&+\Delta^{(1)}-I(S_1\wedge V_1|Q),\\
\Ec^{(1)}_{23} &:       R_{U_1}+R_{T_1}        +R_{U_2} &&\leq E^{(1)}_{23}=I(Y_3\wedge         U_1, T_1,                U_2 | Q, S_1, Z_1, V_1, V_2)      \\
    &&&+\Delta^{(1)}-I(S_1\wedge V_1|Q)-I(S_1, Z_1, V_1\wedge V_2|Q),\\
\Ec^{(1)}_{24} &:       R_{U_1}+R_{T_1}                 &&\leq E^{(1)}_{24}=I(Y_3\wedge         U_1, T_1                     | Q, S_1, Z_1, V_1, V_2, U_2) \\
    &&&+\Delta^{(1)}-I(S_1\wedge V_1|Q)-I(S_1, Z_1, V_1\wedge V_2, U_2|Q),
\end{align*}
\begin{align*}
\Ec^{(1)}_{25} &:               R_{T_1}        +R_{U_2} &&\leq E^{(1)}_{25}=I(Y_3\wedge              T_1,           V_2, U_2 | Q, S_1, Z_1, V_1, U_1 ) -(R'_{V_2})\\
    &&&+\Delta^{(1)}-I(S_1\wedge V_1|Q)-I(S_1, Z_1\wedge U_1|Q),\\
\Ec^{(1)}_{26} &:               R_{T_1}        +R_{U_2} &&\leq E^{(1)}_{26}=I(Y_3\wedge              T_1,                U_2 | Q, S_1, Z_1, V_1, U_1, V_2)       \\
    &&&+\Delta^{(1)}-I(S_1\wedge V_1|Q)-I(S_1, Z_1\wedge U_1|Q)-I(S_1, Z_1, V_1, U_1\wedge V_2|Q),\\
\Ec^{(1)}_{27} &:               R_{T_1}                 &&\leq E^{(1)}_{27}=I(Y_3\wedge              T_1                     | Q, S_1, Z_1, V_1, U_1, V_2, U_2) \\
    &&&+\Delta^{(1)}-I(S_1\wedge V_1|Q)-I(S_1, Z_1\wedge U_1|Q)-I(S_1, Z_1, V_1, U_1\wedge V_2, U_2|Q).
\end{align*}
All the above constraints combined give the region in~\reff{eq:dec1-pe-gen all}.

Subsets of the above achievable region with fewer rate constraints can be obtained as follows:
 \begin{itemize}

 \item
 If $R_{V_1}^{\prime}=R_{V_2}^{\prime}=0$, that is, $V_1$ and $V_2$ are
 not binned against the known interference, then $V_1$ and $V_2$ are correct whenever $Q$ is correct. In
 this case, 16 of the 31 error events listed in Table~\ref{table:Error events at decoder1} are
 impossible (all those for which the bin index in either $V_1$ or $V_2$ is wrong).

 \item
 If $R_{Z_1}^{\prime}=0$ (similar observation can be made if $R_{Z_2}^{\prime}=0$), that is, $Z_1$ is not binned against the known interference, then  $Z_1$ is correct whenever $Q$ and $V_1$ are correct. In this case, the 9
 error events from $\Ec^{(1)}_{13}$ to $\Ec^{(1)}_{21}$ listed in Table~\ref{table:Error events at decoder1}
 are impossible and the achievable region becomes
  \begin{subequations}
 \begin{align}
 R_{V_1}+R_{V_2}+R_{U_1}+R_{T_1}+R_{U_2}&\leq E^{(1)}_{ 0}\\
       R_{U_1}+R_{T_1}+R_{U_2} &\leq
 \min\{E^{(1)}_{ 1},E^{(1)}_{ 2},E^{(1)}_{ 4},E^{(1)}_{ 5}, E^{(1)}_{13},E^{(1)}_{14},E^{(1)}_{16},E^{(1)}_{17},E^{(1)}_{22},E^{(1)}_{23}\} \\
       R_{U_1}+R_{T_1}    &\leq \min\{E^{(1)}_{ 3},E^{(1)}_{ 6}, E^{(1)}_{15},E^{(1)}_{18},E^{(1)}_{24}\}\\
 R_{T_1}+R_{U_2} &\leq \min\{E^{(1)}_{ 7},E^{(1)}_{ 8} E^{(1)}_{19},E^{(1)}_{20},E^{(1)}_{25},E^{(1)}_{26}\}\\
 R_{T_1}         &\leq \min\{E^{(1)}_{ 9} E^{(1)}_{21},E^{(1)}_{27}\},\\ %
       R_{U_2} &\leq \min\{E^{(1)}_{10},E^{(1)}_{11}\}\\ %
       R_{Z_1}        &\leq E^{(1)}_{12}
 \label{eq:dec1-pe-gen all R_{Z1}=0}
 \end{align}
 \end{subequations}
 with only five rate constraints, as for the case of superposition only.
 Notice that the rate bound on $R_{U_2}$ can be removed since an error on
 $U_2$ alone is not an error from the point of view of source~1.

 \item
 Instead of joint decoding of all the messages, one can perform a two-step decoding as follows.
 First step: decode $Q$ and $S_1$ jointly, and then strip them from the received signal. This the first decoding step is successful if
 \begin{align*}
       R_{S_1} &\leq I(Y_3\wedge S_1|Q) \\
       R_{Q}+R_{S_1} &\leq I(Y_3\wedge S_1,Q).
 \end{align*}

 Second step: jointly decode all the other messages. For this second step, one only needs to consider the error
 events from $\Ec^{(1)}_{13}$ to $\Ec^{(1)}_{27}$.

 \end{itemize}

\end{document}